\documentclass[aip,amsmath,amssymb,reprint]{revtex4-1}
\draft
\pdfoutput=1

\usepackage{graphicx}
\usepackage{dcolumn}

\usepackage[export]{adjustbox}
\usepackage[caption=false]{subfig}
\usepackage[capitalize]{cleveref}

\usepackage{siunitx}
\usepackage{xfrac}
\usepackage{multirow}

\graphicspath{{./img/}}
\DeclareGraphicsExtensions{.pdf}

\usepackage{macros}	
\newcommand{\bibfile}{ss-study}

\begin{document}

\title{Limits of Adiabatic Clocking in Quantum-dot Cellular Automata}

\author{Jacob Retallick}
\email{jret@ece.ubc.ca}
\affiliation{Department of Electrical and Computer Engineering, University of British Columbia}

\author{Konrad Walus}
\email{konradw@ece.ubc.ca}
\affiliation{Department of Electrical and Computer Engineering, University of British Columbia}

\date{\today}


\begin{abstract}
  Ultimate bounds on the maximum operating frequency of networks of quantum dot cellular automata devices have yet to be established. We consider the adiabaticity of such networks in the two-state approximation where clocking is achieved via modulation of the inter-dot tunneling barriers. Estimates of the maximum operating frequency that would allow a 99\% probability of observing the correct logical output are presented for a subset of the basic components used in QCA network design. Simulations are performed both in the coherent limit and for a simple dissipative model. We approach the problem of tunnel-based clocking from the perspective of quantum annealing, and present an improved clocking schedule allowing for faster operation. Using an analytical solution for driven QCA wires, we show that the maximum operating frequency in the coherent limit falls off with the square of the wire length, potentially limiting the size of clocked regions. 
\end{abstract}

\pacs{}

\maketitle


\section{Introduction}

In recent years, there has been great interest in technologies that extend beyond the projected scale limits of conventional CMOS, ranging from new transistor designs with alternate channels \cite{nourbakhsh2016, qiu2017} to entirely novel computational architectures \cite{bhowmik2013, fong2016, wolkow2014}. Quantum-dot Cellular Automata (QCA) encodes binary information in the distribution of charges in devices or \emph{cells} composed of arrays of quantum dots \cite{lent1993, tougaw1994}. Coulombic interactions between occupying charges facilitate coupling between the charge states of neighbouring cells. Arrangements of these cells can be designed with ground states that encode familiar logic gates \cite{amlani1999}. Among the most promising potential implementations for QCA are mixed-valence molecular devices \cite{lent2003, christie2015} and patterned dangling bonds on hydrogen passivated silicon \cite{wolkow2014, huff2018}. Each QCA cell occupies only a few \SI{}{\nano\meter^2} in area, potentially offering high device densities of \SI{e14}{\centi\meter^{-2}}. Significant challenges must be solved for any realistic QCA implementation, such as limiting device power at high density using reversible gates \cite{timler2002, lent2006}, designing robust wire crossings \cite{abedi2015} and clocking networks \cite{campos2016, blair2018}, and interfacing with the existing CMOS architecture.

\begin{figure}
  \newcommand{\Height}{.2\linewidth}
  \subfloat[2-State]{\includegraphics[height=\Height]{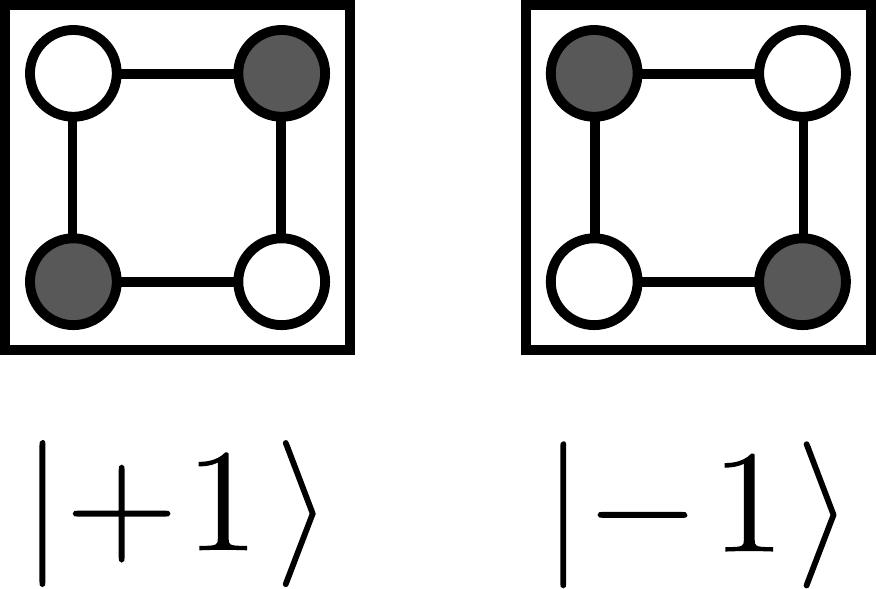}}\hspace{5ex}
  \subfloat[3-State]{\includegraphics[height=\Height]{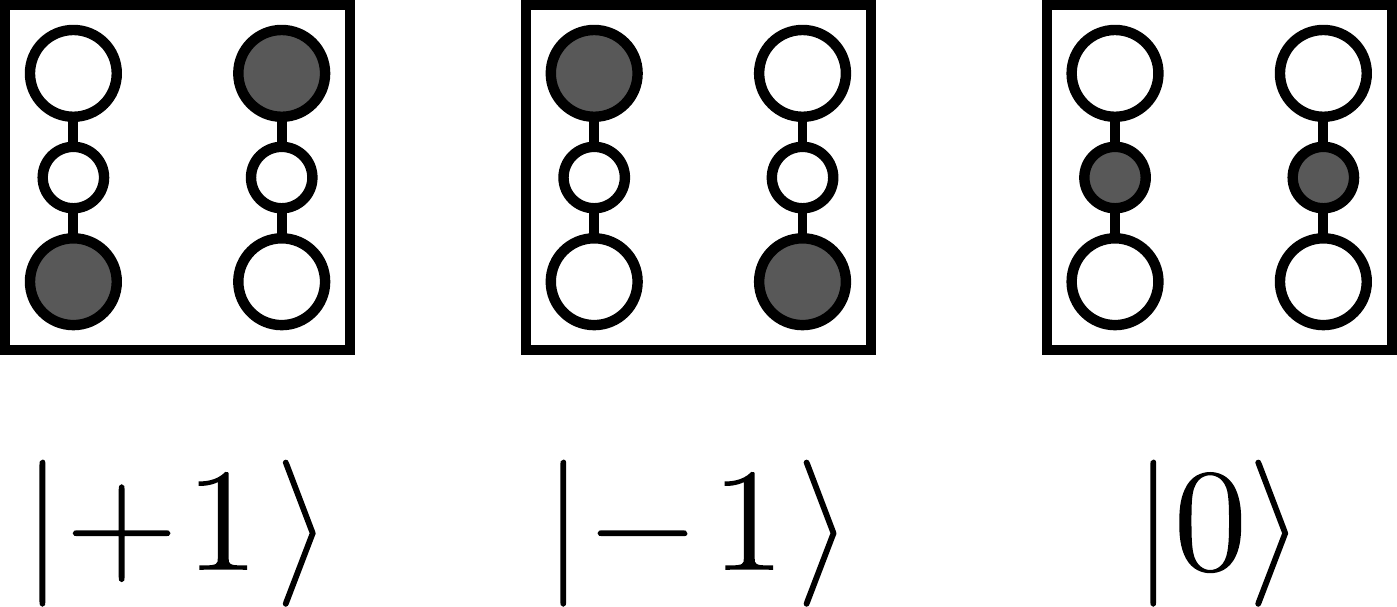}}
  \caption{Schematics and polarizations for common QCA devices. The shaded quantum dots are occupied by additional charges. Inter-dot tunneling paths are indicated by the lines.}
  \label{fig: qca-cells}
\end{figure}

\begin{figure}
  \newcommand{\Height}{.35\linewidth}
  \subfloat[Schematic of 4-phase clocking]{\includegraphics[height=\Height]{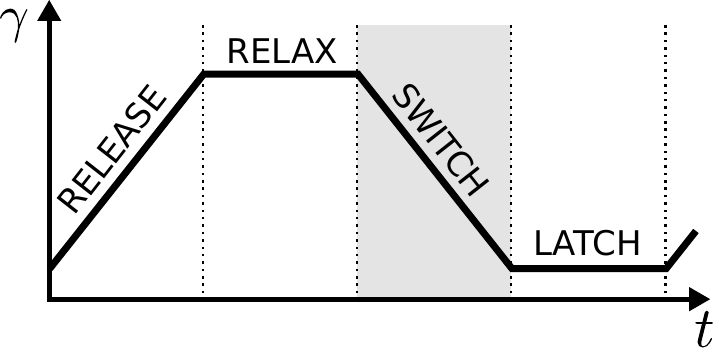}}\quad
  \subfloat[XOR]{\includegraphics[height=\Height]{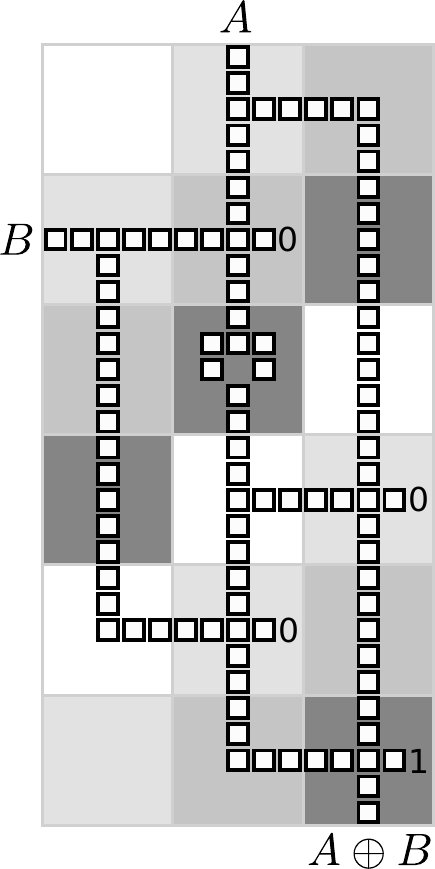}}
  \caption{In 4-phase clocking, regions of QCA devices are driven between unpolarized \emph{relaxed} and polarized \emph{latched} states: eg. by modulating their inter-dot tunneling rates, $\gamma$. Adjacent clock zones are $\sfrac{\pi}{2}$ phase shifted to enforced directional information flow. An example QCA network and arrangement of clock zones is shown in (b) with the clock phases indicated by the shade. We are primarily interested in the \emph{switching} phase, in which the polarization ground state needs to be achieved.}
  \label{fig: clock-schem}
\end{figure}

The operation of QCA networks requires the generation of 4-phase \emph{clocking fields} which control information flow by sequentially activating regions of the network called \emph{clock zones} \cite{toth1999}.  There are two species of QCA devices usually considered: 2-state and 3-state. Schematics of these devices and the clocking protocol are shown in \cref{fig: qca-cells} and \cref{fig: clock-schem}. In 2-state QCA, the two polarization states are defined by occupation of the antipodal sites and the clocking field is interpreted as a modulation of the inter-dot tunneling barriers \cite{timler2002}. In 3-state QCA, we introduce additional dots associated with a non-interacting or inactive \emph{null} state \cite{blair2018}. By applying a relative voltage to the null dots we control the energy of the null state and can activate/deactivate cells in a clock zone. The influence of the form of the clocking fields remains largely unstudied and an ultimate upper-bound on device operation frequency is still to be determined.

In this work, we investigate the adiabaticity and maximum operating frequency of a subset of the basic building blocks of QCA networks. In particular, we consider the frequency below which a QCA logic gate gives the correct output with at least 99\% likelihood. 
An estimate of the maximum operating frequency of a QCA wire based on the limits of heat sink power dissipation has previously been considered by Timler and Lent \cite{timler2002}. For reasonable assumptions of the interaction energies and device densities for molecular QCA, and assuming a maximum heat sink dissipation of \SI{100}{\watt\cdot\centi\meter^{-2}}, they give an operating frequency 1-3 orders of magnitude lower than the intrinsic frequency. For comparison with Timler's results and in order to show an evident connection to existing understandings from quantum annealing, we consider only 2-state QCA; however, the methods extend naturally to 3-state QCA with only minor modification. In \cref{sec: qca-sim}, we describe our QCA simulation methods. \cref{sec: clocking} discusses tunnel-based clocking in detail, including the link to quantum annealing in transverse Ising spin-glasses and some considerations for choosing a clocking schedule. Finally, \cref{sec: coherent,sec: dissipative} show performance results for both full coherence and in the presence of a thermal bath.

\section{Simulating QCA Dynamics}
\label{sec: qca-sim}

We consider some network of $N$ QCA cells under the influence of some set of fixed polarization \emph{driver} cells representing network inputs. In the two-state approximation, the Hamiltonian can be expressed as \cite{tougaw1996}
\begin{equation}\label{eq: H-qca}
  \HH(t) = -\tfrac{1}{2} \sum_{i=1}^N \gamma_i \hs{x}{i} + \tfrac{1}{2} \left[\sum_{i=1}^N h_i \hs{z}{i} - \sum_{\pairs{ij}}^N E_k^{ij} \hs{z}{i} \hs{z}{j} \right],
\end{equation}
with $\gamma_i$ the tunneling energies, ${h_i = \sum_D E_k^{i,D} P_D}$ the electrostatic biases due to driver cell polarizations $P_D$, $E_k^{ij}$ the kink energies between cells, and $\hs{a}{i}$ the Pauli operators associated with the $i^{th}$ cell. The dynamics in the presence of a dissipative thermal bath are taken to be described by a Liouville-von Neumann equation with a relaxation term \cite{timler2002}:
\begin{equation}\label{eq:lvn}
 \frac{d}{dt} \hrho(t) = -\frac{i}{\hbar} \comm{\HH(t)}{\hrho(t)} - \frac{1}{\tau} \left[\hrho(t) - \hrhos(t, \hrho) \right],
\end{equation}
with density operator $\hrho$, steady state density operator $\hrhos$, and relaxation time $\tau$. The exact forms of $\hrhos$ that we consider will be discussed in \cref{sec: dissipative}. We express \cref{eq:lvn} in dimensionless form with $s = ft$ the normalized time for \emph{switching rate} $f$ and $\HHd(s)=\sfrac{\HH}{\Eps}$ the dimensionless Hamiltonian for \emph{energy scale} $\Eps$ which we equate to the nearest neighbour kink-energy:
\begin{equation}\label{eq:lvn-d}
 \runrate \frac{d}{ds} \hrho(s) = -i\comm{\HHd(s)}{\hrho(s)} - \dissip \left[\hrho(s) - \hrhos(s, \hrho) \right].
\end{equation}
Here $\runrate = \sfrac{f}{f_0}$ and $\dissip = \sfrac{1}{f_0\tau}$ define the relative switching and dephasing/relaxation rates with respect to the \emph{intrinsic frequency} $f_0 = \sfrac{\Eps}{\hbar}$ of the coherent dynamics of $\HH$. For reference, an $\Eps$ of 100 meV corresponds to an intrinsic frequency of ${f_0 \approx 150 \text{ THz}}$; the kink energies typically discussed in nanoscale QCA architectures range from $10^1 - 10^3$ meV. We will drop the $\Eps$ for conciseness. Note that with 4-phase clocking the actual operating frequency of the network is $\runrate/4$.

\subsection{Coherence Vector Formalism}
\label{sec: cvf}

Due to suspected limited correlations in QCA networks, it has been proposed to express the dynamics in terms of the coherence vector formalism \cite{toth2001, karim2014a}. The density operator can be expressed in the basis of the generators of SU($2^N$) as
\begin{equation} \label{eqn:cvf-rho}
\hrho(s) = \frac{1}{2^N} \brak{\One + \sum_i \lmu{a}{i} \hs{a}{i} + \sum_{\pairs{ij}} \Kmn{ab}{ij} \hs{a}{i} \hs{b}{j} + \cdots},
\end{equation}
where we employ Einstein summation for the subscripts over the set $\set{x,y,z}$. The dynamics of $\hrho$ are equivalent to those of the real-valued coefficients: $\lmu{a}{i} = \expect{\hs{a}{i}}$, $\Kmn{ab}{ij} = \expect{\hs{a}{i} \hs{b}{j} }$, etc. These coefficients are classified by the number of cells they consider: each cell is assigned a \textit{coherence vector} $\bm{\lambda}_i$ containing the $3$ single-point expectation values $\lmu{a}{i}$; each pair of cells gets a \textit{two-point correlation tensor} $\bm{\Kmn{ij}{}}$ containing the $3^2$ two-point correlations $\Kmn{ab}{ij}$; and so on. The power of this formalism comes in the capacity to limit the number of terms used to approximate $\hrho$. The dynamics of any one of these terms can be computed as
\begin{equation} \label{eqn:cvf-svd}
    \runrate \frac{d}{ds} \sve{k}(s) = i \Expect{\comm{\HHd(s)}{\svo{k}}} - \dissip \brak{ \sve{k}(s) - \svs{k}(s, \hrho)},
\end{equation}
with $\svo{k}$ any of the operators in \cref{eqn:cvf-rho}, $\sve{k}$ its expected value, and $\svs{k} = \tr \hrhos \svo{k}$.
Only the lowest order approximation of the dynamics is usually considered: the so-called Inter-cellular Hartree Approximation (ICHA) \cite{lent1993}. The ICHA includes only the coherence vectors, excluding two-point and higher correlations. The resulting system of equations is given by 
\begin{equation} \label{eqn:icha}
  \runrate \frac{d}{ds} \bm{\lambda}_i = \bm{\Gamma}_i \times \bm{\lambda}_i - \dissip (\bm{\lambda}_i - \bm{\eta}_i),
\end{equation}
with $\bm{\Gamma}_i = \bsm{ -\gamma_i, & 0, & h_i - \sum_{n \neq i} \Ek{ni} \lmu{z}{n}}$ and $\bm{\eta}_i$ the local dissipation vector. Absent dissipation, this has the nice property that each cell can be seen to evolve under an instantaneous Hamiltonian dependent on the $\lmu{z}{i}$ values of neighbouring cells:
\begin{equation}
  \HH_i = -\frac{1}{2} \gamma_i \hs{x}{i} + \frac{1}{2} \htl{i} \hs{z}{i},
\end{equation}
with instantaneous effective bias $\htl{i} = h_i - \sum_{n \neq i} \Ek{ni}\lmu{z}{n}$. Absent dissipation, \cref{eqn:icha} is stationary only if $\bm{\lambda}_i \propto \bm{\Gamma}_i$; otherwise, coherent oscillations occur as has been previously studied \cite{taucer2015}. This can be seen by taking the derivative of the $\lmu{y}{i}$ equation in \cref{eqn:icha}:
\begin{equation}\label{eqn: icha-osc}
    \runrate^2 \frac{d^2}{ds^2} \lmu{y}{i} = -(\htl{i}^2 + \gamma_i^2) \lmu{y}{i} + O(\gamma_n \lmu{y}{n} ),
\end{equation}
where the remaining terms tend to be small near the end of the switching phase. For sufficiently fast $\runrate$, the $\lmu{y}{i}$ can remain non-zero and will then oscillate with angular frequencies $\csa_i^2 = (\htl{i}^2+\gamma_i^2)/\runrate^2$. We note then the oscillation in $\lmu{x}{i}$ and $\lmu{z}{i}$ via
\begin{equation}   
  \runrate \frac{d}{ds} \lmu{x}{i} = -\htl{i}\lmu{y}{i}, \qquad \runrate \frac{d}{ds} \lmu{z}{i} = -\gamma_i \lmu{y}{i}.
\end{equation}
Dissipation must be employed to dampen out these oscillations which cannot be justified in the coherent limit.

\subsection{Solver Details}

In general, we are interested in an initial value problem of the form
\begin{equation}
  \dds \bm{y}(s) = g(s,\bm{y}) / \runrate  \qquad \bm{y}(0) = \bm{y}_0
\end{equation}
where $g(s,\bm{y})$ is given by either \cref{eq:lvn-d} or \cref{eqn:icha}. We use T\'oth's approach to define the initial state\cite{toth2001}: starting with the ground state of $\HHd(0)$, we use a Newton-Raphson method to find a nearby root of $g(s, \bm{y})$. The time dependent solution is computed using a Dormand-Prince (RK45) method for \cref{eq:lvn-d} and a first order backward differentiation formula method (BDF1) for the ICHA \cite{scipy}.


\section{Choosing a Clocking schedule}
\label{sec: clocking}

In addition to the choice of the steady state solution, the schedule of the tunneling barriers during clocking, the \textit{clocking schedule}, will also affect the performance of the network. The tunnel barrier modulation in 2-state QCA clocking is achieved by allowing $\gamma_i$ in \cref{eq: H-qca} to vary with time: $\gamma_i \to \gamma_i(s)$. As the tunneling barriers are raised, $\gamma_i(s)$ goes from some large $\gamma_i(0) \gg \Eps$ to an effectively zero $\gamma_i(1) \ll \Eps$ at which point $\HHd$ is approximately classical. 
The quantum adiabatic theorem \cite{born1928} guarantees that a network initialised in the ground state of $\HHd(0)$ will reach the desired ground state of $\HHd(1)$ for sufficiently slow $\runrate$. However, \textit{sufficiently slow} relates to the rate of change of $\HHd$ and hence on the schedule of $\gamma_i(s)$. 
QCADesigner, a popular QCA design and simulation tool, uses a sinusoidal schedule which is approximately linear over the switching regime \cite{walus2004}. Linear clocking schedules have also been considered for the 3-state QCA model \cite{pidaparthi2018}. There has of yet been no expansive study of the influence of the clocking schedule on QCA performance; however, similar considerations appear in the study of quantum annealing.


\subsection{Quantum Annealing Framework}
\label{subsec: qa-formalism}

In quantum annealing, the ground state of some challenging and interesting Hamiltonian is found by slowly transforming an initially simple quantum system into the one of interest \cite{kadowaki1998}. The Hamiltonian is often expressed as
\begin{equation}\label{eqn: sch}
  \HHd(s) = -\tfrac{1}{2} A(s)\HX + \tfrac{1}{2}B(s) \HP,
\end{equation}
where $\HX = \sum_i \hs{x}{i}$ and $\HP = \sum_i h_i \hs{z}{i} + \sum_\pairs{ij} J_{ij} \hs{z}{i}\hs{z}{j}$ define a transverse Ising spin-glass problem with $\HP$ the problem of interest. Clearly such problems include the two-state approximation in QCA. The $A(s)$ and $B(s)$ here define the annealing schedule. In tunnel barrier modulated QCA, we have $B(s)=1$ and take $\gamma_i(s) = A(s)$ to be the same for all cells in a clock zone. Before we consider a candidate clocking schedule, we first need to define our performance metrics and establish some necessary criteria for $A(s)$.

\subsection{Performance Metrics}

In order to quantify the clocking performance, we define the following metrics:
\begin{subequations}
\newcommand{\lbl}[1]{\text{\emph{#1:}}}
\label{eq: mets}
\begin{align}
 &\lbl{Adiabaticity} &\quad& \Met_A(s) = \tr \hrho(s) \proj{g}(s) \label{eq: met_A}\\
 &\lbl{Classical} &\quad& \Met_{cl} = \tr \hrho(1) \proj{\PP} \label{eq: met_cl}\\
 &\lbl{Logical} &\quad& \Met_L = \frac{1}{2^{|\Omega|}} \prod_{i \in \Omega} (1 + \lmu{z}{i}(1)P_i) \label{eq: met_L}
\end{align}
\end{subequations}
For adiabatic clocking, the system should remain near the ground state at all times. $\Met_A(s)$ describes the overlap between the system state and the space of potentially degenerate ground states of $\HHd(s)$, given as the expected value of the projection operator ${\hat{P}_g(s) = \sum_d \ket{\Psi_g^d}\bra{\Psi_g^d}}$. Clocking ideally results in the system reaching the ground state of the classical Hamiltonian $\HP$. $\Met_{cl}$ describes the probability of this outcome, with projection $\proj{\PP}$. If the classical ground state is non-degenerate and there are no correlations in $\hrho(1)$, $\Met_{cl}$ simplifies to 
\begin{equation} \label{eq: Mcl}
  \tilde{\Met}_{cl} \approx \frac{1}{2^N}\prod_{i=1}^N (1+\lmu{z}{i}(1)P_i),
\end{equation}
with $P_i = \tr \proj{\PP} \hs{z}{i}$ the polarization of each cell in the classical ground state. In principle, the polarizations of certain output cells may be logically correct even if the state fails to reach the ground state. By restricting the product in \cref{eq: Mcl} to a finite subset, $\Omega$, of the cells, we can define a logical performance, $\Met_L$. In this work, we consider a ``high performance'' target of $\Met_{cl} \geq 0.99$.

\subsection{Quality of the Ground State}
\label{subsec: gs-quality}

\begin{figure}
\newcommand{\Height}{.28\linewidth}
 \subfloat[Wire]{\includegraphics[height=\Height, valign=c]{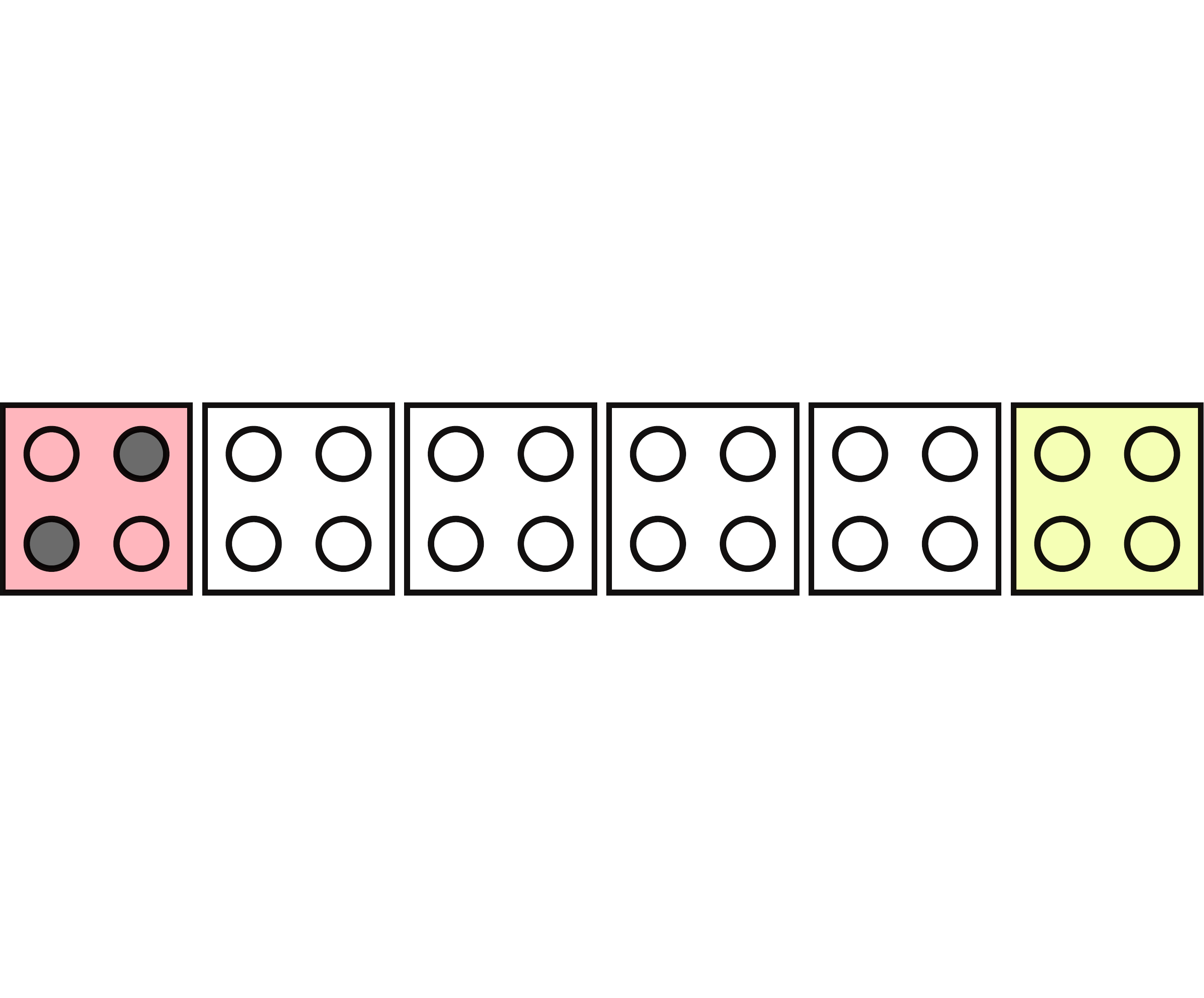}}\quad
 \subfloat[Inverter]{\includegraphics[height=\Height, valign=c]{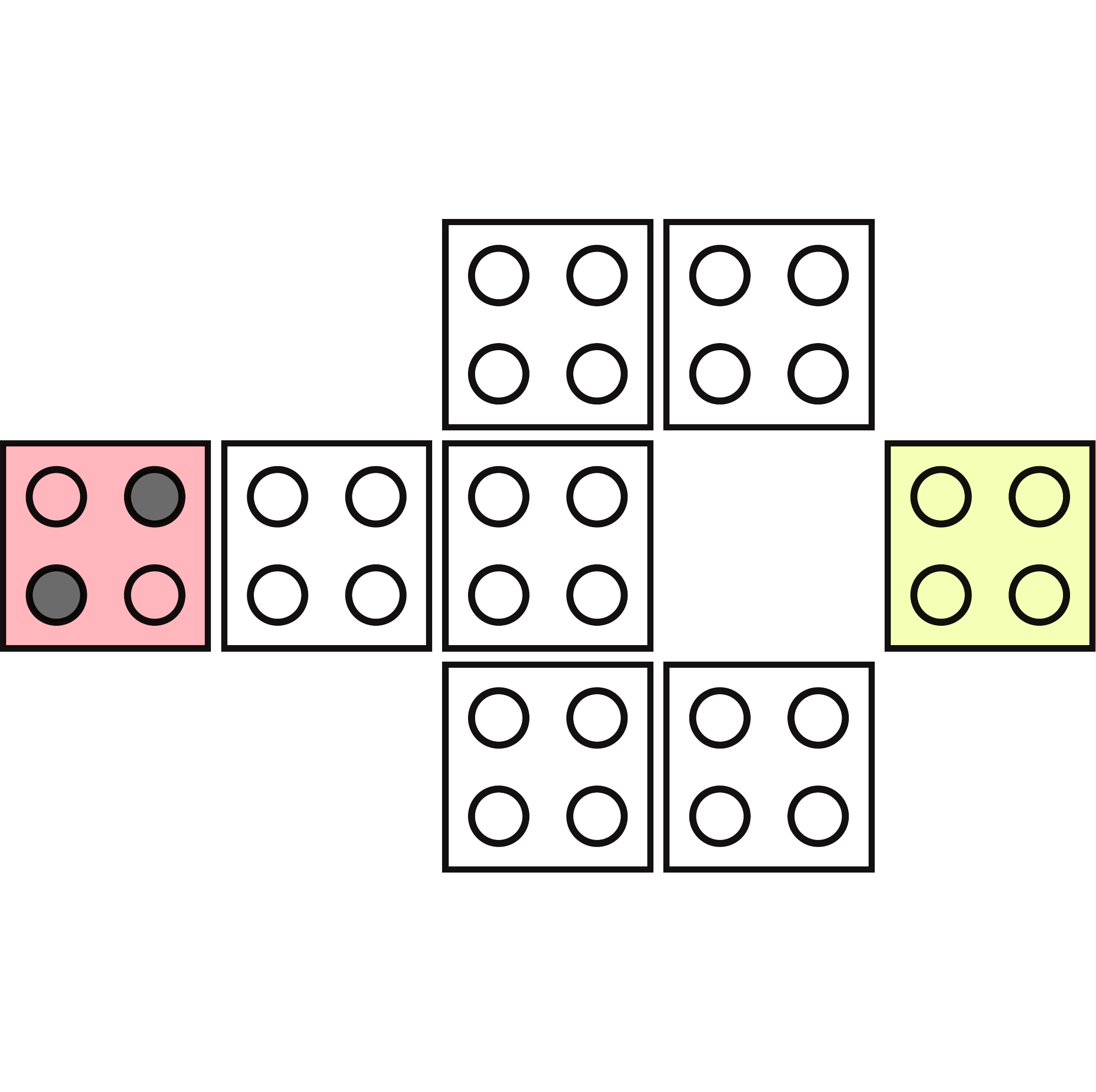}}\quad
 \subfloat[Majority Gate]{\includegraphics[height=\Height, valign=c]{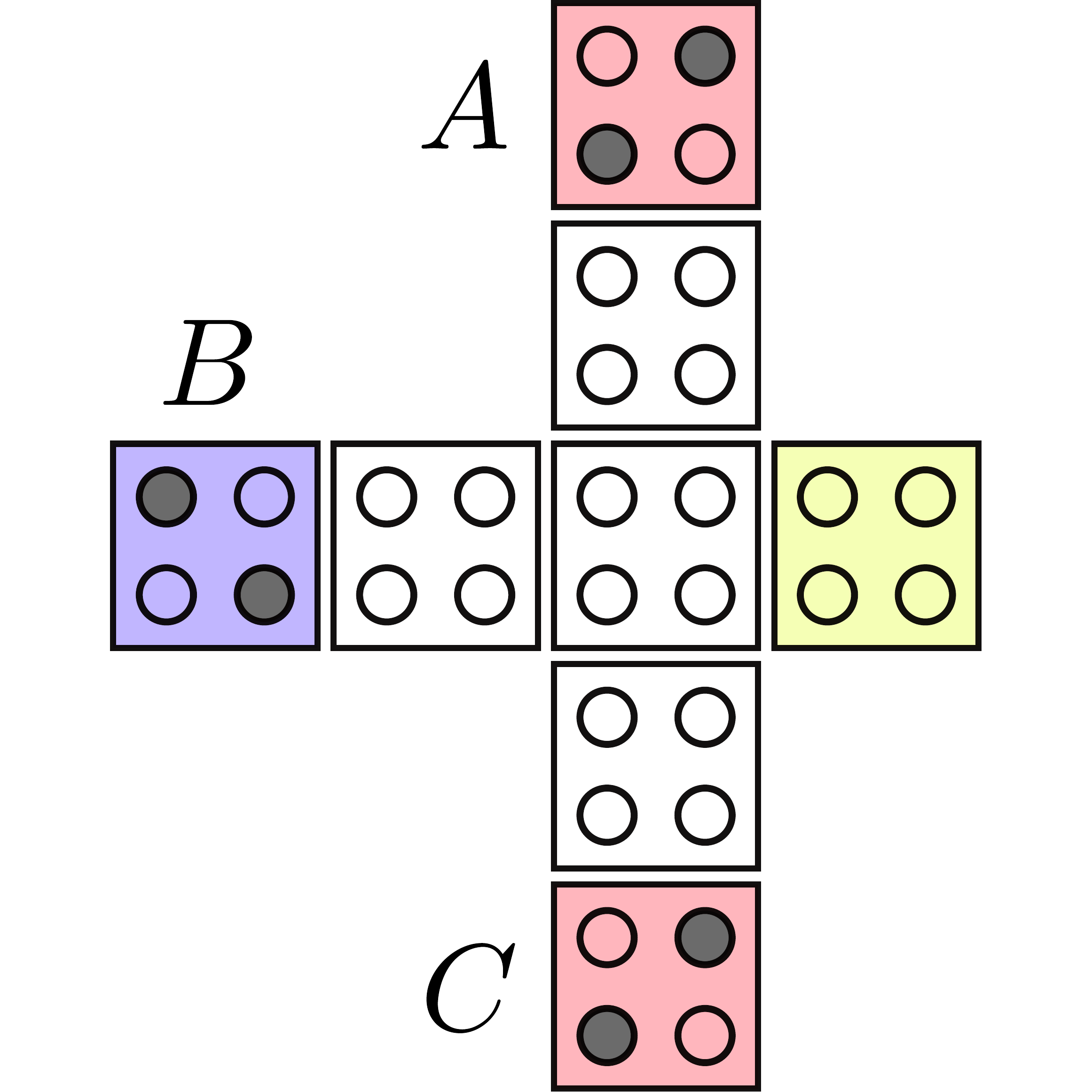}}
 \caption{Schematics for simulated devices. Red and blue shaded cells indicate inputs with fixed polarizations of $\pm 1$. The rightmost cells produce the logical outputs. We denote by ``Wire-N'' a wire of length $N$ and \mbox{``Maj-ABC''} a majority gate with binary inputs as indicated: ex. \mbox{Maj-101} shown.}
 \label{fig:qca-schem}
\end{figure}

\begin{figure}
  \newcommand{\Width}{.34\linewidth}
  \subfloat[Device dimensions]{\includegraphics[width=\Width, valign=c]{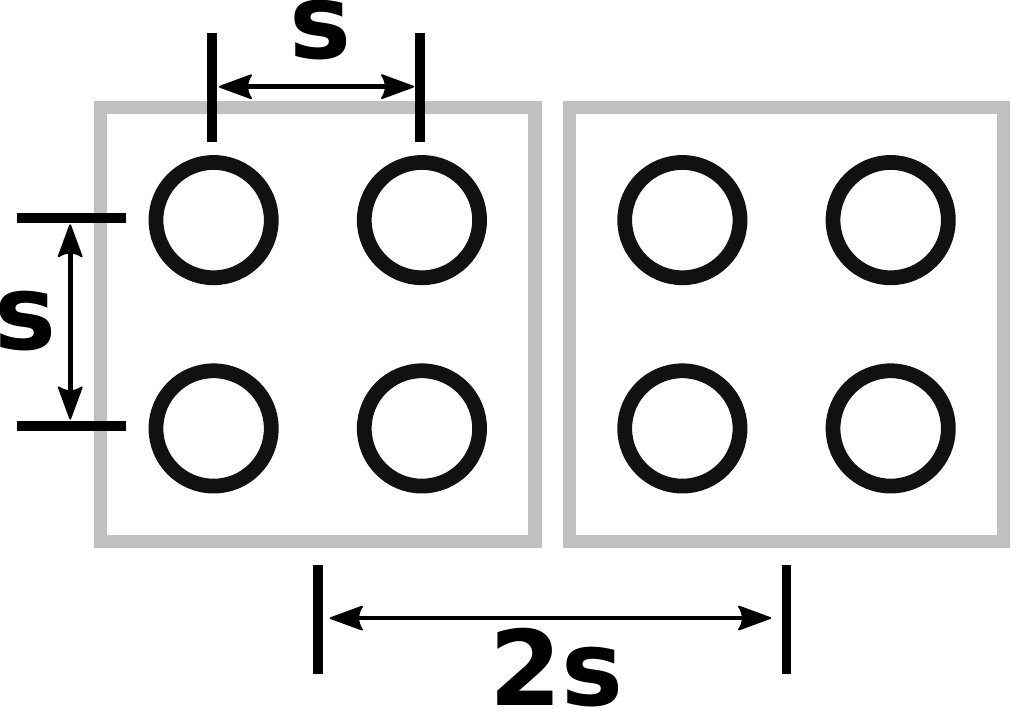}
  \vphantom{\includegraphics[width=\Width, valign=c]{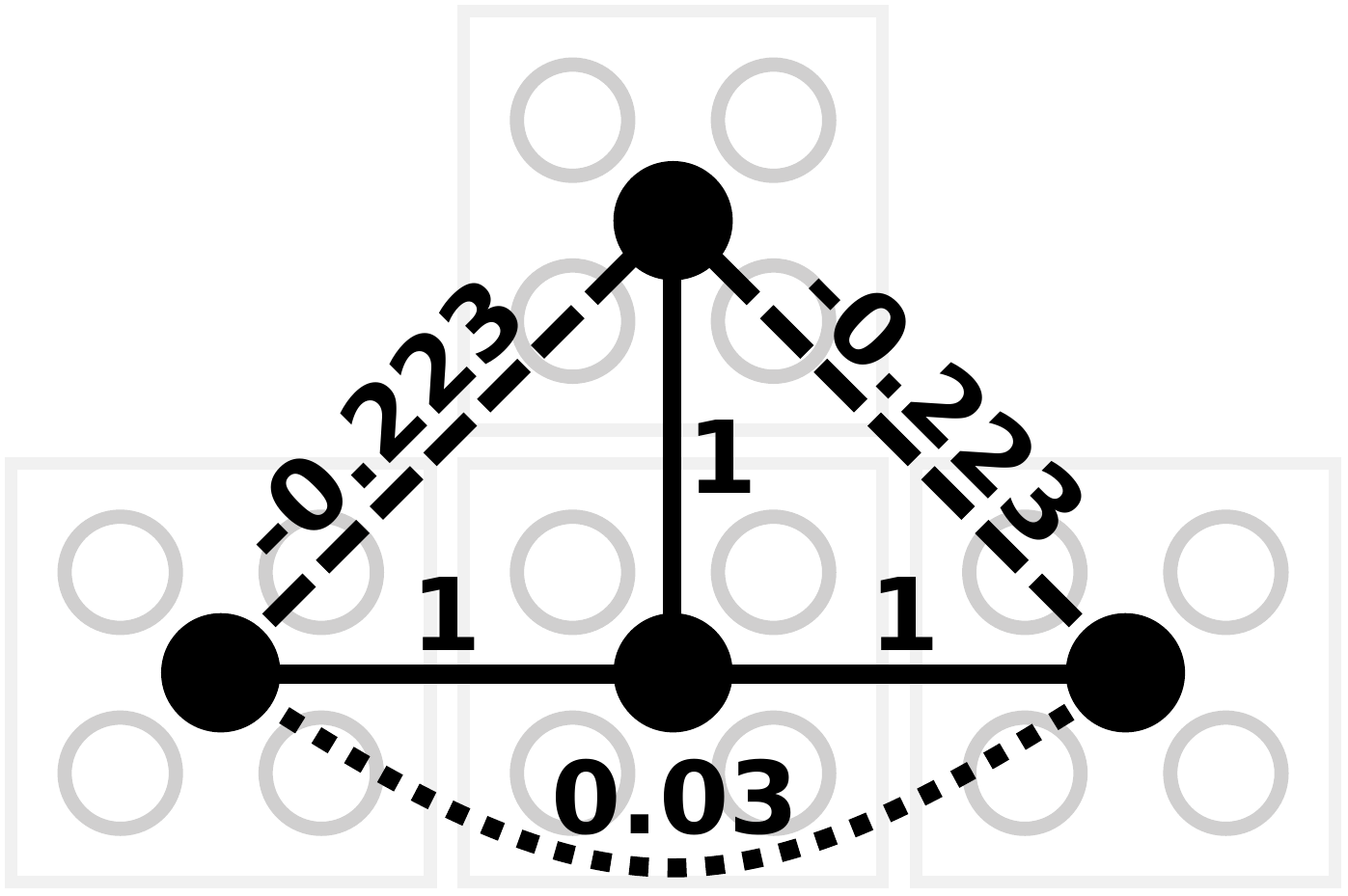}}} \quad
  \subfloat[Kink energies]{\includegraphics[width=\Width, valign=c]{Ekinks}}
  \caption{QCA Device dimensions and computed kink energies relative to $\Eps$. We ignore the -1.03 and weaker interactions.}
  \label{fig: schematics}
\end{figure}

For typical QCA networks, we will have a non-degenerate ground state. We can use non-degenerate perturbation theory to determine the ``quality'' of the initial and final ground states:
\begin{subequations}
  \label{eqn: qual}
  \begin{align}
    \Met_0 = \expect{\proj{X}}(0)  &= 1 - \atb_0^{-2} F_0 + O \pa{ \atb_0^{-3} },\\
    \Met_1 = \expect{\proj{\PP}}(1) &= 1 - \atb_1^{2} F_1 + O \pa{ \atb_1^{3} },
  \end{align}
\end{subequations}
where $P_X$ is the projection operator onto the ground state of $-\HX$, expectation values are for the ground state of $\HHd(s)$, $\atb_0$ and $\atb_1$ are the initial and final values of the ratio $\atb(s) =  \sfrac{A(s)}{B(s)}$, and $F_0$ and $F_1$ are network-specific parameters dependent only on $(h_i, \Ek{ij})$. Note that $\Met_1$ is precisely the limit of $\Met_{cl}$ for slow $\runrate$; hence, we should choose $\atb_1$ such that $\Met_1>0.99$. For \cref{eq: H-qca}, we find
\begin{equation}
    F_0 = \tfrac{1}{4} \sum_{i=1}^N |h_i|^2 + \tfrac{1}{16} \sum_\pairs{ij}^N |\Ek{ij}|^2 \:, \quad
    F_1 = \tfrac{1}{4} \sum_{i=1}^N \frac{1}{\htl{i}^2},
\end{equation}
with $\htl{i} = h_i - \sum_{j \neq i} \Ek{ij} P_j$ calculated for the polarizations of the ground state of $\HP$.
For our performance target, we obtain a necessary constraint on $\atb_1$:
\begin{equation}
  \atb_1^{-1} \geq \max \sqrt{\sfrac{F_1}{1-Q^*}},
\end{equation}
with some $\Met^* > 0.99$. We consider the devices shown in \cref{fig:qca-schem}. Schematics of the device interactions are shown in \cref{fig: schematics} with the computed $F_0$ and $F_1$ shown in \cref{table: F-params}. Our constraint is approximately $\atb_1 \leq \sfrac{1}{15}$. We can make an observation regarding the dependence of $\Met_{cl}$ on the parameters $(\runrate, \alpha_0, \alpha_1)$. Assuming again a non-degenerate ground state of $\HP$, we have
\begin{equation} \label{eq: Pp-prox}
  \proj{\PP} = \Met_1^{-1} \proj{g}(1) - \tfrac{1}{2} \atb_1 \sum_i \frac{1}{\htl{i} P_i} \acomm{\proj{\PP}}{\hs{x}{i}} + O(\atb_1^2).
\end{equation}
We are interested in the regime where $\Met_A(1)$ is close to 1 and hence $\hrho(1) \approx \Met_A(1) \proj{g}$. We observe that $\Met_1^{-1} = \Met_1 + 2F_1 \atb_1^2 + O(\atb_1^3)$. Upon multiplying \cref{eq: Pp-prox} by $\hrho(1)$ and taking the trace, $\tr \hrho(1) \acomm{\proj{\PP}}{\hs{x}{i}}$ becomes $-\Met_A(1) \Met_1 \atb_1 / \htl{i} P_i$ and the second order term becomes third order. We are left with
\begin{equation}
  \Met_{cl} \approx \Met_A(1) \Met_1.
\end{equation}
Further, if we assume a large $\Met_A(1)$ is independent of small $\atb_1$, we arrive at the useful expression
\begin{equation} \label{eq: Qcl}
  \Met_{cl}(\runrate, \alpha_0, \alpha_1) \approx \Met_{cl}(\runrate, \alpha_0, 0) \Met_1(\alpha_1).
\end{equation}
In this work, we will use QCADesigner's default value of $\atb_1 = \sfrac{1}{20}$, giving $\Met^* = .9943$ which satisfies our performance constraint.

\begin{table}
  \caption{Gate parameters for \cref{eqn: qual} computed for the kink energies given in \cref{fig: schematics}}
  \label{table: F-params}
  \begin{ruledtabular}
    \begin{tabular}{c@{\hspace{1ex}}|ccccc}
      Device & Wire-N & Inverter & Maj-111 & Maj-101 & Maj-110 \\
      \hline
      $F_0$ & \multirow{2}{*}{$\tfrac{1}{16}(N+3)$} & 0.581 & 1.012 & 1.012 & 1.012\\
      $F_1$ & & 2.217 & 1.141 & 2.259 & 1.735
    \end{tabular}
  \end{ruledtabular}
\end{table}


\subsection{Candidate Clocking schedules}
\newcommand{\kay}{k}

\begin{figure}
\newcommand{\Height}{.35\linewidth}
 \subfloat[Linear: \cref{eq: linear}]{\includegraphics[height=\Height]{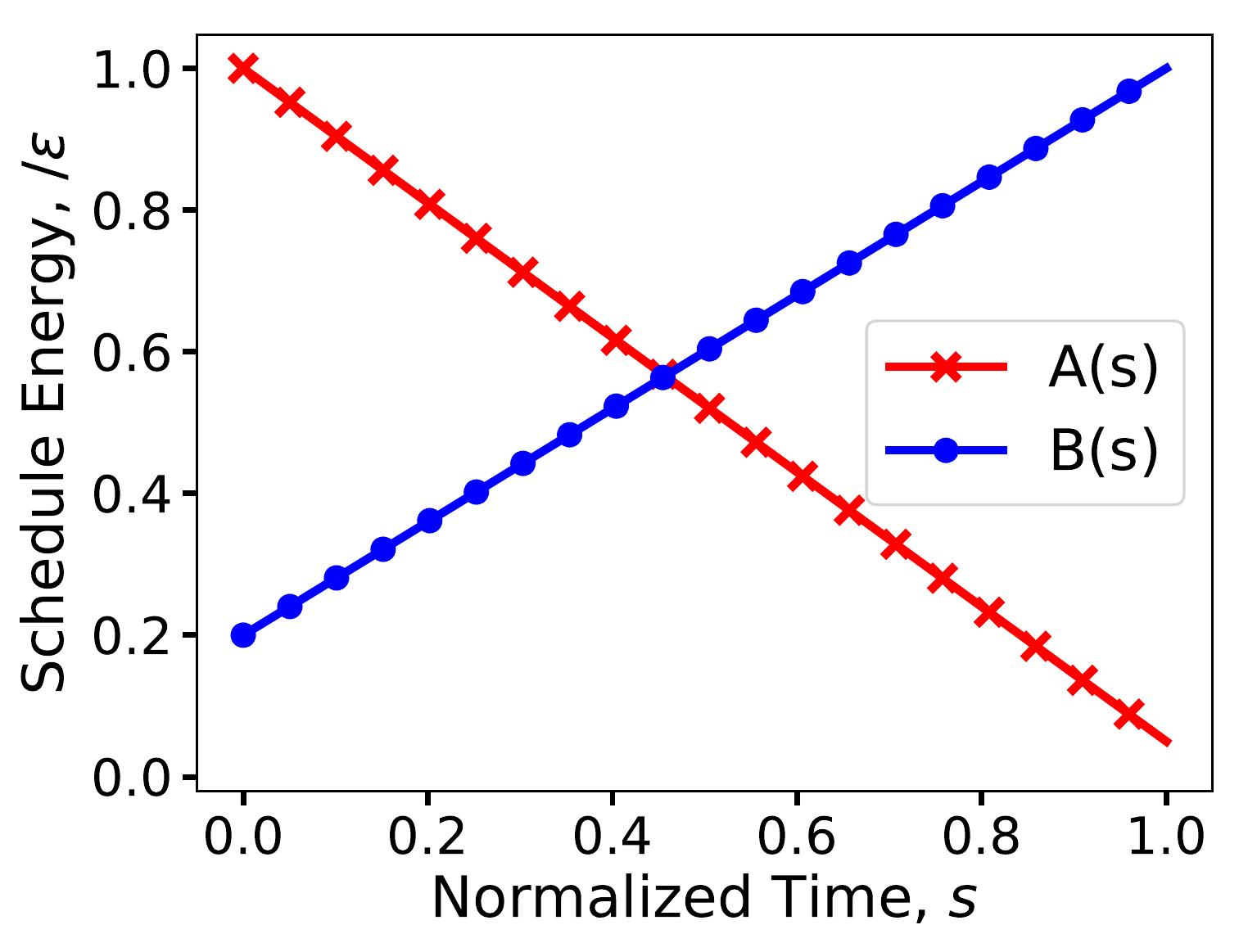}}\quad
 \subfloat[Spectrum]{\includegraphics[height=\Height]{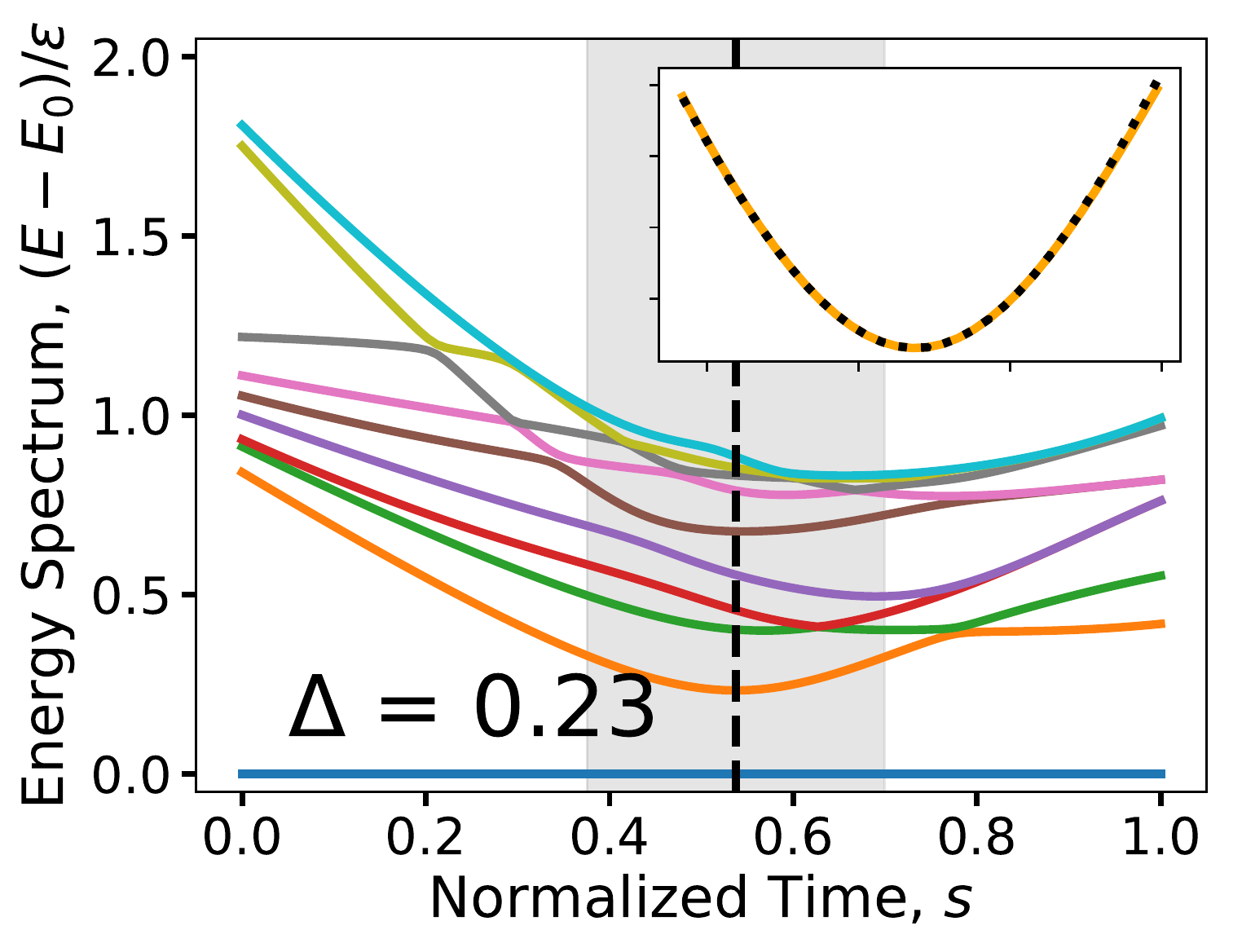}}\\
 \subfloat[Quasi-Linear: \cref{eq: quasi}]{\includegraphics[height=\Height]{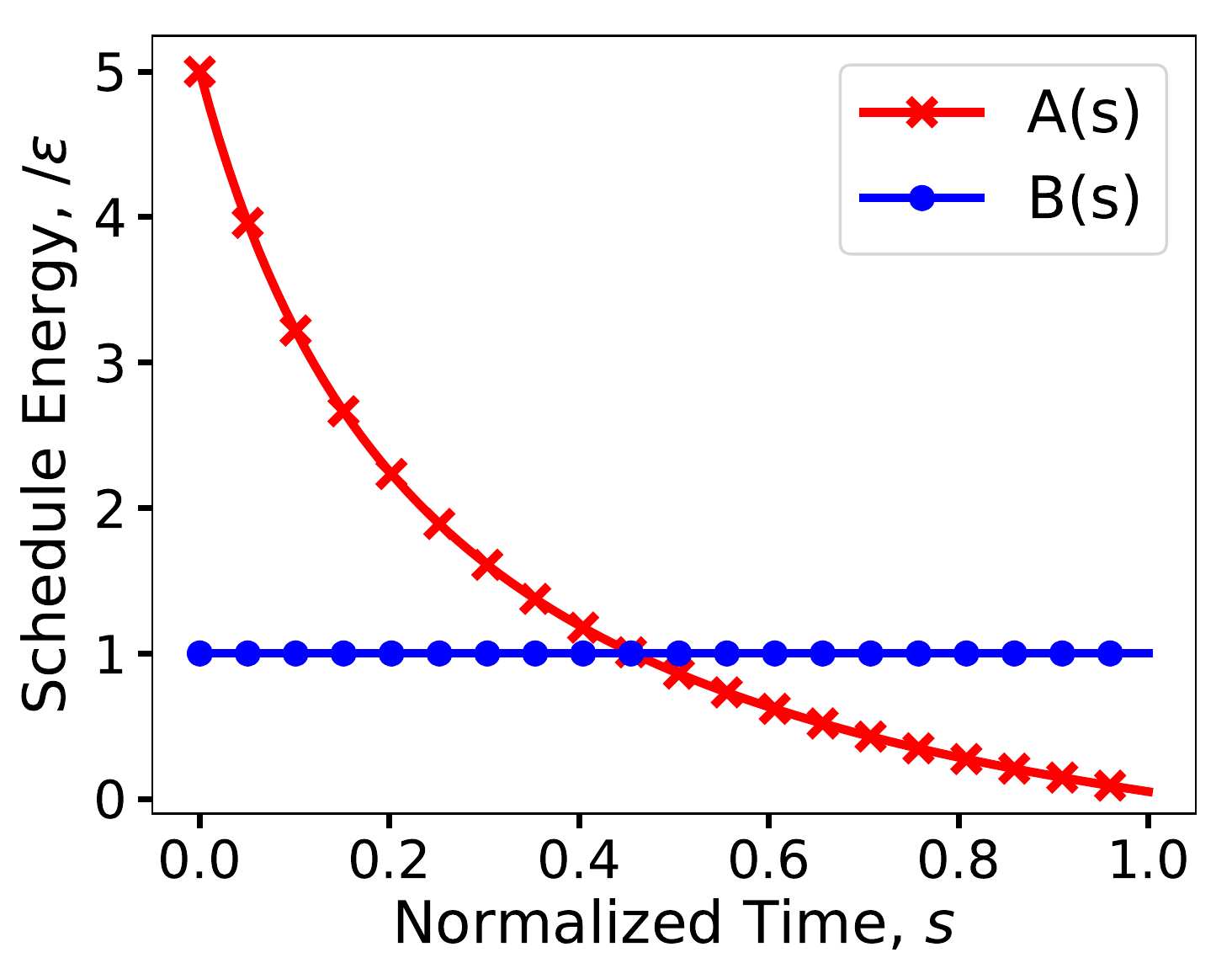}}\quad
 \subfloat[Spectrum]{\includegraphics[height=\Height]{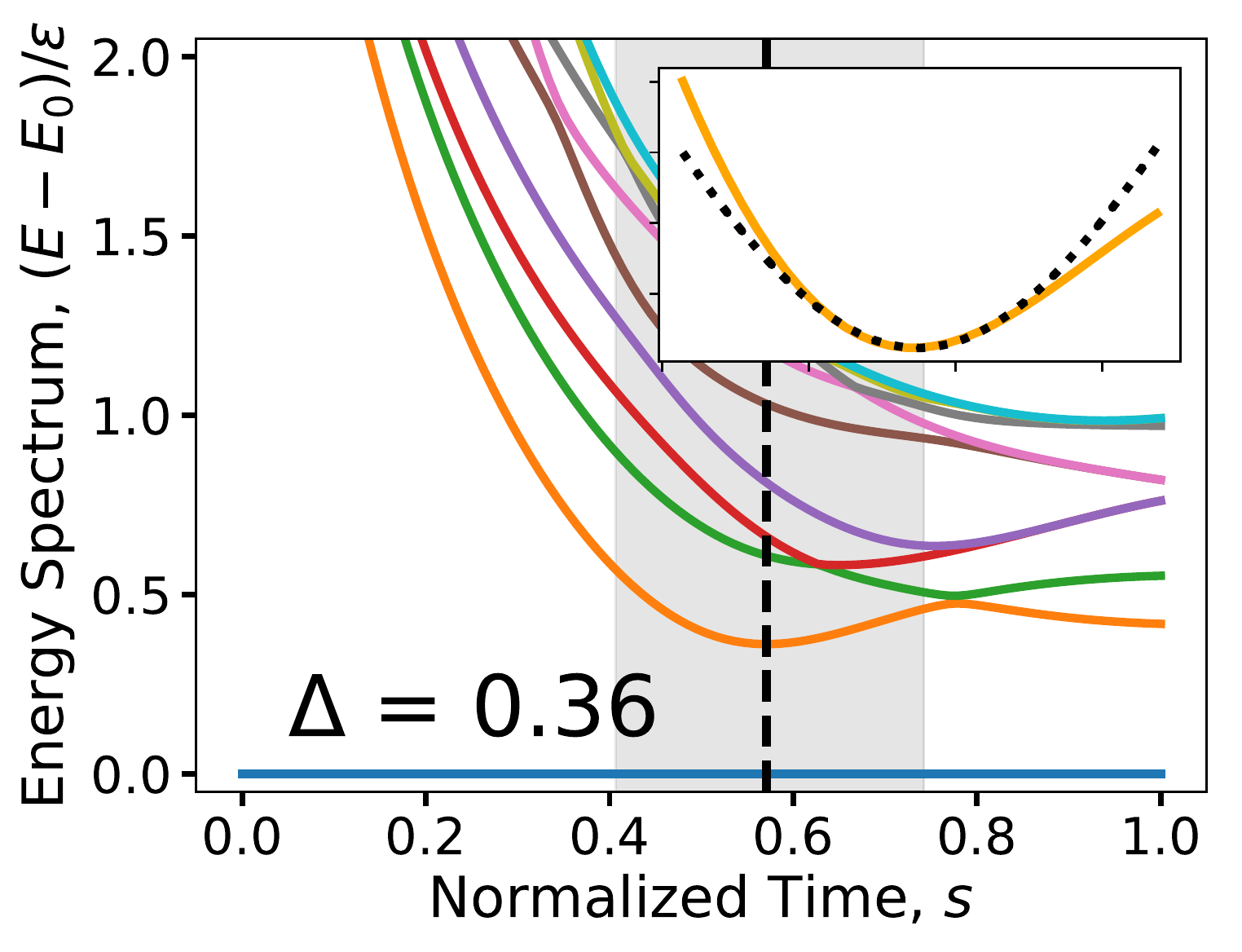}}\\
 \subfloat[Sinusoidal: \cref{eq: sinus}]{\includegraphics[height=\Height]{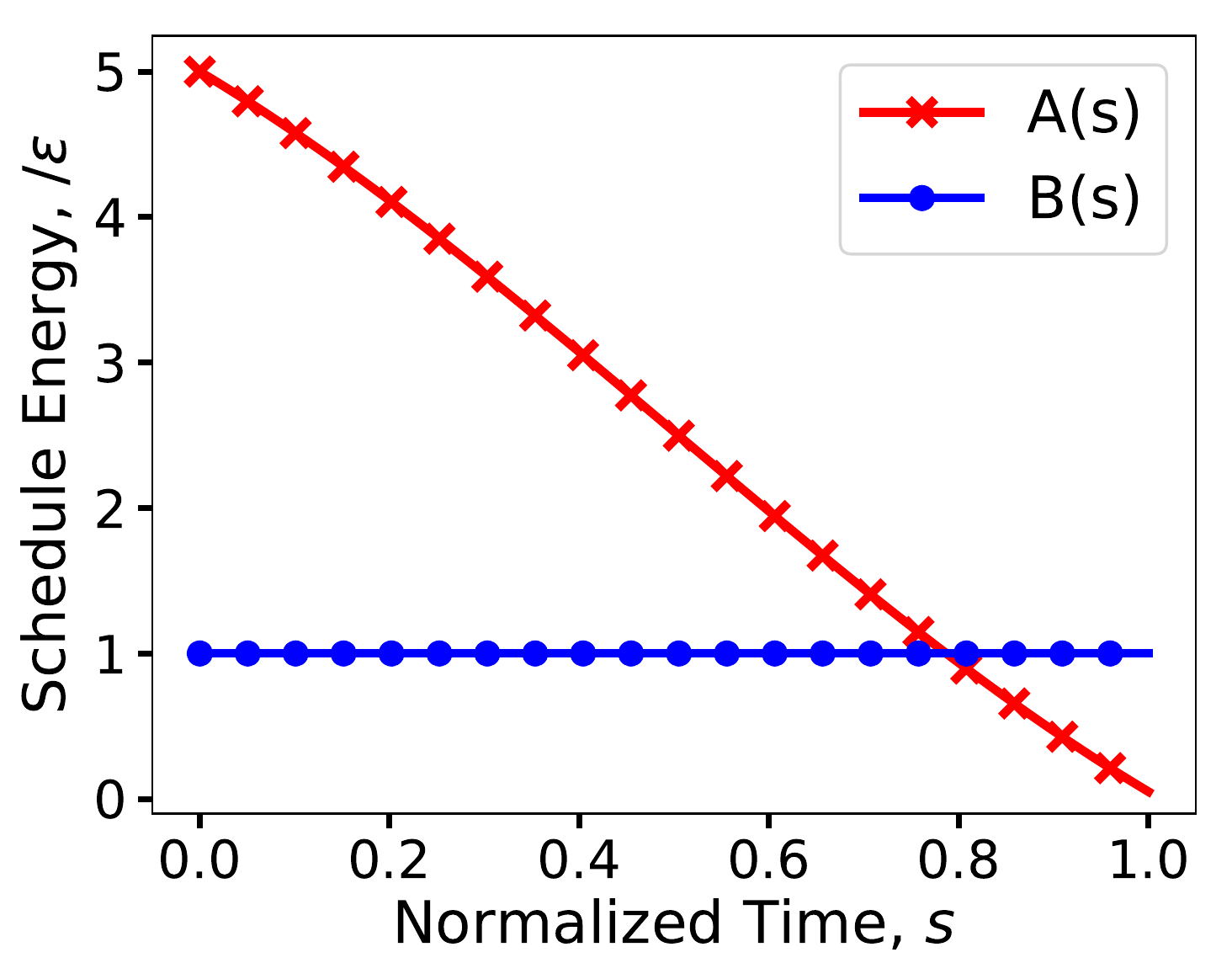}}\quad
 \subfloat[Spectrum]{\includegraphics[height=\Height]{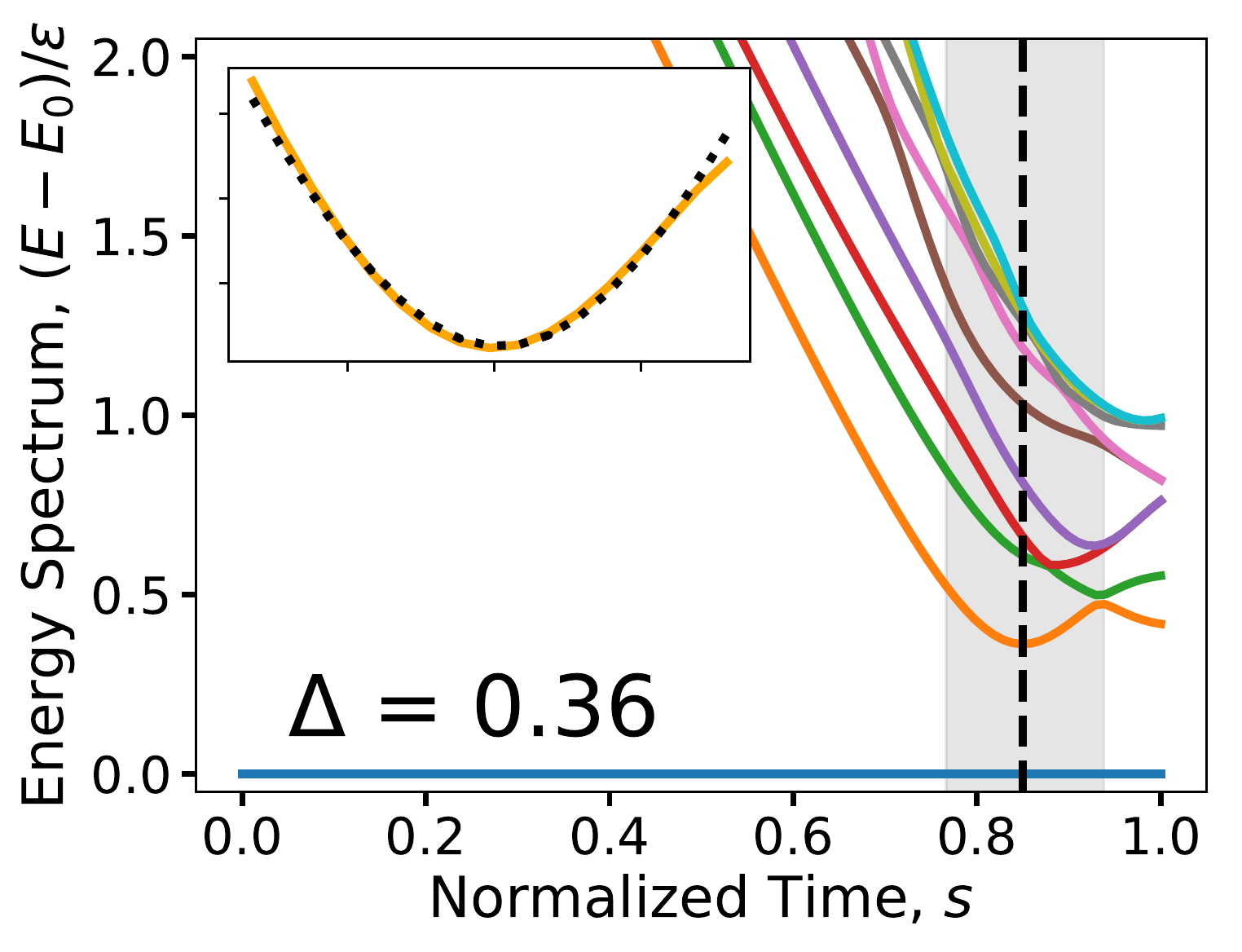}}
 \caption{Different clocking schedules and the 10 lowest energy eigenvalues for a QCA inverter with $\atb_0=5$, $\atb_1 = \sfrac{1}{20}$. The location and size of the minimum gap is indicated for comparison. The gap between the ground and first excited states was fit using \cref{eq: hyper} within $\Delta s = 0.07$ of the minimum to extract the width parameter $W$. The fit is shown in the inset over the shaded region representing $\pm W$ about the minimum. }
 \label{fig: clocking-spectra}
\end{figure}

\begin{figure}
 \newcommand{\WW}{.48\linewidth}
 \subfloat[Wire-5]{\includegraphics[width=\WW]{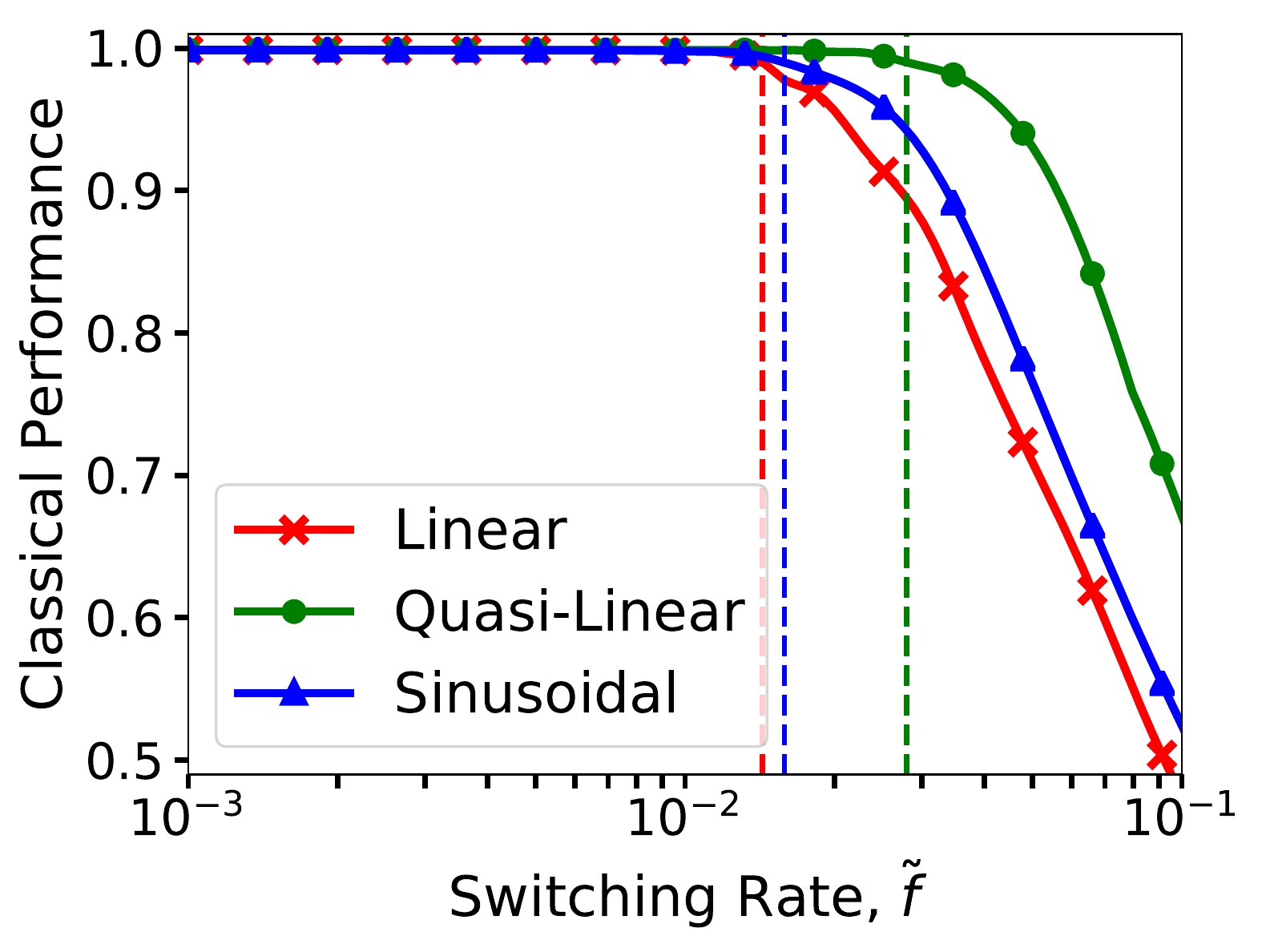}}\quad
 \subfloat[Inverter]{\includegraphics[width=\WW]{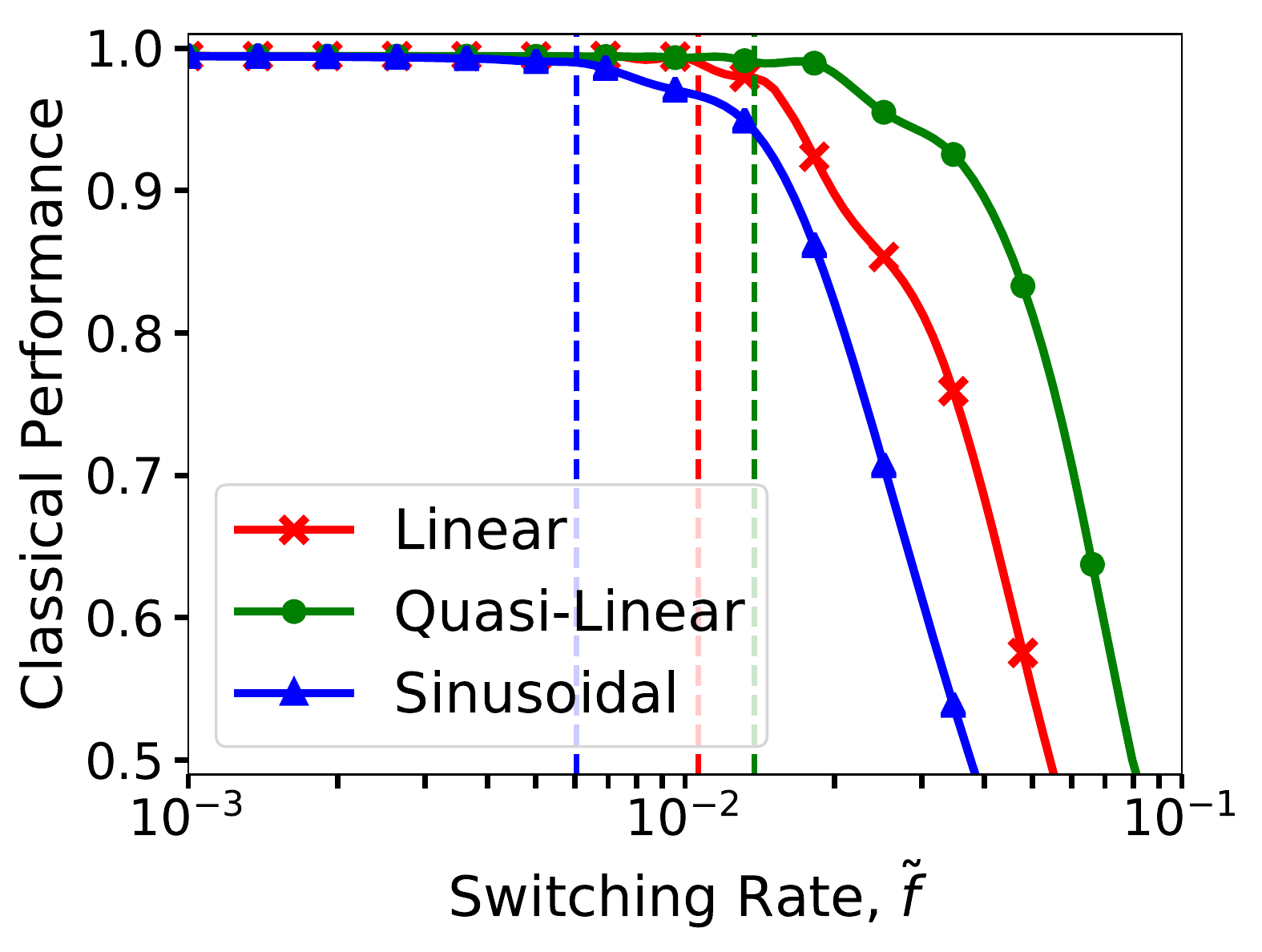}}\\
 \subfloat[Maj-111]{\includegraphics[width=\WW]{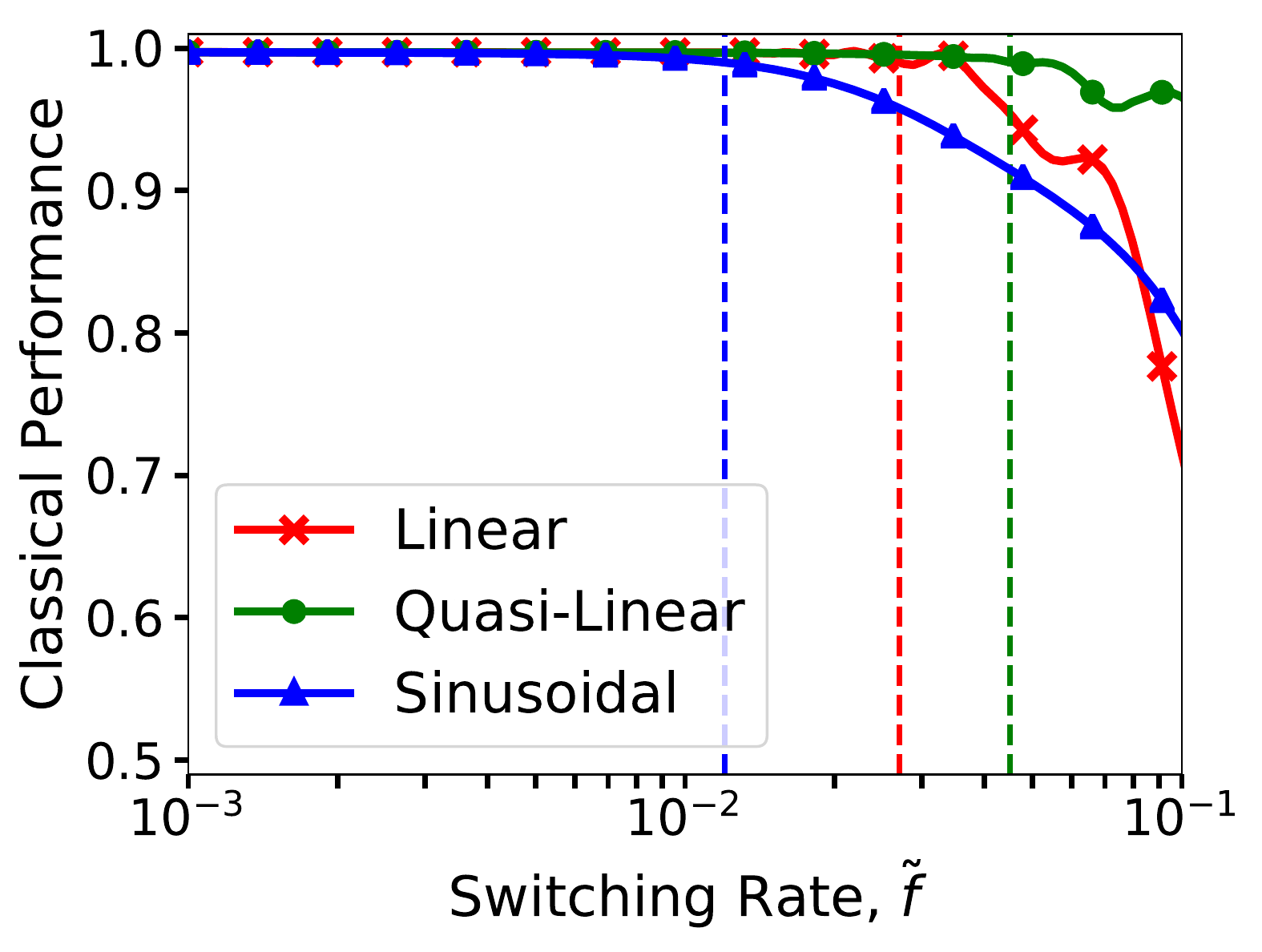}}\quad
 \subfloat[Maj-101]{\includegraphics[width=\WW]{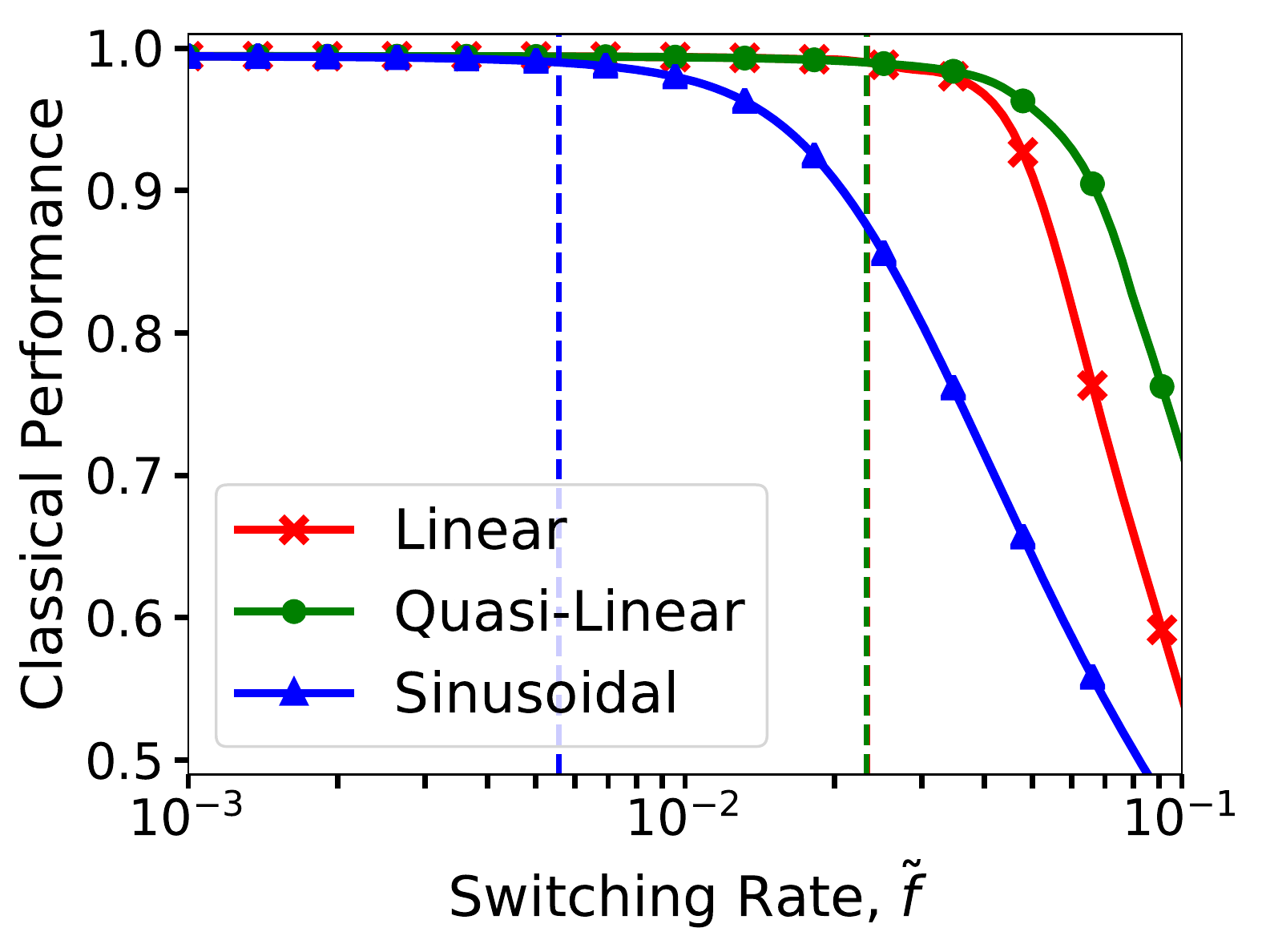}}
 \caption{Comparison of classical performances for the different clocking schedules: $\atb_0=5$, $\atb_1 = \sfrac{1}{20}$. Dashed lines mark the maximum switching rate, below which $\Met_{cl} \geq 0.99$. Schedule smoothing employed: see \cref{subsec: cohosc}.}
 \label{fig: clock-comparison}
\end{figure}

An optimal choice of clocking schedule will depend on details of the environmental interaction and will not be addressed here. Instead, we introduce a schedule which mimics the standard linear schedule used in quantum annealing studies. We assume the following: (1) the kink energies are fixed by the network geometry and hence $B(s)=1$; (2) the tunneling energies cannot be made infinite or zero; and (3) the rate of change of the eigenstates for the linear schedule is a good choice for quantum annealing. The eigenstates of \cref{eqn: sch} depend only on the ratio $\alpha(s)$. We define our clocking schedules by their initial and final ratios $\atb_0$ and $\atb_1$. We consider for comparison an appropriate \textbf{Linear} schedule with $B(1)=1$:
\begin{subequations} \label{eq: linear}
 \begin{align}
   A_L(s) &= 1-(1-\alpha_1)s,\\
   B_L(s) &= 1-(1-\sfrac{1}{\alpha_0})(1-s).
 \end{align}
\end{subequations}
For $B(s)=1$ we define an analogous schedule with the same eigenstates: $A_Q(s) = \sfrac{A_L(s)}{B_L(s)}$,
\begin{equation} \label{eq: quasi}
 A_Q(s) = 1 + (\alpha_0-1) \frac{1-\kay s}{1+(\alpha_0-1)s},
\end{equation}
which we refer to as the \textbf{Quasi-Linear} schedule, as it is perhaps the closest we can get to the linear schedule under the given constraints. The $\kay$ here satisfies $A_L(\sfrac{1}{\kay}) = B_L(\sfrac{1}{\kay})$ and has value $\kay = 1 + (1-\alpha_1)/(1-\atb_0^{-1})$. Finally, the \textbf{Sinusoidal} schedule in QCADesigner has the form
\begin{equation} \label{eq: sinus}
 A_S(s) = \frac{\alpha_0-\alpha_1}{\sqrt{2}} \cos \left( \frac{\pi}{2}(s+\sfrac{1}{2}) \right) + \frac{\alpha_0+\alpha_1}{2}.
\end{equation}
Schematics of these clocking schedules and the corresponding low energy eigenspectra of an inverter are shown in \cref{fig: clocking-spectra}. A first order performance characteristic for quantum annealing can be extracted from the spectrum. If we fit a hyperbola of the form
\begin{equation} \label{eq: hyper}
    \Delta(s) = \Delta_0\sqrt{1 + \tfrac{1}{W^2}(s-s_0)^2}
\end{equation}
to the minimum energy gap, referred to as the \emph{level crossing}, between the ground and first excited states, where $\Delta_0$ is the minimum gap and $W$ is a measure of the gap width, then the Landau-Zener diabatic transition probability \cite{landau1932, zener1932} can be expressed in the form 
\begin{equation} \label{eqn:lz}
  P_{g \to e} = e^{-2\pi \Delta_0 W /4\runrate}.
\end{equation}
The probability of remaining in the ground state through the minimum gap then depends on the product of the gap size and its width. Fits of \cref{eq: hyper} in the vicinity of the minimum gaps of the inverter in \cref{fig: clocking-spectra} are summarized in \cref{table: lz-inv-fits}. If the crossing were well described by the Landau-Zener model we should expect the switching rate needed to achieve a given $\Met_{cl}$ to be proportional to $\Delta_0 W$; however, we have not established that the eigenstates involved in the crossing satisfy the conditions inherent to the Landau-Zener approximation. Further, the hyperbolic fit is not particularly convincing for the Quasi-Linear schedule and we are ignoring other details of the spectrum. Nevertheless, we should at least take this result as inspiration that the Quasi-Linear schedule might outperform both the Linear and Sinusoidal schedules. 
The classical performance was computed for our simple QCA components using \cref{eq:lvn-d} for the different clocking schedules. Performance comparisons are shown in \cref{fig: clock-comparison}. In most cases, we see that the Quasi-Linear schedule is significantly better and will therefore  be used in all results that follow.

\begin{table}
  \caption{Fit parameters for the hyperbolic approximation of the minimum gap of the inverter for the different clocking schedules. Errors indicate $2\sigma$ estimates from the fit covariance matrix where significant.}
  \label{table: lz-inv-fits}
  \begin{ruledtabular}
    \begin{tabular}{c@{\hspace{2ex}}|ccc}
      Clocking schedule   & $\Delta_0$  &   $W$   & $\Delta_0 W$ \\
      \hline
      Linear & $0.233$ & $0.162$ & $0.037$ \\
      Quasi-Linear & $0.362$ & $0.168 \pm 0.005$ & $0.062 \pm 0.002$ \\
      Sinusoidal & $0.362$ & $0.087 \pm 0.002$ & $0.030 \pm 0.001$
    \end{tabular}
  \end{ruledtabular}
\end{table}
%


\subsection{Initial Coherent Oscillations}
\label{subsec: cohosc}

\newcommand{\Ao}{A_0}
\newcommand{\omgr}{\omega_r}

\begin{figure}
  \newcommand{\Height}{.34\linewidth}
  \subfloat[Single-Point Expectations]{\includegraphics[height=\Height]{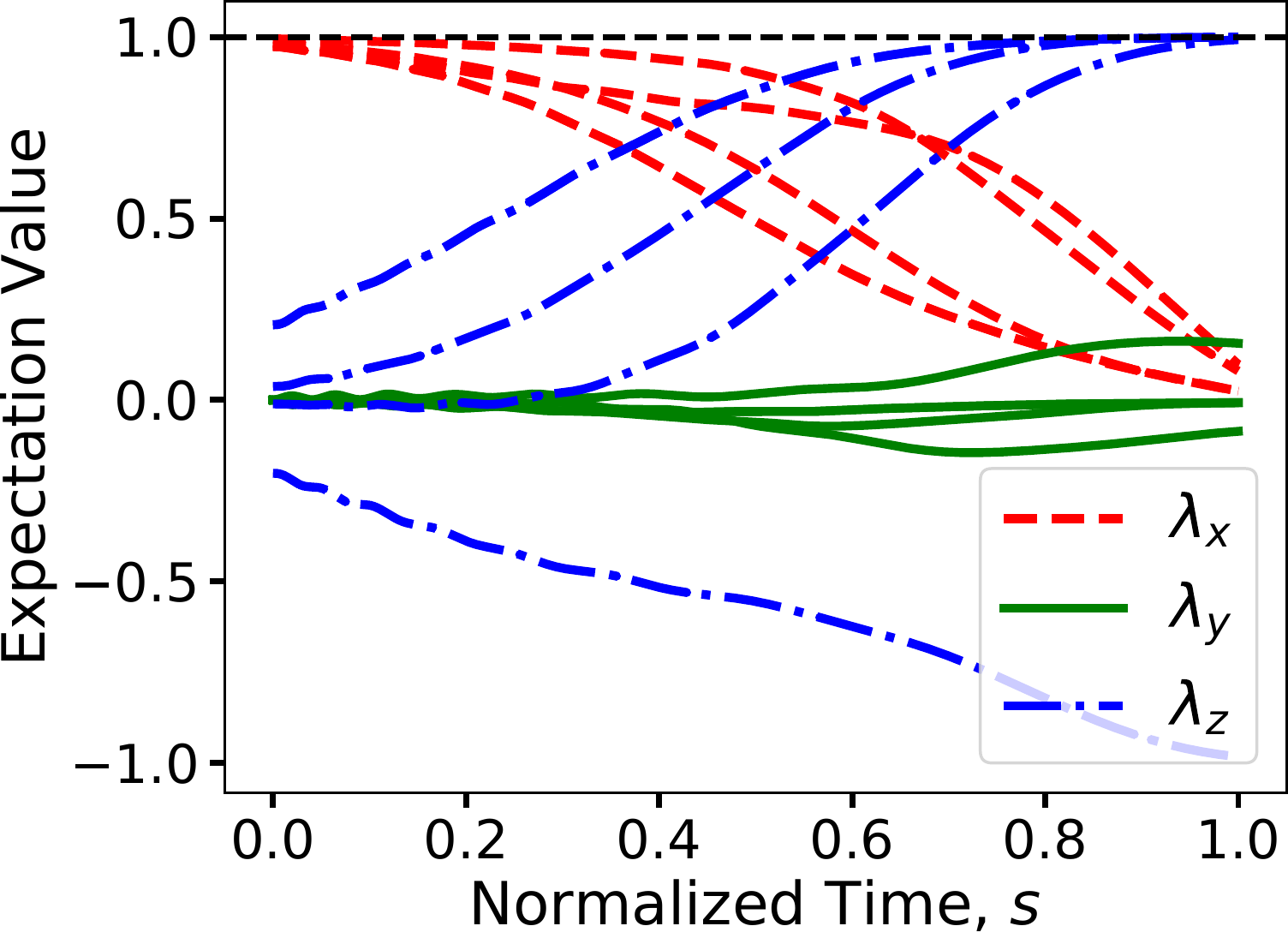}}\quad
  \subfloat[Adiabaticity]{\includegraphics[height=\Height]{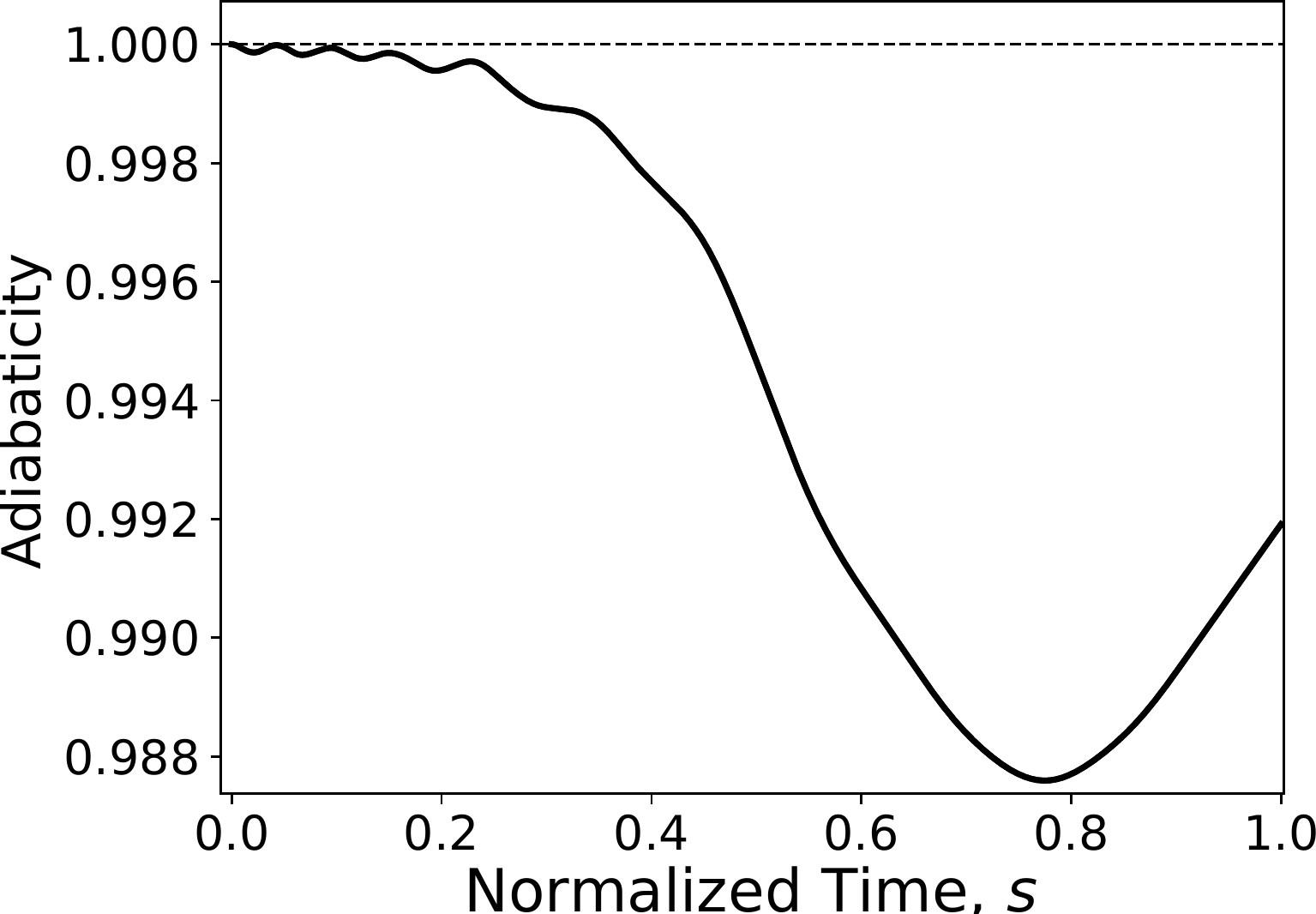}}
  \caption{Simulation of Maj-101 using the Quasi-Linear schedule with $\runrate = 3\sci{-2}$, $\atb_0 = 5$, $\atb_1 = \sfrac{1}{20}$. Initial coherent oscillations arise due to the changing Hamiltonian. These oscillations tend to weaken during clocking; however, a lowered adiabaticity when approaching the critical regime of avoided level crossings can propagate to a reduced final performance.}
  \label{fig: maj101-rabi}
\end{figure}

\begin{figure}
  \newcommand{\WW}{.48\linewidth}
  \subfloat[Unmodified]{\includegraphics[width=\WW]{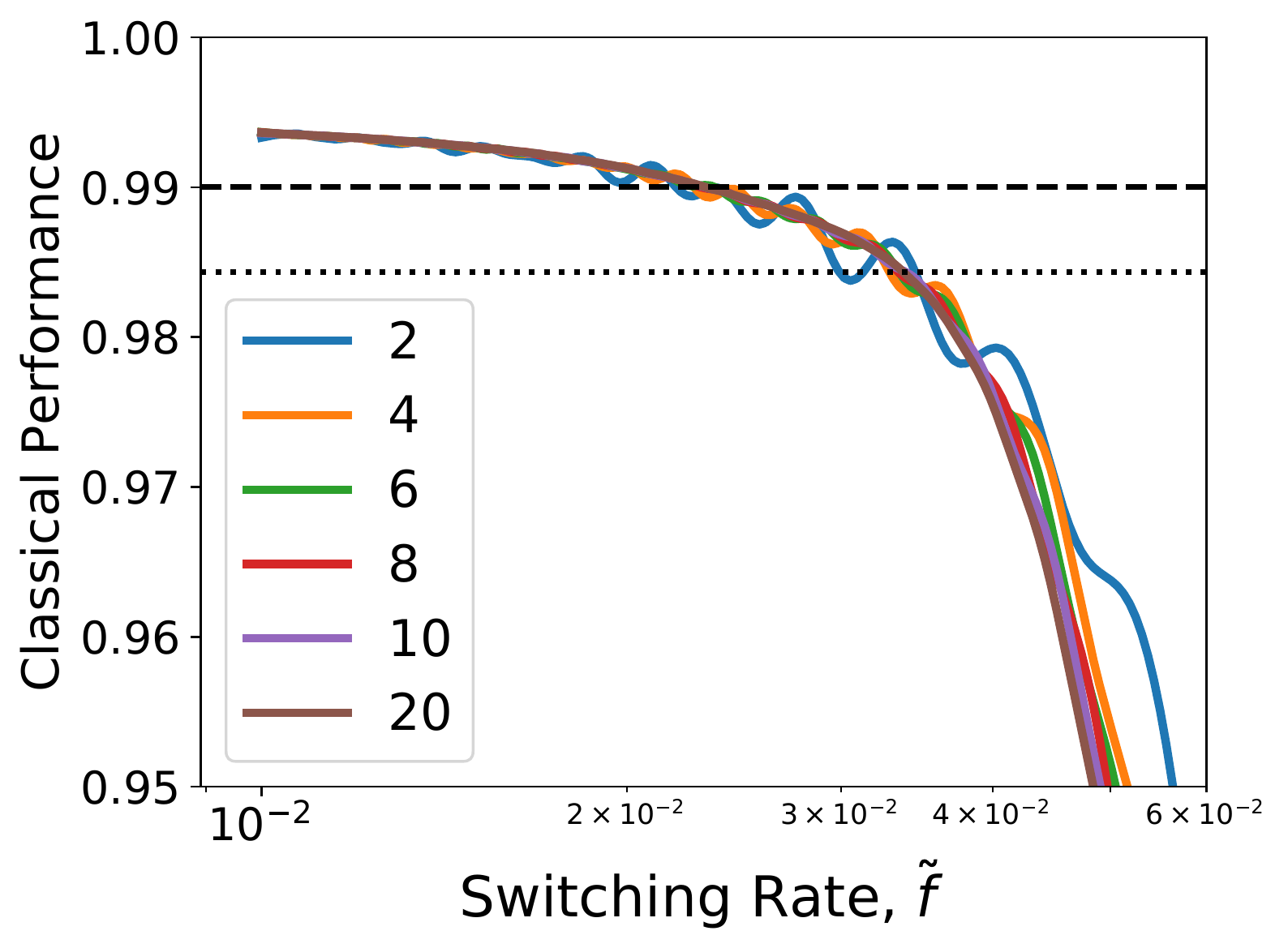}}\quad
  \subfloat[Smoothed]{\includegraphics[width=\WW]{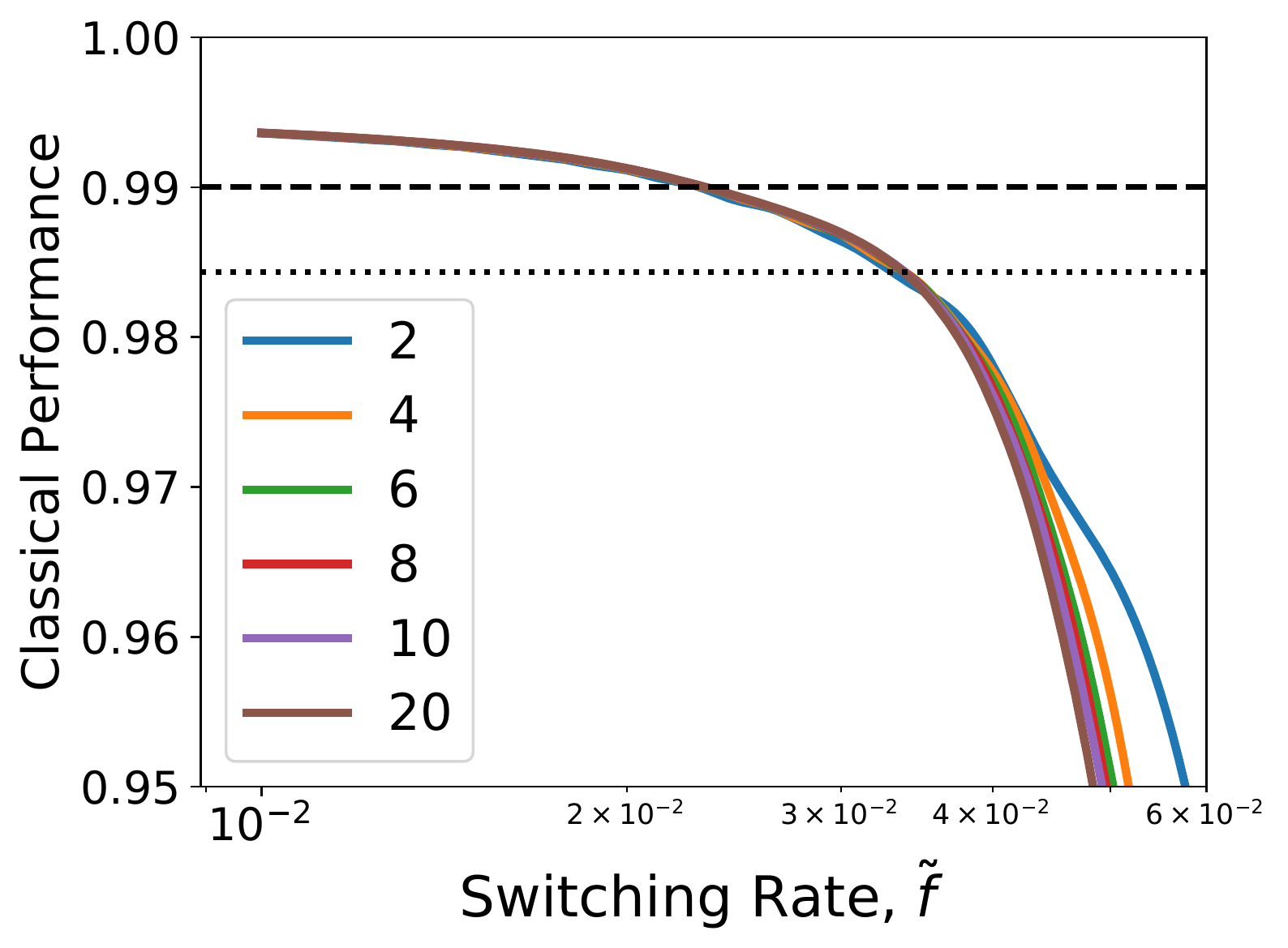}}
  \caption{Classical performance of Maj-101 for different $\atb_0$ values with and without initial smoothing. Results use the Quasi-Linear schedule with $\atb_1 = \sfrac{1}{20}$. The dashed line indicates the $\Met_{cl} = 0.99$ performance threshold, with the dotted line the effective threshold for $\alpha_1 = 0$ using \cref{eq: Qcl}. Note the more prominent oscillations for small $\atb_0$.}
  \label{fig: rabi-osc}
\end{figure}

Even if we initialize the system in the ground state of $\HHd(0)$, we will observe Rabi-like oscillations induced by the clocking field. These oscillations manifest in our performance metrics as can be seen in \cref{fig: maj101-rabi,fig: rabi-osc}. We can approximate the magnitude of oscillations using time dependent perturbation theory for the ground state and first excited state of $\HHd(0)$, ignoring degeneracy for this simple analysis. If $A(s)$ has an initial linear component, we get a transition probability into the excited state of
\begin{equation}
  P_{g \to e} (s) \propto \frac{|\dds{A}(0)|^2}{\Ao^4} \cdot \pa{\frac{\runrate}{\Ao}} \sin\pa{\omgr s},
\end{equation}
where $\Ao = A(0)$ and $\omgr \approx \sfrac{\Ao}{\runrate}$. If $A(s)$ has no initial linear component, we calculate
\begin{equation}
  P_{g \to e} (s) \propto \frac{|\frac{d^2}{ds^2}A(0)|^2}{\Ao^4} \cdot \pa{\frac{\runrate}{\Ao}}^3 \sin\pa{\omega_r s}.
\end{equation}
Importantly, if $\sfrac{\runrate}{\Ao} \ll 1$ we can effectively negate the initial oscillations by modifying the initial clocking schedule in order to remove any linear component. We employ a smoothing procedure in which we apply the map
\begin{equation}
  s' = s \cdot \pa{1 - e^{-s^2/2 \sigma^2}}.
\end{equation}
This map has a few useful properties: (1) It leaves the initial value of the schedule unchanged; (2) it cancels the linear component of the initial schedule; and (3), it only affects the schedule over a period of $\sim 2\sigma$, meaning we can remove the initial oscillations without affecting the later schedule. We set $\sigma = 2 \cdot 2\pi\sfrac{\runrate}{\Ao}$ to smooth over two periods of the oscillations. The effect is clear in \cref{fig: rabi-osc}b.

\subsection{Choosing the Initial Clocking Field}

\begin{figure}[t]
  \newcommand{\WW}{.48\linewidth}
  \subfloat[Wire-5]{\includegraphics[width=\WW]{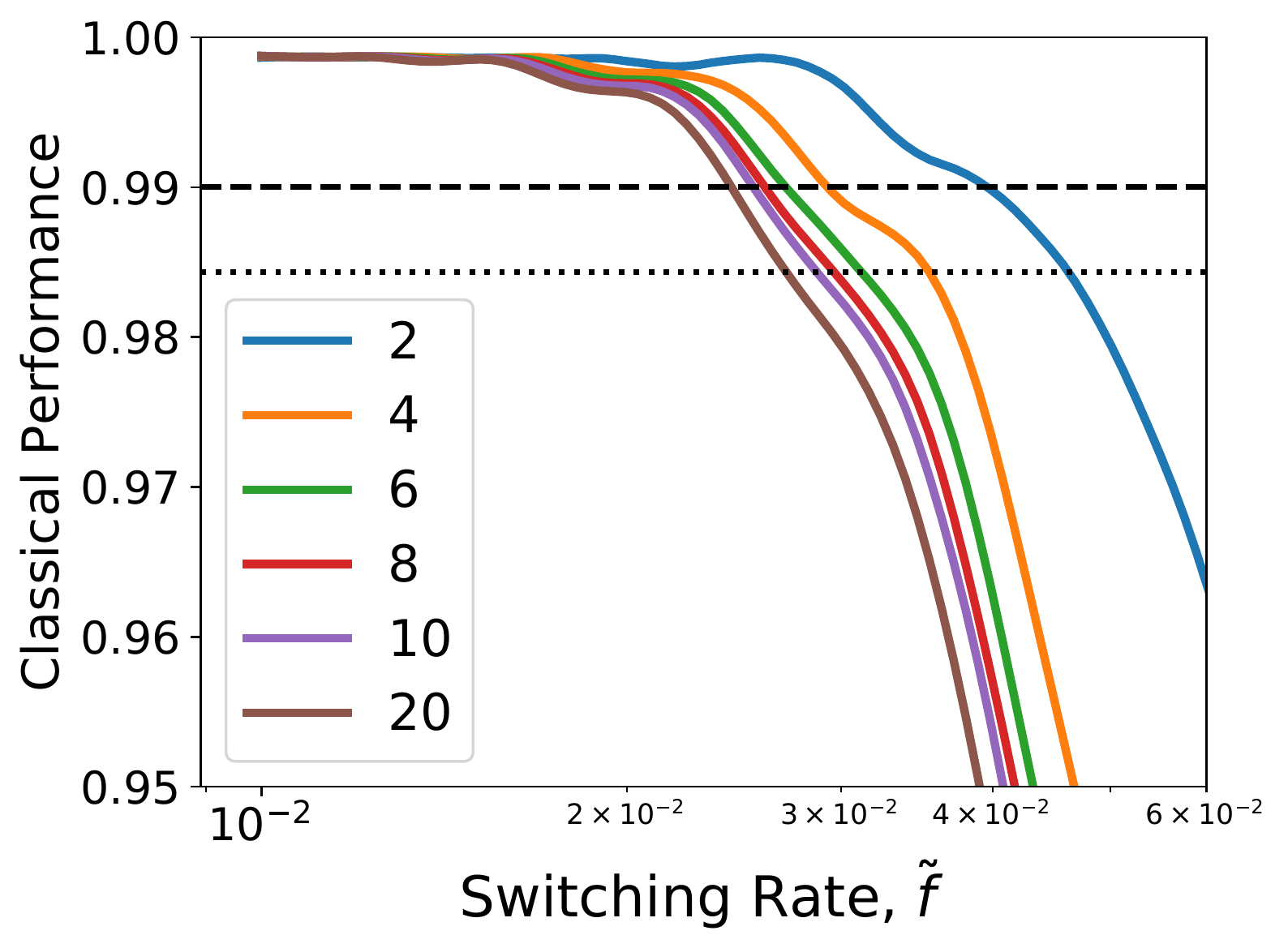}}\quad
  \subfloat[Inverter]{\includegraphics[width=\WW]{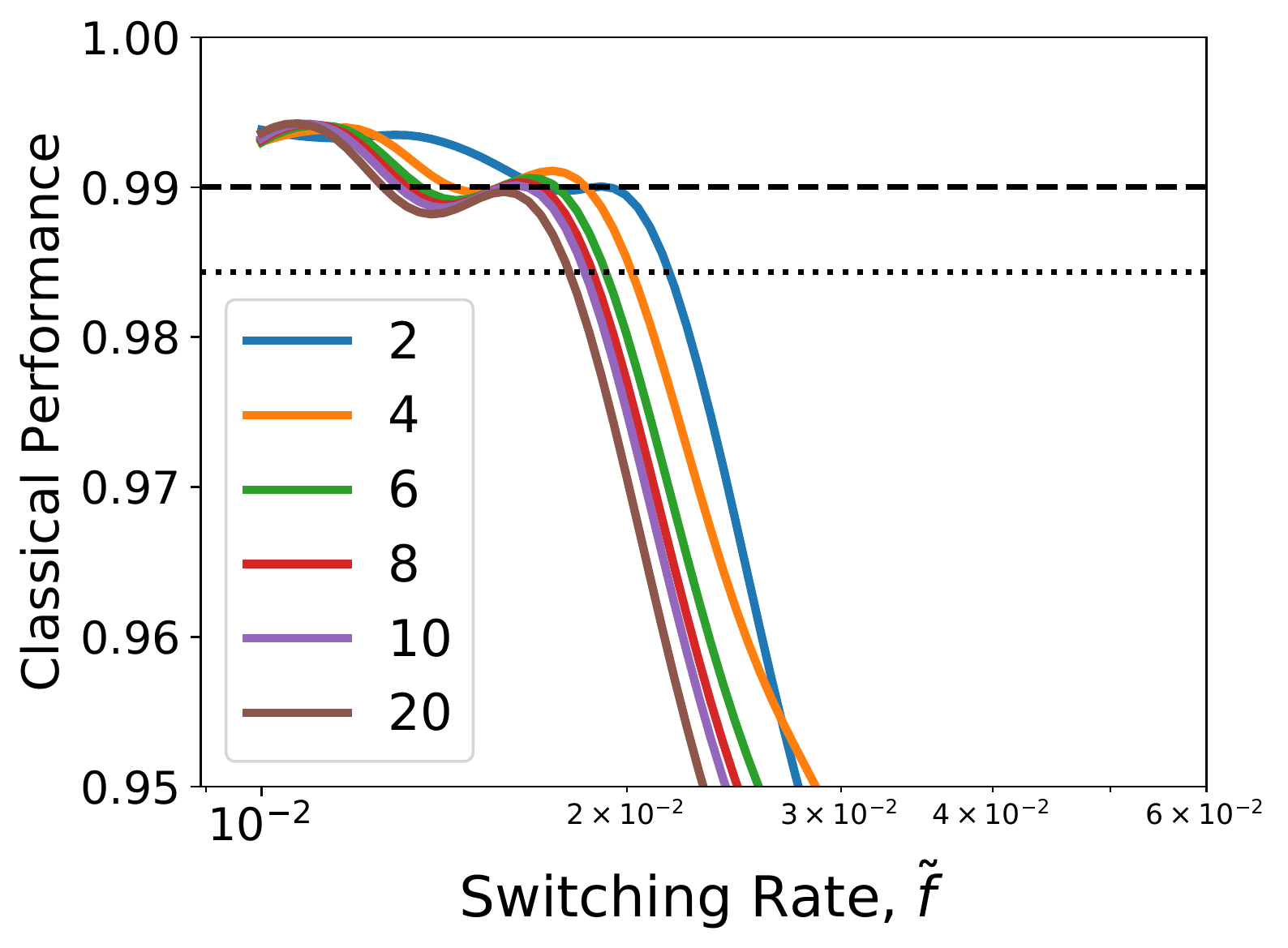}}
  \caption{Classical performance for different $\atb_0$ values with $\atb_1 = \sfrac{1}{20}$. In cases where there is a clear avoided level crossing in the low energy spectrum, smaller $\atb_0$ values yield higher operating switching rates. We use the left-most intercept when defining maximum switching rate to account for potential performance oscillations. The lower $F_0$ value of the wire gives it a higher $\Met_1$ upper bound.}
  \label{fig: a0-comp}
\end{figure}

\begin{figure}
  \newcommand{\Width}{.48\linewidth}
  \subfloat[Wire-5]{\includegraphics[width=\Width]{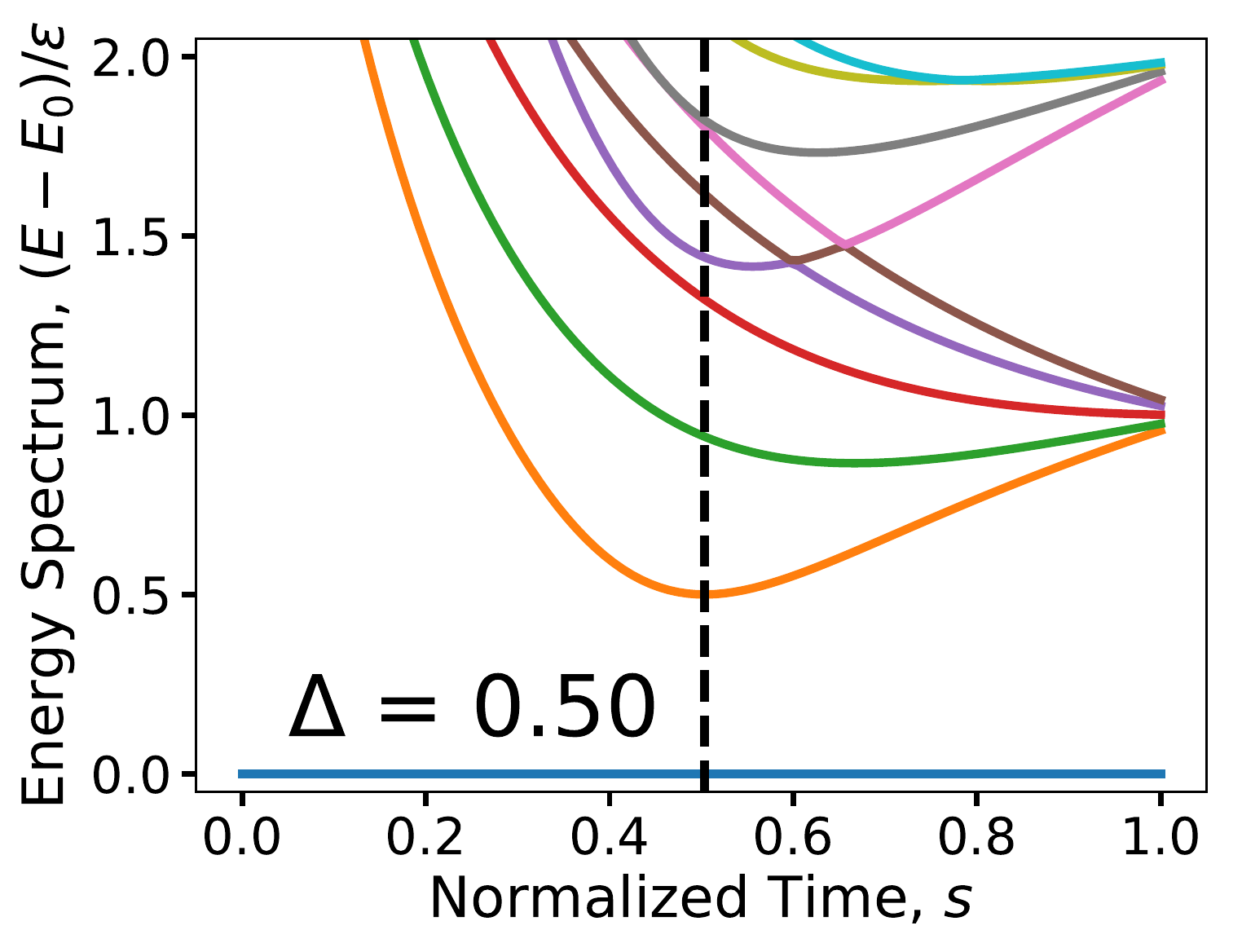}}\quad
  \subfloat[Inverter]{\includegraphics[width=\Width]{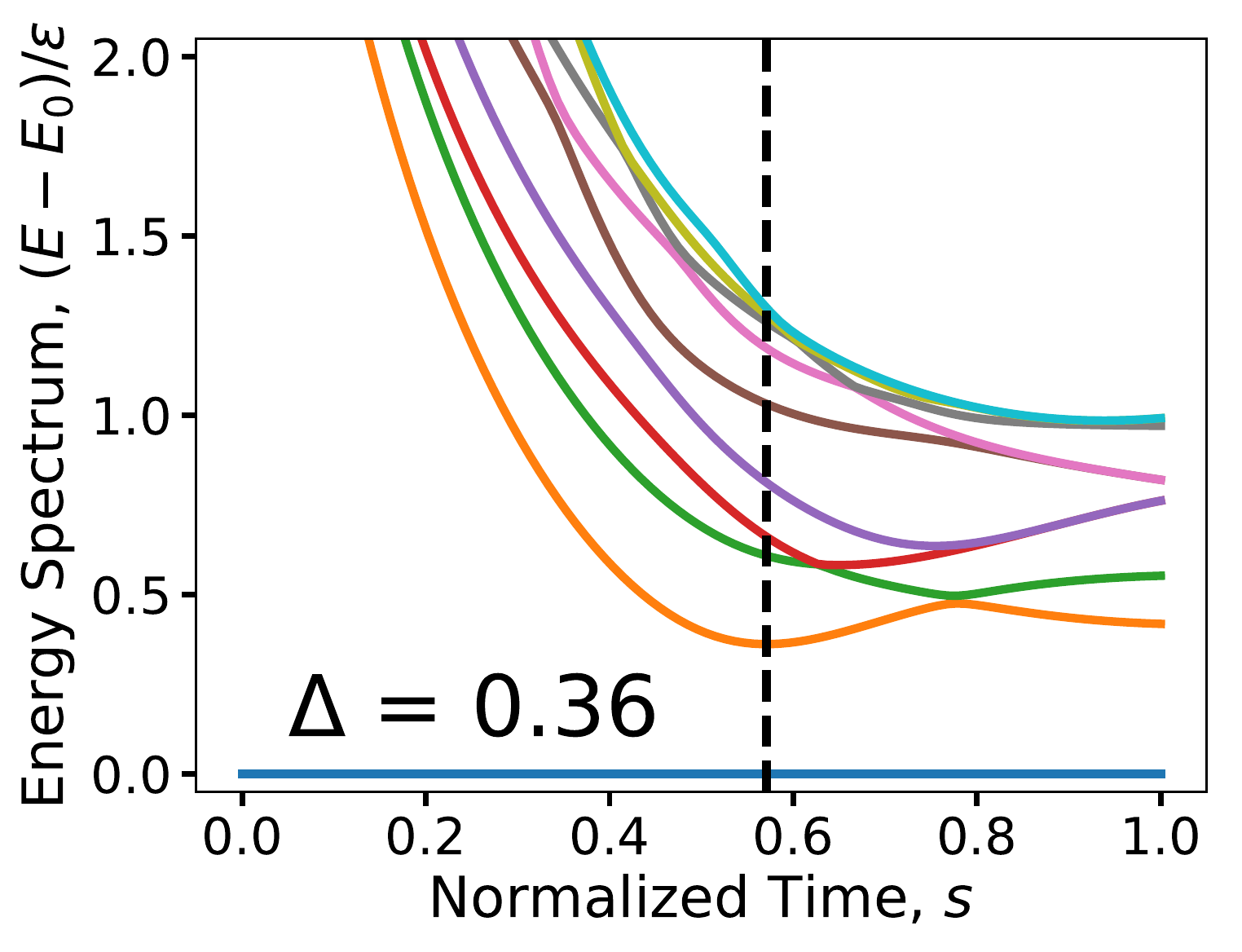}}\\
  \subfloat[Maj-111]{\includegraphics[width=\Width]{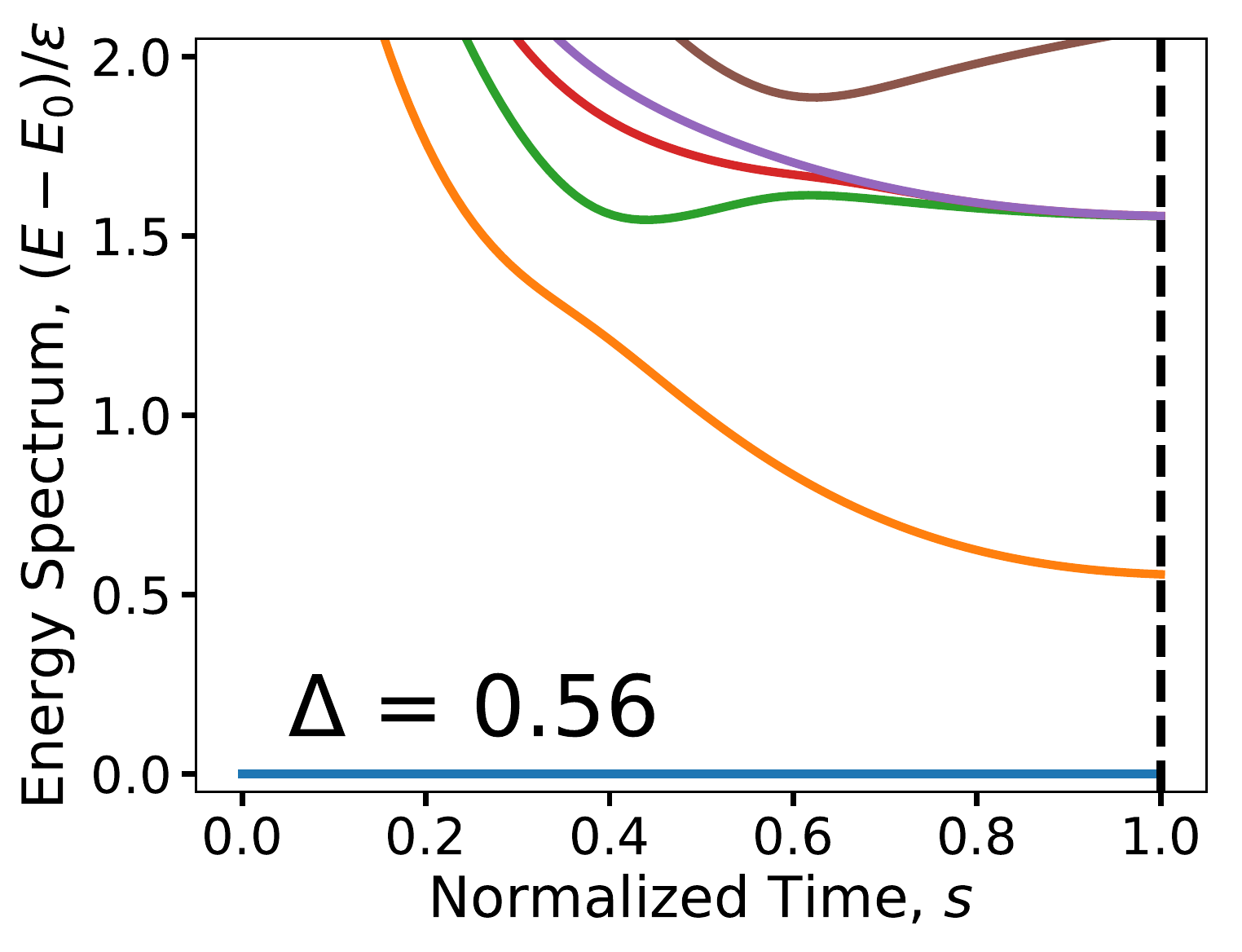}}\quad
  \subfloat[Maj-101]{\includegraphics[width=\Width]{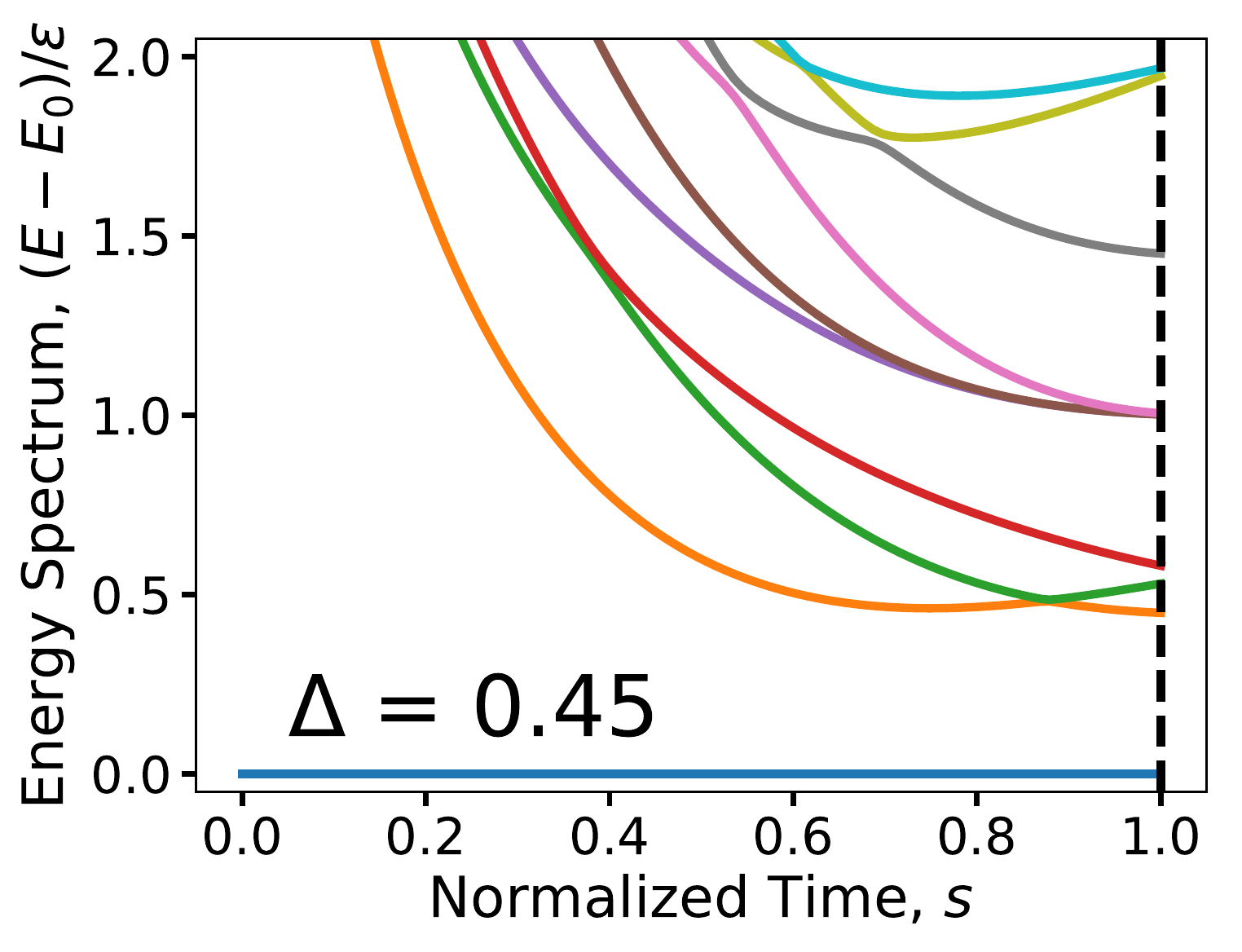}}
  \caption{First 10 energy levels of the simulated devices. The location and size of the minimum gap is indicated.}
  \label{fig: spectra}
\end{figure}

For Maj-101, the choice of $\atb_0$ did not significantly influence the classical performance. This will not generally be the case. In \cref{fig: a0-comp}, we see the performance for both the inverter and wire for different initial clocking values. In these cases, the lower the value of $\atb_0$, the higher the performance. If we assume the only important feature of the clock to be a single Landau-Zener-like avoided level crossing, we can solve \cref{eqn:lz} to find the maximum switching rate that allows a given adiabaticity,
\begin{equation}
  \runrate_{max} \approx \frac{\pi \Delta_0 W}{-2\ln(1-\Met_A)},
\end{equation}
where the gap width, $W$, is inversely proportional to the slope of the clocking field near the minimum gap. For the Quasi-Linear schedule, the slope satisfies
\begin{equation}
  \dds{A_Q}(s) = \frac{\atb_0^2}{\brak{1+(\atb_0-1)s}^2} \dds{A_Q}(1),
\end{equation}
where $\dds{A_Q}(1) = -(1-\sfrac{\atb_1}{\atb_0}) \approx 1$ for small $\atb_1$. We then expect $\runrate_{max} \propto \brak{s_* + (1-s_*)/\atb_0}^2$ with $s_*$ the location of the minimum gap. The spectra for most of the simulated devices are shown in \cref{fig: spectra}. For majority gates, the minimum gap occurs in the classical limit, $s_*=1$, hence we expect no $\atb_0$ dependence. If the minimum gap occurs at some earlier $s_*$, we expect $\runrate_{max}$ to decrease with increasing $\atb_0$ as observed.

QCADesigner uses a value of $\atb_0=5$ and from these considerations we might naively assume we should use an even smaller value of $\atb_0$; however, we are ignoring some important caveats. We initialize the system near the ground state of $\HHd(0)$. If $\atb_0$ is small, the initial state already has a significant projection onto the classical ground state. In that sense, much of the work the switching is supposed to achieve must have been done by whatever process set up the initial state. Indeed from \cref{eqn: qual} we see that $\atb_0=5$ gives a projection onto the ground state of $-\HX$ of only $\Met_0 = 0.96$ in the worst case of a majority gate. In addition, the cell polarizations, $\lmu{z}{}$, will have initial values on the order of $\atb_0^{-1}$, as high as $0.2$ in \cref{fig: maj101-rabi}, pointing to a second issue: with small $\atb_0$, ``deactivated'' cells may remain partially polarized, perhaps enough that the next clock zone may bias our outputs. Both of these issues are resolved by using a higher value of $\atb_0$; however, we see in \cref{fig: a0-comp} that there is only minimal decrease in performance beyond $\atb_0=5$. 
We conclude then that we cannot justify using a smaller value of $\atb_0$, nor would we expect using a larger value to influence performance within the scope of our analysis.

\section{Coherent Behaviour of QCA Components}
\label{sec: coherent}

Here we consider the behaviour of our QCA components excluding any dissipation. In all results that follow, we use clocking ratios of $\atb_0 = 5$ and $\atb_1 = \sfrac{1}{20}$. Unless otherwise indicated, the Quasi-Linear schedule is used.

\subsection{Wire Analysis}
\label{subsec: wire-performance}

\newcommand{\lzf}{\csa}

\begin{figure}
 \newcommand{\Height}{.35\linewidth}
 \subfloat[Classical performance using Quasi-Linear clocking]{\includegraphics[height=\Height]{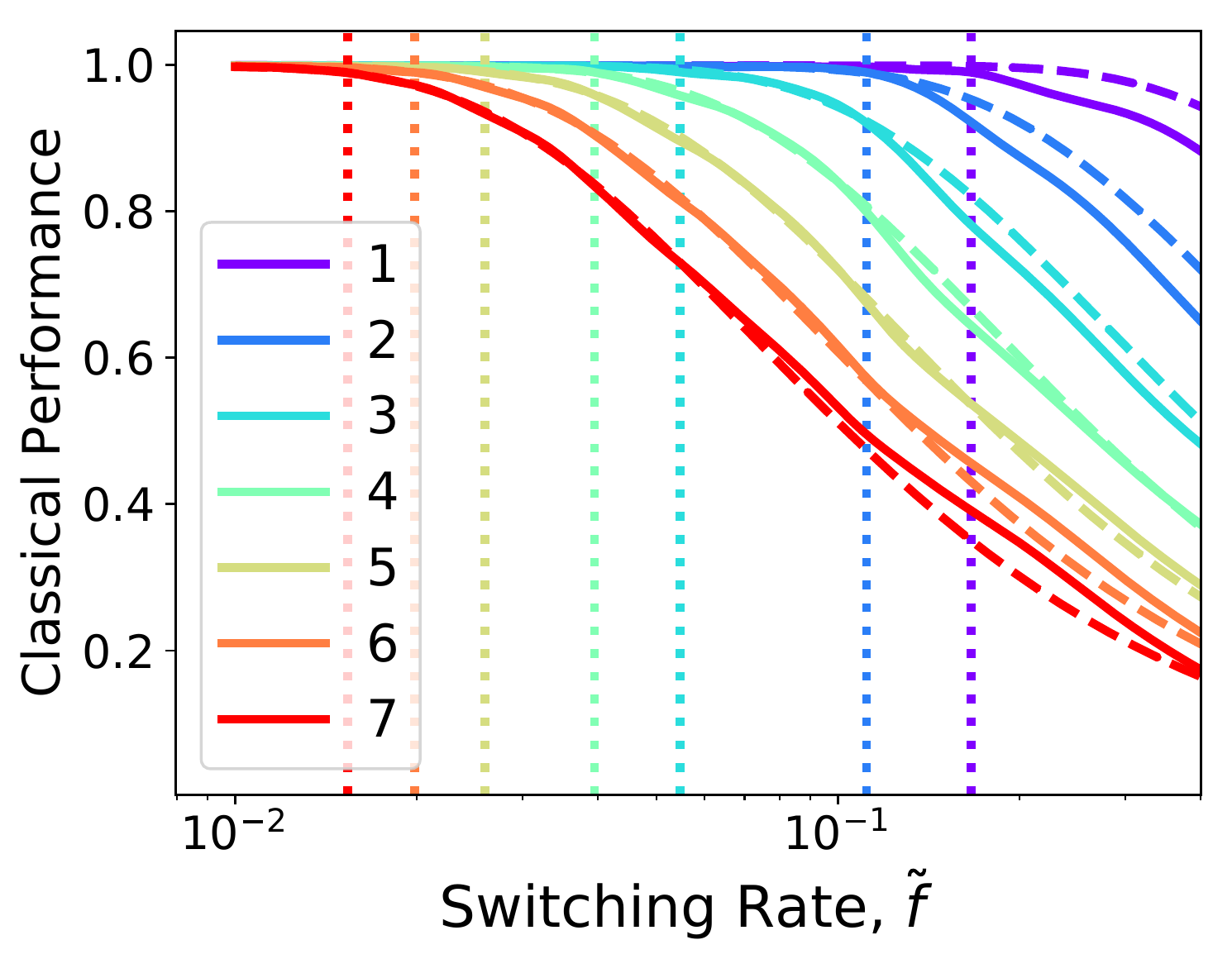}}\quad
 \subfloat[Maximum switching rates for different clocking schedules.]{\includegraphics[height=\Height]{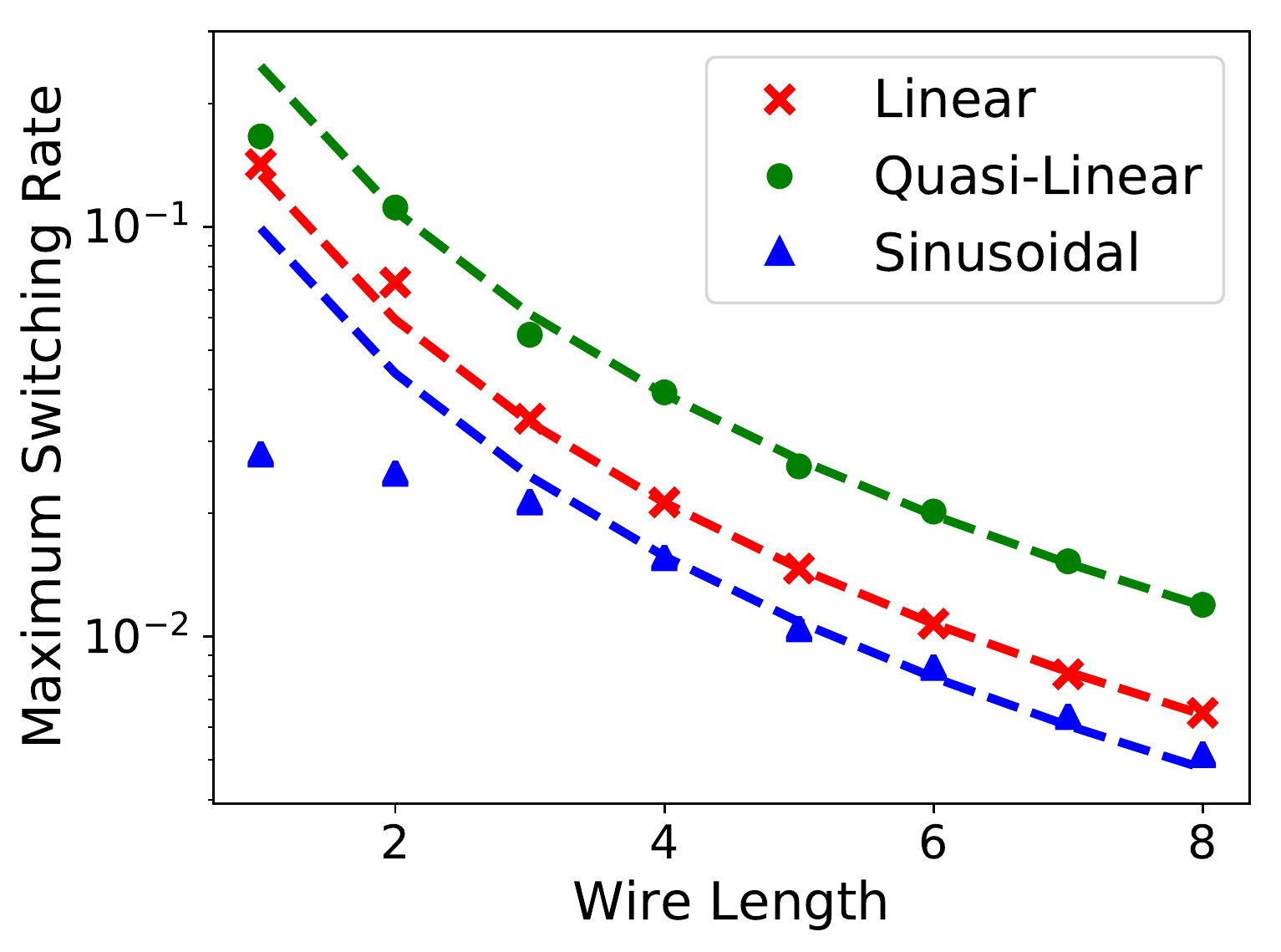}}
 \caption{Performance of wires of different lengths. Dashed lines indicate predictions from \cref{eqn: fmax-wire} with fit $\lzf$.}
 \label{fig: wire-performance}
\end{figure}

In \cref{fig: wire-performance}, we consider the performance of wires of various lengths. For the special case of a left-driven wire as in \cref{fig:qca-schem}, the Hamiltonian is of the form
\begin{equation}
  \HHd_{W}(s) = -\frac{1}{2} A(s) \sum_{i=1}^N \hs{x}{i} + \frac{1}{2} B(s) \Big [ \hs{z}{1} - \sum_\pairs{ij}^N \hs{z}{i} \hs{z}{j} \Big ].
\end{equation}
Using a slight modification to the Jordan-Wigner transform approach used for the unbiased wire\cite{dzi2005}, we can obtain an analytic description of the low energy spectrum for biased wires in terms of the set of eigenenergies
\begin{equation}
    \epsilon_k = \sqrt{ B^2(s) \sin^2(q_k) + \brak{A(s)-B(s)\cos(q_k)}^2  },
\end{equation}
with $k$ in $1,\cdots,N$ and pseudo-momenta $q_k = \sfrac{k\pi}{(N+1)}$. Each energy level of $\HHd_W(s)$ is found by summing over a subset of the eigenenergies $\epsilon_k$. For large $N$, each of the $\epsilon_k$ can be fit to the hyperbolic form \cref{eq: hyper} with the minimum occurring for $A(s_*) = B(s_*)$ and parameters:
\begin{equation} \label{eqn: wire-pars}
  \Delta_0 \approx B(s_*) q_k, \qquad W = \frac{\Delta_0}{\Delta m} \approx \frac{B(s_*) q_k}{\Delta m},
\end{equation}
where $\Delta m = \tfrac{d}{ds}|A-B|(s_*)$ is the difference in schedule rates at the gap. Importantly, for a wire of length $N$, both $\Delta_0$ and $W$ are approximately proportional to $q_k \propto (N+1)^{-1}$. The Landau-Zener approximation then predicts an adiabaticity of 
\begin{equation}\label{eqn: Qa-lz-wire}
  \Met_A(N, \runrate) \approx 1-e^{-\sfrac{2\pi\lzf}{4\runrate(N+1)^2}}.
\end{equation}
If there were only one excited state, $\sfrac{\nu}{(N+1)^2}$ would correspond to $\Delta_0 W$ for $k=1$ and large $N$; however, as there are $N$ excited wire states, we will just take $\nu$ to be some parameter dependent on the clocking schedule. Using \cref{eq: Qcl} and solving \cref{eqn: Qa-lz-wire} for $\runrate$ , the maximum switching rate for 99\% classical performance is then
\begin{equation} \label{eqn: fmax-wire}
  \runrate_{max}(N) \approx -\frac{\pi \lzf}{2\log(1-\sfrac{0.99}{Q_1(N)})(N+1)^2},
\end{equation}
Simulations were run for wires of various lengths and their maximum switching rates extracted. The results are shown in \cref{fig: wire-performance}. These rates were then fit with \cref{eqn: fmax-wire} for $N \geq 4$ to obtain $\lzf_L \approx 1.60$, $\lzf_Q \approx 2.93$, and $\lzf_S \approx 1.18$ for our respective Linear, Quasi-Linear, and Sinusoidal schedules. It is clear that this simple Landau-Zener model gives a good estimate of wire performance, at least beyond $N=3$. Using \cref{eqn: wire-pars} and considering only the first excited state, we can estimate $\lzf \approx \lzf_1$:
\begin{table}
 \caption{Comparison of fit and analytical estimates of the Landau-Zener $\lzf$ parameter for wires. Errors indicate $2\sigma$ deviations from the fit covariance matrix.}
 \label{table: lz-wire}
 \begin{ruledtabular}
   \begin{tabular}{c@{\hspace{2ex}}|cc@{\hspace{5ex}}}
     Clocking schedule  &  $\lzf_{fit}$  &  $\lzf_1$ \\
     \hline
     \rule{0pt}{3ex}
     Linear        & $1.60 \pm 0.02$  &  $1.80$ \\
     Quasi-Linear  & $2.93 \pm 0.07$  &  $3.19$  \\
     Sinusoidal    & $1.18 \pm 0.04$  &  $1.99$
  \end{tabular}
 \end{ruledtabular}
\end{table}
\begin{equation}
  \lzf_1 = (N+1)^2 \left. \Delta_0 W \right|_{k=1} \approx \frac{\pi^2}{\Delta m} B^2(s_0).
\end{equation}
The other excited states serve as additional channels for diabatic transitions and hence we should expect $\lzf < \lzf_1$. For all our schedules, we can find expressions for $\lzf_1$:
\begin{subequations}
    \begin{flalign}
      \csa_1^L &= \pi^2\frac{(1-\sfrac{\alpha_1}{\alpha_0})^2}{\brak{2-(\alpha_1+\sfrac{1}{\alpha_0})}^3} = 1.80, \\
      \csa_1^Q &= \pi^2\frac{\kay + (\alpha_0-1)}{\kay^2(\alpha_0-1)} = 3.19,
      \\
      \csa_1^S &= \frac{4\pi}{\sqrt{(\alpha_0-\alpha_1)^2 + 4(\alpha_0-1)(1-\alpha_1)}} = 1.99.
    \end{flalign}
\end{subequations}
For comparison, the fit $\lzf$ and analytical estimates of $\lzf_1$ are listed in \cref{table: lz-wire}. We see that $\lzf<\lzf_1$ as expected. Further, with the exception of the Sinusoidal schedule, the first excited state gives a good estimate for $\lzf$. We arrive at two important observations: (1) QCA wires seem to adhere to the simple Landau-Zener model for adiabaticity, even considering only the first excited state; and (2), in the absence of additional factors, the $(N+1)^{-2}$ scaling of $\runrate_{max}$ means the maximum switching rate of any QCA network will quickly be limited by the longest wire to be clocked. The Landau-Zener character of wires, in addition to the ease of calculating the energy spectrum, suggests an obvious target for future investigation of an optimised wire clocking schedule.

\subsection{QCA Building Blocks}

\begin{figure}
  \newcommand{\WW}{.48\linewidth}
  \newcommand{\dpack}[1]{\includegraphics[width=\WW]{freq_sweep_#1}}
  \subfloat[Wire-5]{\dpack{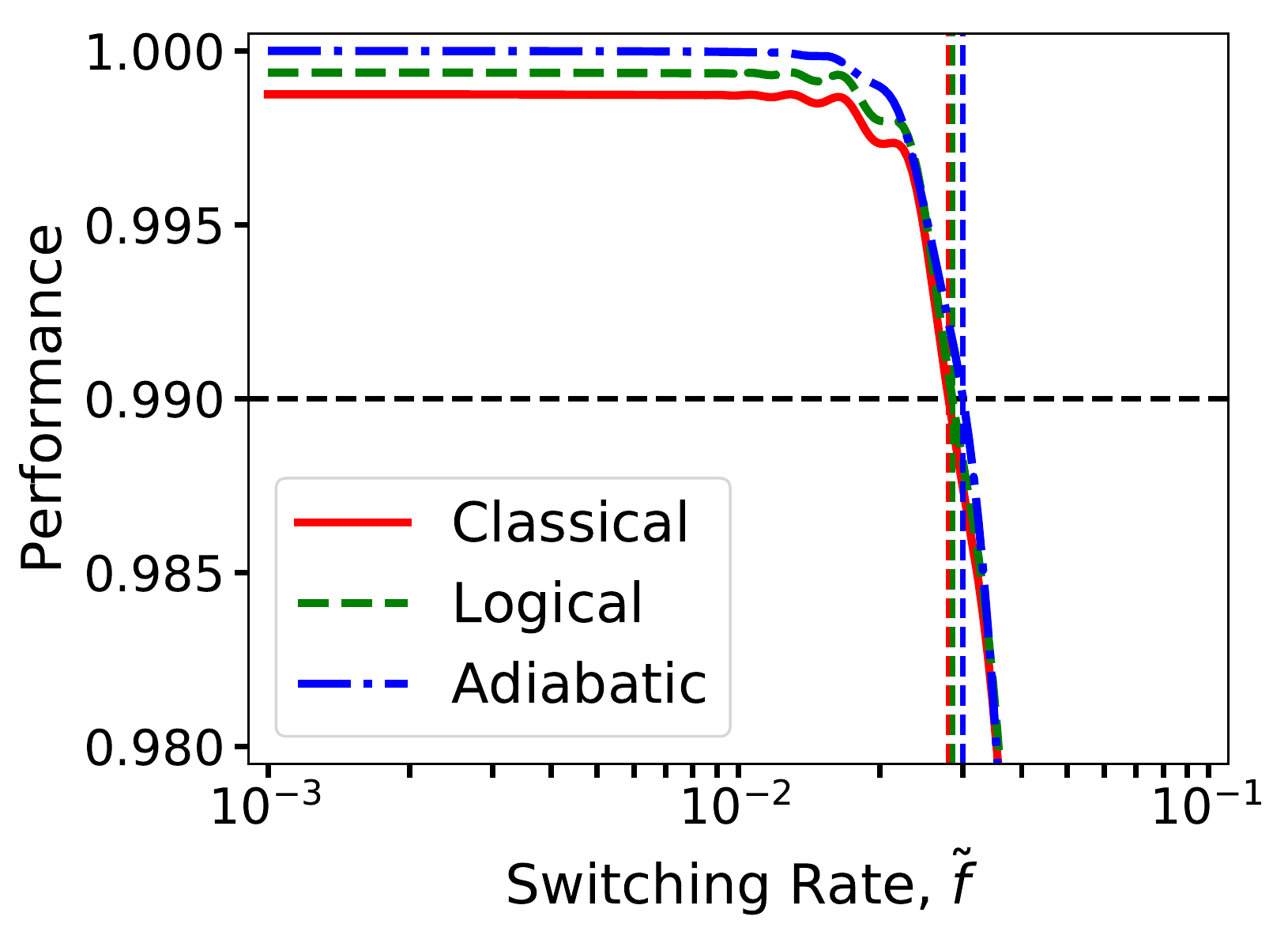}}\quad
  \subfloat[Inverter]{\dpack{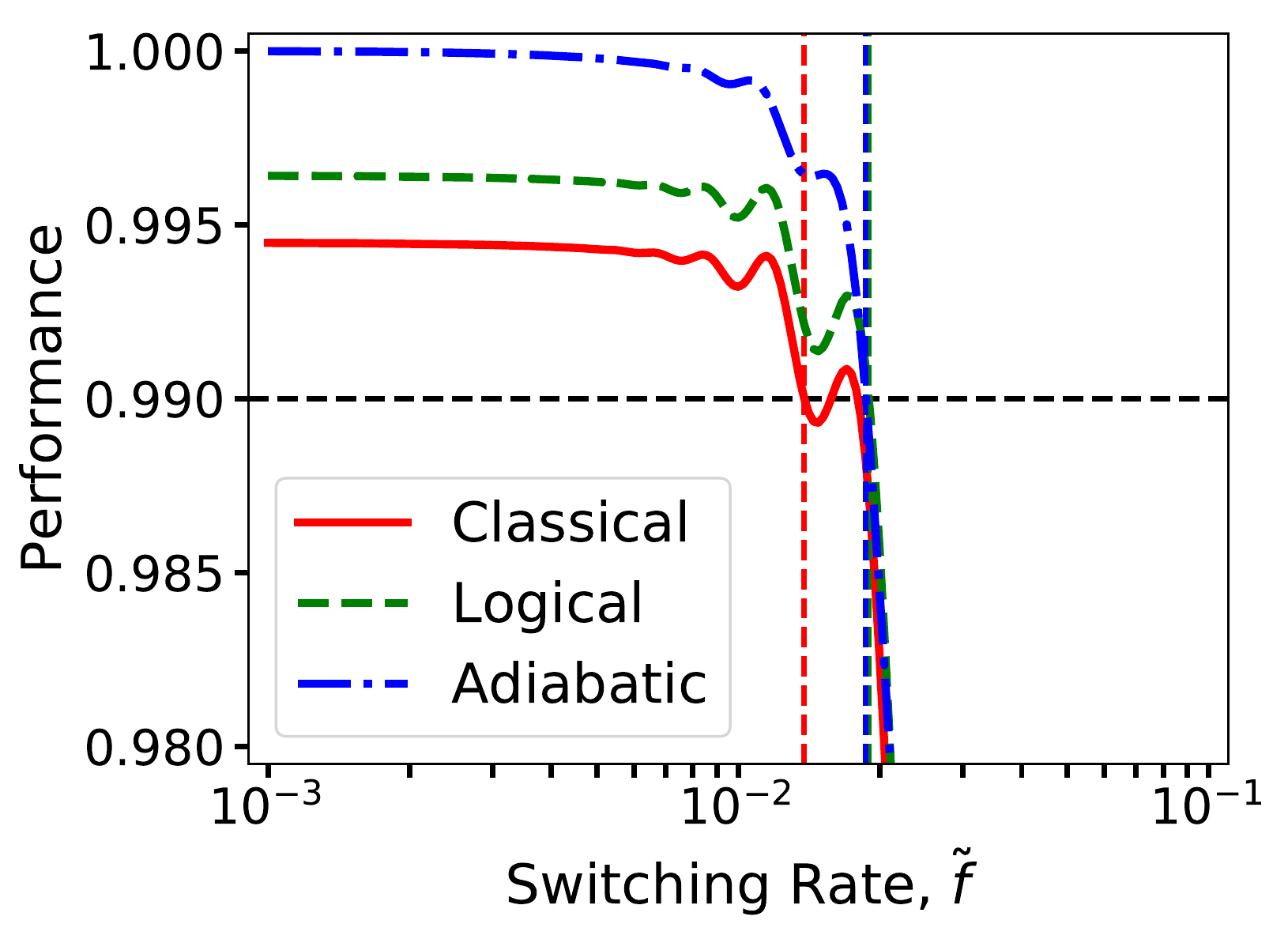}}\\
  \subfloat[Maj-111]{\dpack{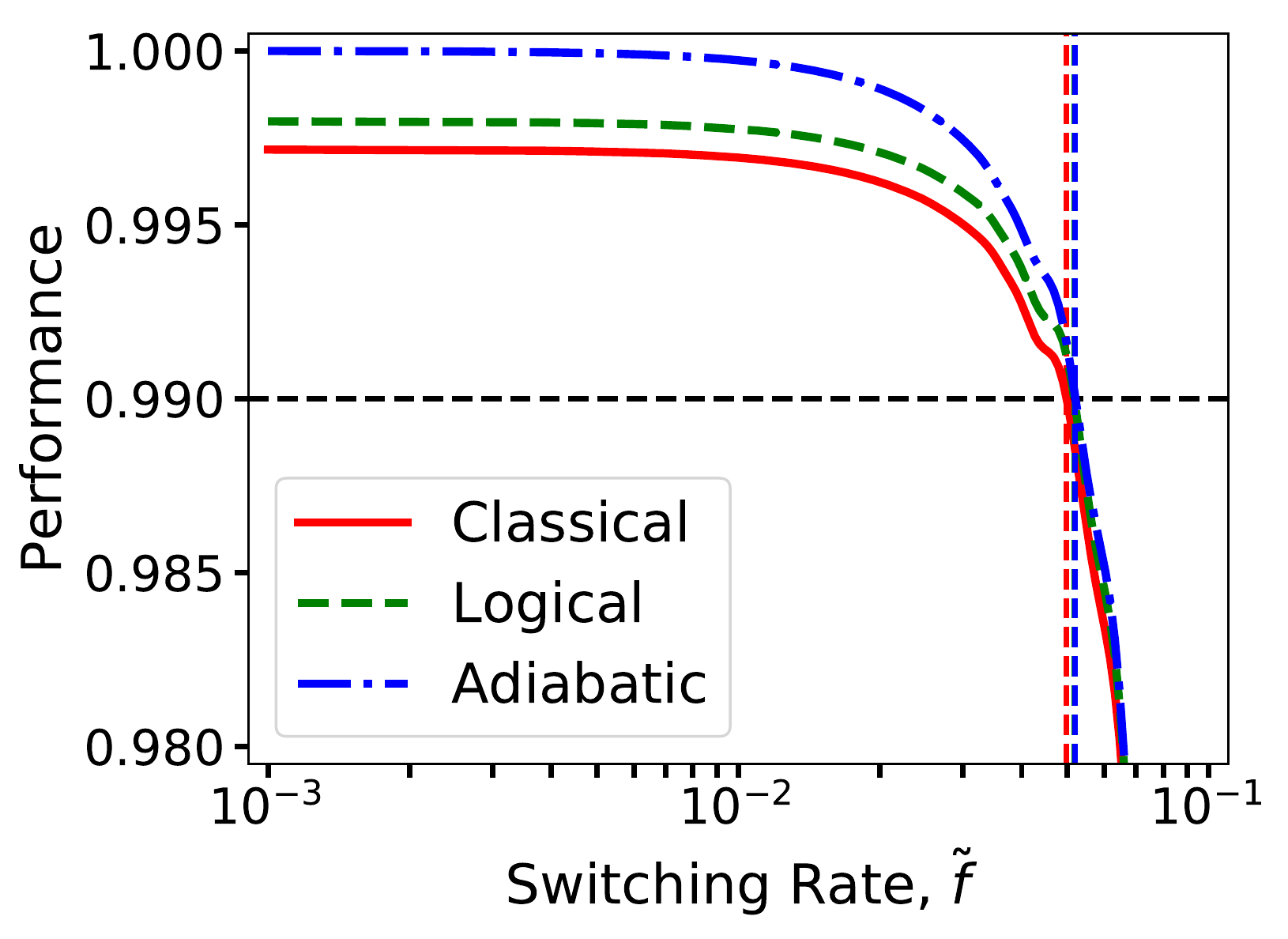}}\quad
  \subfloat[Maj-101]{\dpack{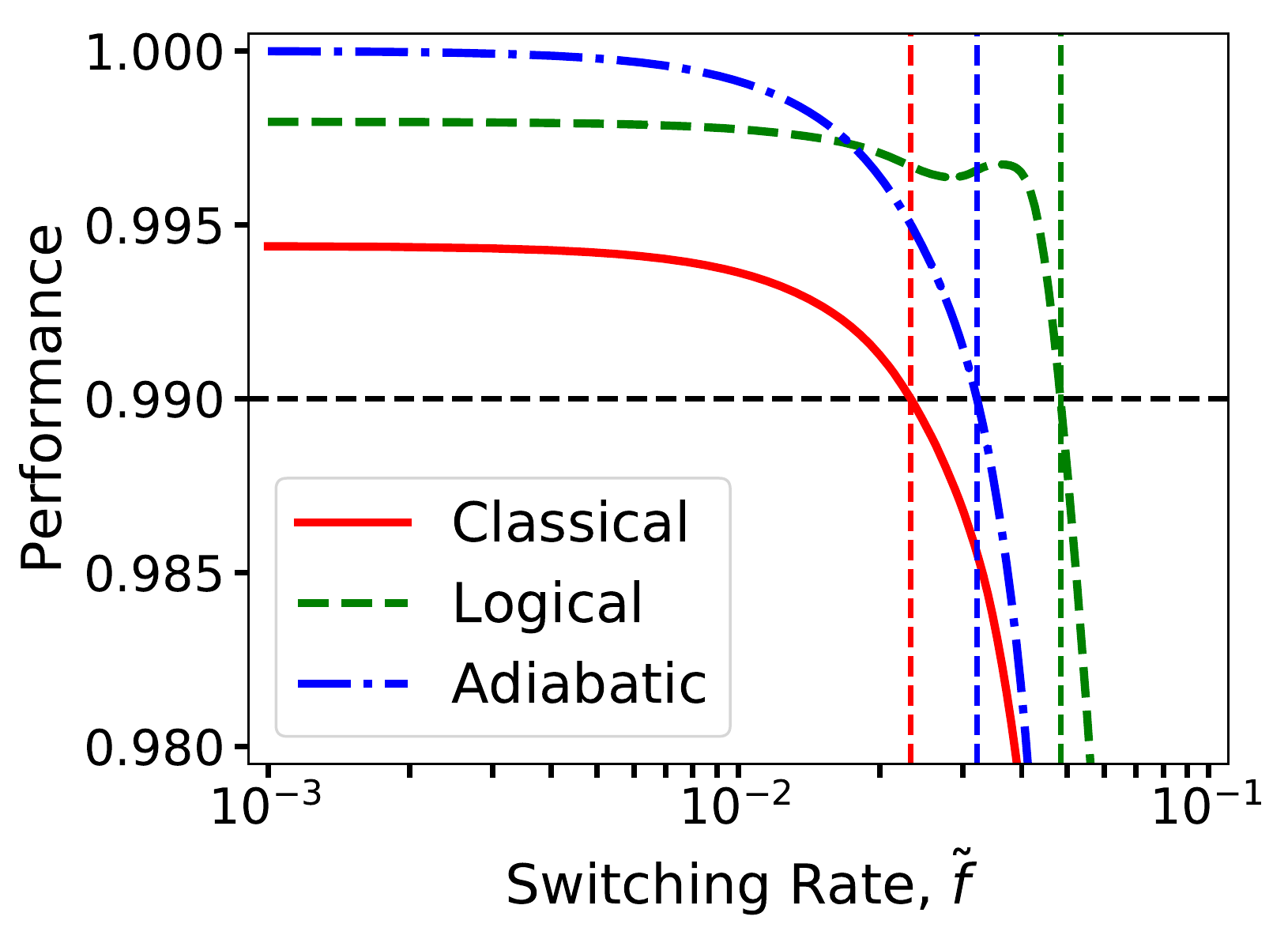}}
  \caption{Performance metrics for the QCA building blocks. 99\% thresholds for the different metrics are indicated.}
  \label{fig: coherent-performance}
\end{figure}

\begin{figure}
 \newcommand{\Height}{.35\linewidth}
 \subfloat[Maj-110 spectrum and performance metrics.]{
  \includegraphics[height=\Height]{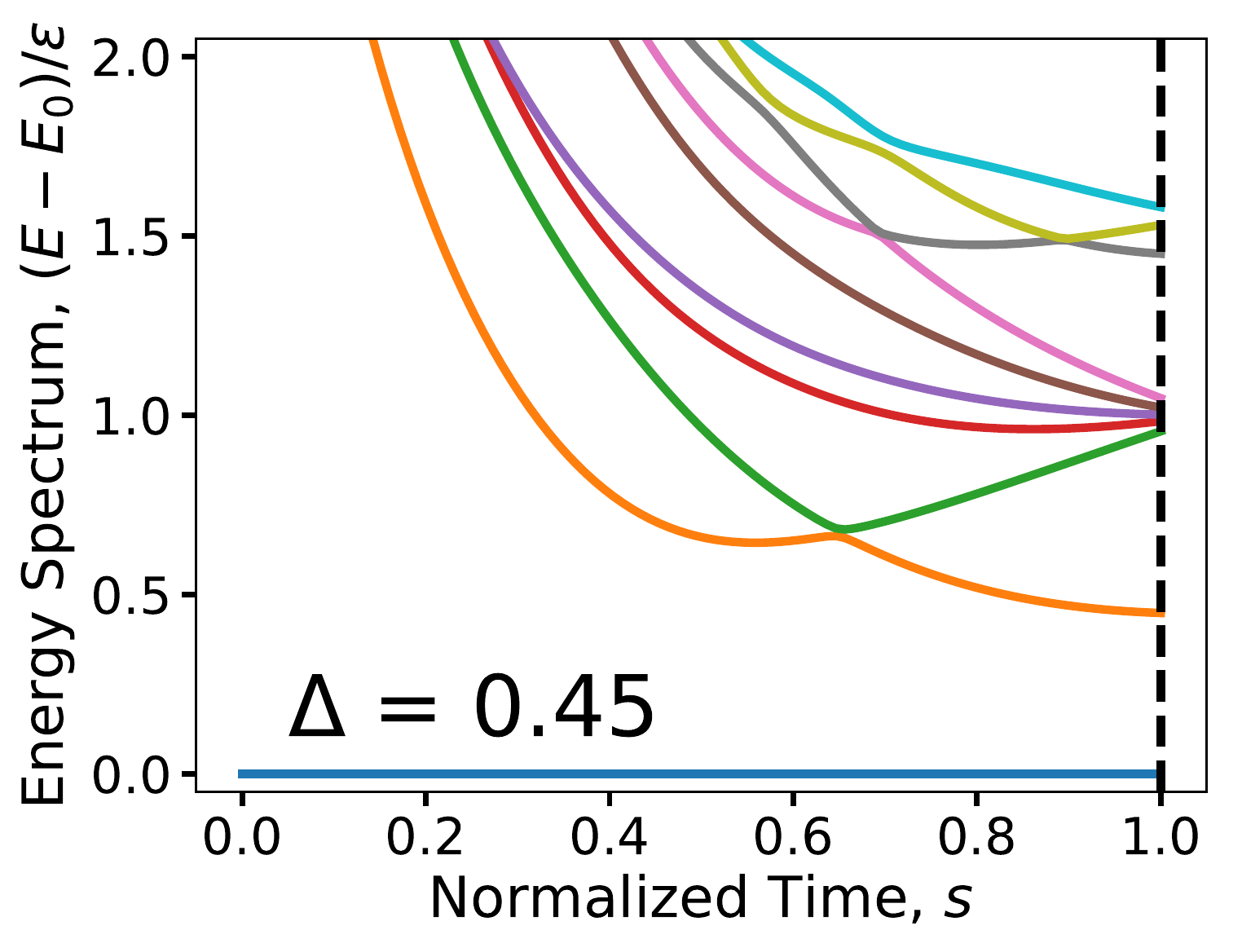} \quad \includegraphics[height=\Height]{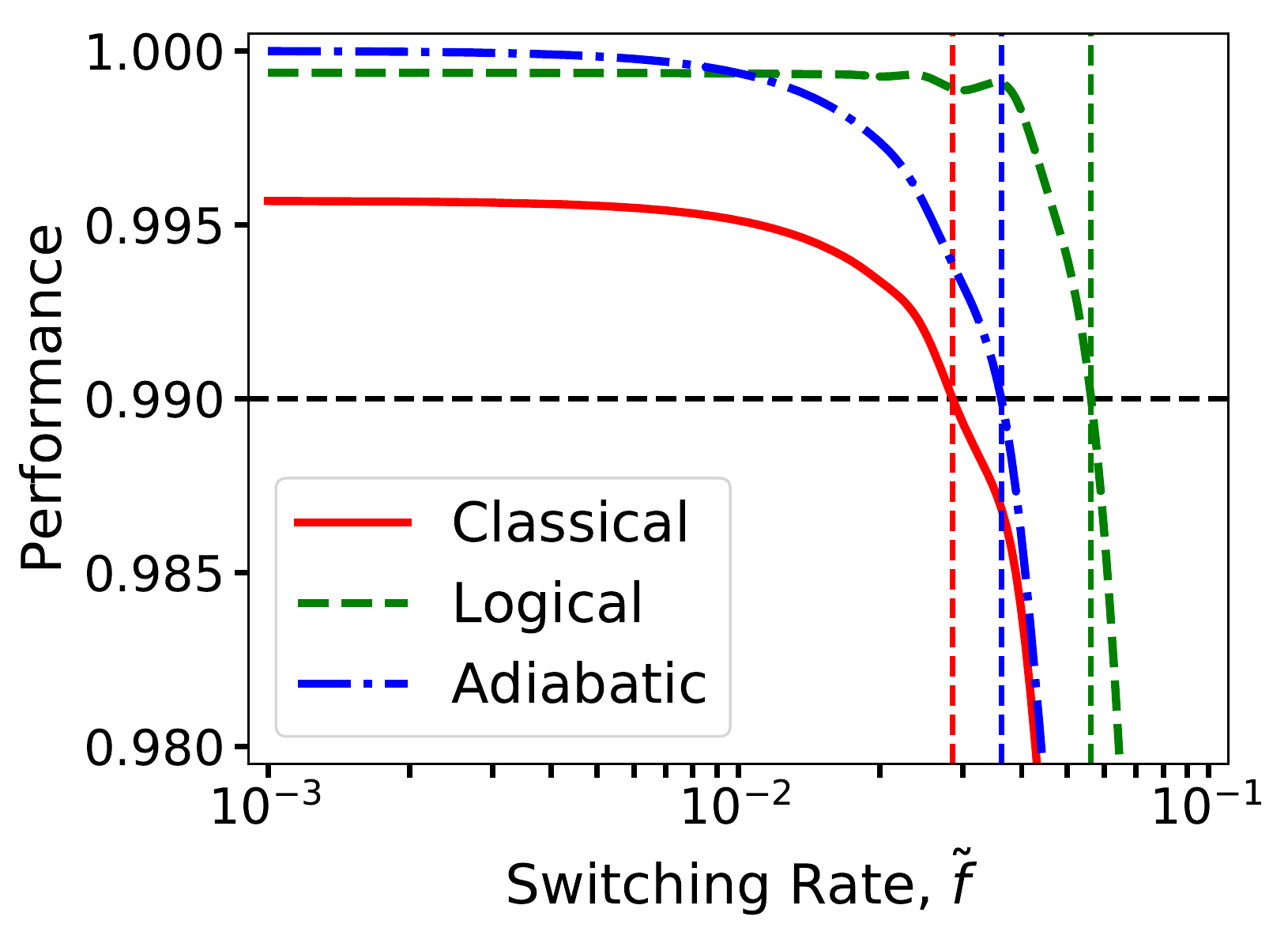}}\\
  \subfloat[Ground State]{\includegraphics[height=\Height]{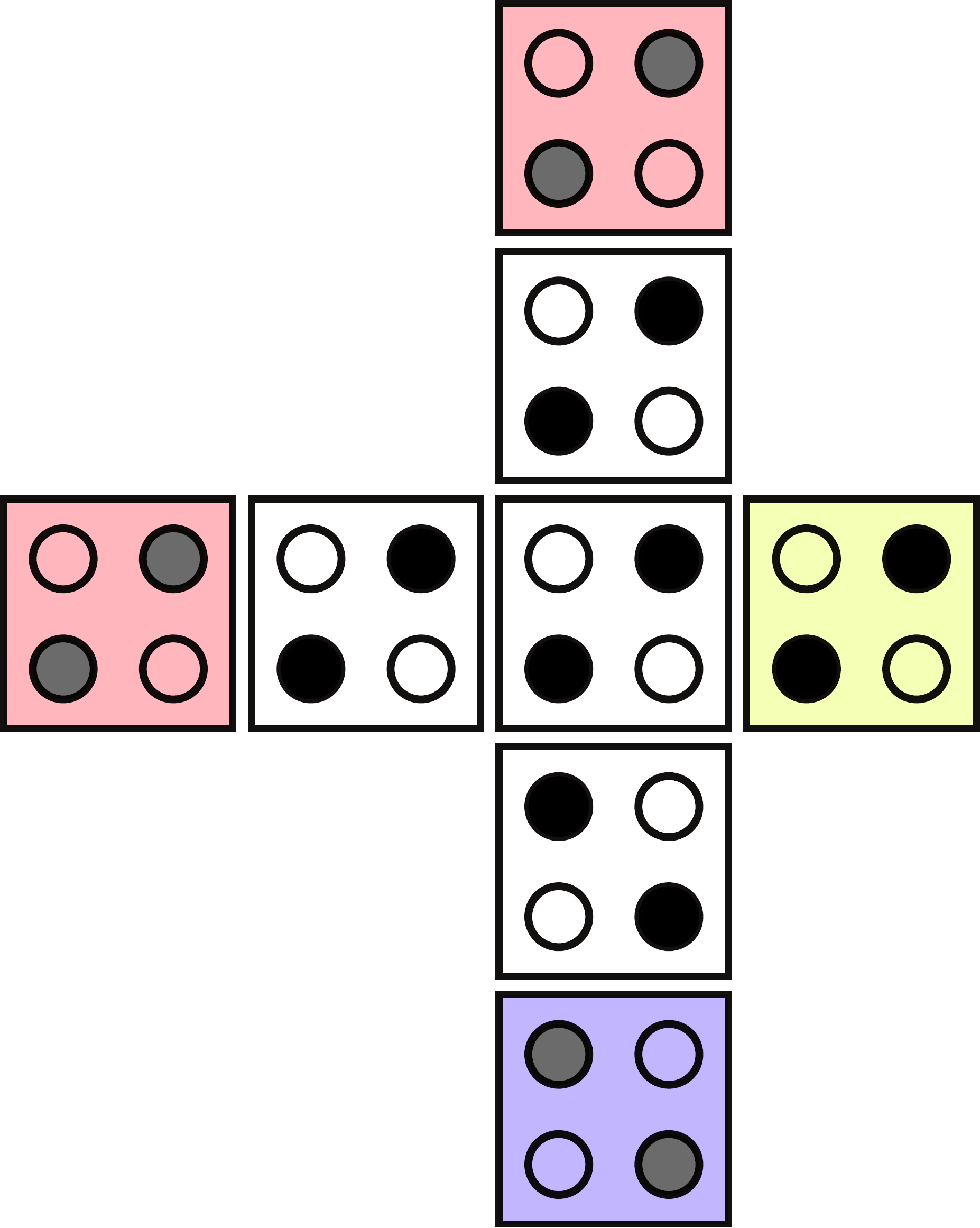}}\hspace{2cm}
  \subfloat[Excited State]{\includegraphics[height=\Height]{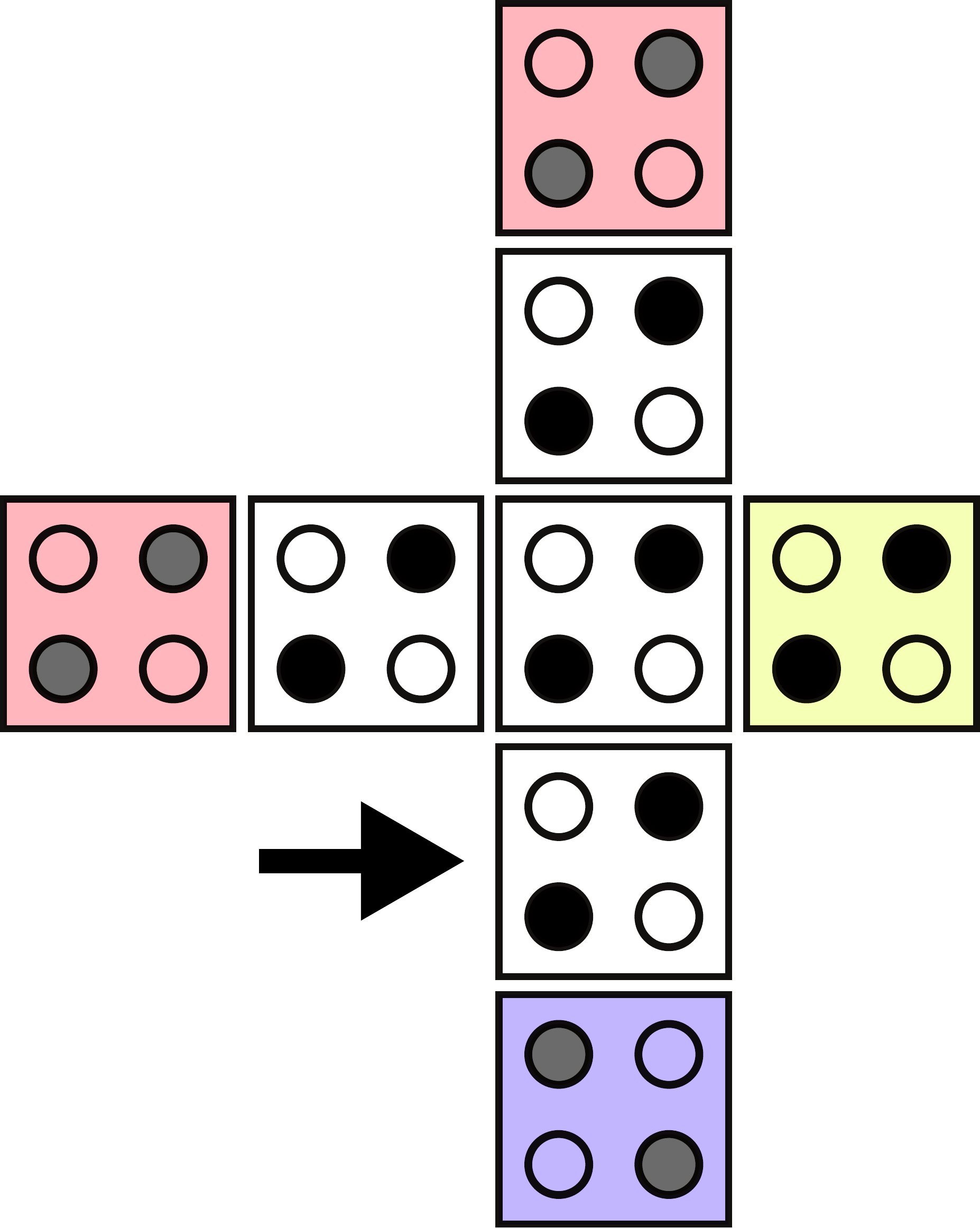}}
  \caption{The first excited state for Maj-110 has a logical output which agrees with the ground state, resulting in an increased $\Met_L$ metric over $\Met_{cl}$ and higher switching rates.}
  \label{fig: maj110}
\end{figure}

We consider now the coherent performance of our set of basic QCA components. We transition from using the classic performance to using the logical performance, \cref{eq: met_L}, for the component output cells. This is perhaps a more meaningful metric for real device performance and will also allow more direct comparison with the ICHA results in the next section. \cref{fig: coherent-performance} shows all the performance metrics for each device as a function of the switching rate. There are a number of things to note:
(1) in the limit of high adiabaticity, or slow switching, the metrics approximately differ only by the ratios determined in \cref{subsec: gs-quality}:
\begin{equation}
  \sfrac{\Met_{cl}}{\Met_A} \approx \Met_1, \qquad \sfrac{\Met_L}{\Met_A} \approx 1-\tfrac{1}{4} \sfrac{\atb_1^2}{|\htl{n}|^2},
\end{equation}
with $\htl{n}$ the effective bias for the output cell in the classical limit;
(2) the remaining features and oscillations in the performance arise later in the simulations and are not a consequence of the initial Rabi oscillation discussed in \cref{subsec: cohosc};
(3) the predicted maximum switching rate for wires doesn't significantly depend on the choice of metric, meaning our discussions in the previous section also apply to the logical performance;
and (4), we observe that in certain cases the logical performance gives significantly higher maximum switching rate than either of the other metrics. Note that a network is logically correct either when in the ground state or when in an excited state which happens to have correct outputs. This is precisely the case for Maj-101 and Maj-110; the latter is illustrated in more detail in \cref{fig: maj110}. A summary of the maximum operating frequencies for the different devices and metrics in included in \cref{table: coherent-metrics}. Because of their simplicity, we should expect wires to be the highest performing devices of a given size. Our inverters have an input-to-output path length of 5 cells and maximum operating frequencies similar but slightly lower than those of Wire-5. Majority gates have a path length of 3 cells from each input, which we can compare against the results for Wire-3. This suggests our bounds for wires found in the previous section may serve as upper bounds on more complicated components through an effective input-to-output path length. Another necessary observation is that majority gates, unlike wires and inverters, do not have a prominent hyperbolic minimum gap in their spectra. We should not be surprised then that the minimum gap is not a good predictor for their performance. A more detailed analysis of the low energy spectrum would yield a better prediction.

\begin{table}
    \caption{Maximum operating frequencies, $\runrate_{max}/4$, for our basic QCA circuits based on different metrics. We include also Wire-3 which has a comparable size scale to the majority gates. Minimum gaps are included for reference. }
    \label{table: coherent-metrics}
    \begin{ruledtabular}
      \newcommand{\hsp}{\hspace{2ex}}
      \begin{tabular}{c@{\hsp}|ccc@{\hsp}|c@{\hsp}}
          Device & Adiabatic & Classical & Logical  & $\Delta_0$ \\
          \hline
          \rule{0pt}{3ex}
          \textbf{Wire-5}      & $7.5\sci{-3}$  & $7.0\sci{-3}$  & $7.1\sci{-3}$  & $0.500$  \\
          \textbf{Inverter}    & $4.7\sci{-3}$  & $3.5\sci{-3}$  & $4.7\sci{-3}$  & $0.362$  \\
          \hline
          \rule{0pt}{3ex}
          Wire-3              & $15.6\sci{-3}$ & $15.0\sci{-3}$ & $15.1\sci{-3}$  & $0.707$ \\
          \textbf{Maj-111}     & $13.0\sci{-3}$  & $12.5\sci{-3}$  & $12.9\sci{-3}$  & $0.556$ \\
          \textbf{Maj-101}     & $8.0\sci{-3}$  & $5.8\sci{-3}$ & $12.1\sci{-3}$  & $0.449$  \\
          \textbf{Maj-110}     & $9.1\sci{-3}$  & $7.1\sci{-3}$  & $14.1\sci{-3}$  & $0.448$ 
      \end{tabular}
    \end{ruledtabular}
\end{table}

\section{Dissipative Behaviour}
\label{sec: dissipative}

We will consider two approaches to modelling a simple dissipation mechanism for environmental interactions: \textit{spectral relaxation}, in which the density operator relaxes to some $\hrhos(s)$ dependent only on the eigenspectrum of $\HHd(s)$ as in \cite{taucer2015}; and \textit{mean field relaxation}, where the coherence vectors of each cell relax to a local dissipation vector given by the instantaneous state of the network \cite{mahler1998}. For spectral relaxation, we consider three different potential steady states:
\begin{subequations}
 \label{eq: rho-ss}
 \begin{align}
   &\text{Boltzmann}\cite{taucer2015}:\quad     &\sskt(s) &= \tfrac{1}{\ZZ} \gibbs{\HHd (s)}\\
   &\text{Ground}:\quad       &\ssgr(s) &=  \tfrac{1}{\ZZ} \proj{G}(s) \\
   &\text{Classical}\cite{karim2014b}:\quad    &\sscl(s) &= \tfrac{1}{\ZZ} \proj{\PP}
 \end{align}
\end{subequations}
where $\proj{G}(s)$ and $\proj{\PP}$ are the projectors onto the eigenspaces of the respective ground states of $\HHd(s)$ and $\HP$ (see \cref{eqn: sch}), and $\ZZ$ is an appropriate normalization constant such that $\tr \hrhos(s) = 1$. Each of these spectral steady states has a local dissipation vector:
\begin{figure}
    \newcommand{\WW}{.48\linewidth}
    \subfloat[$\ssgr$]{\includegraphics[width=\WW]{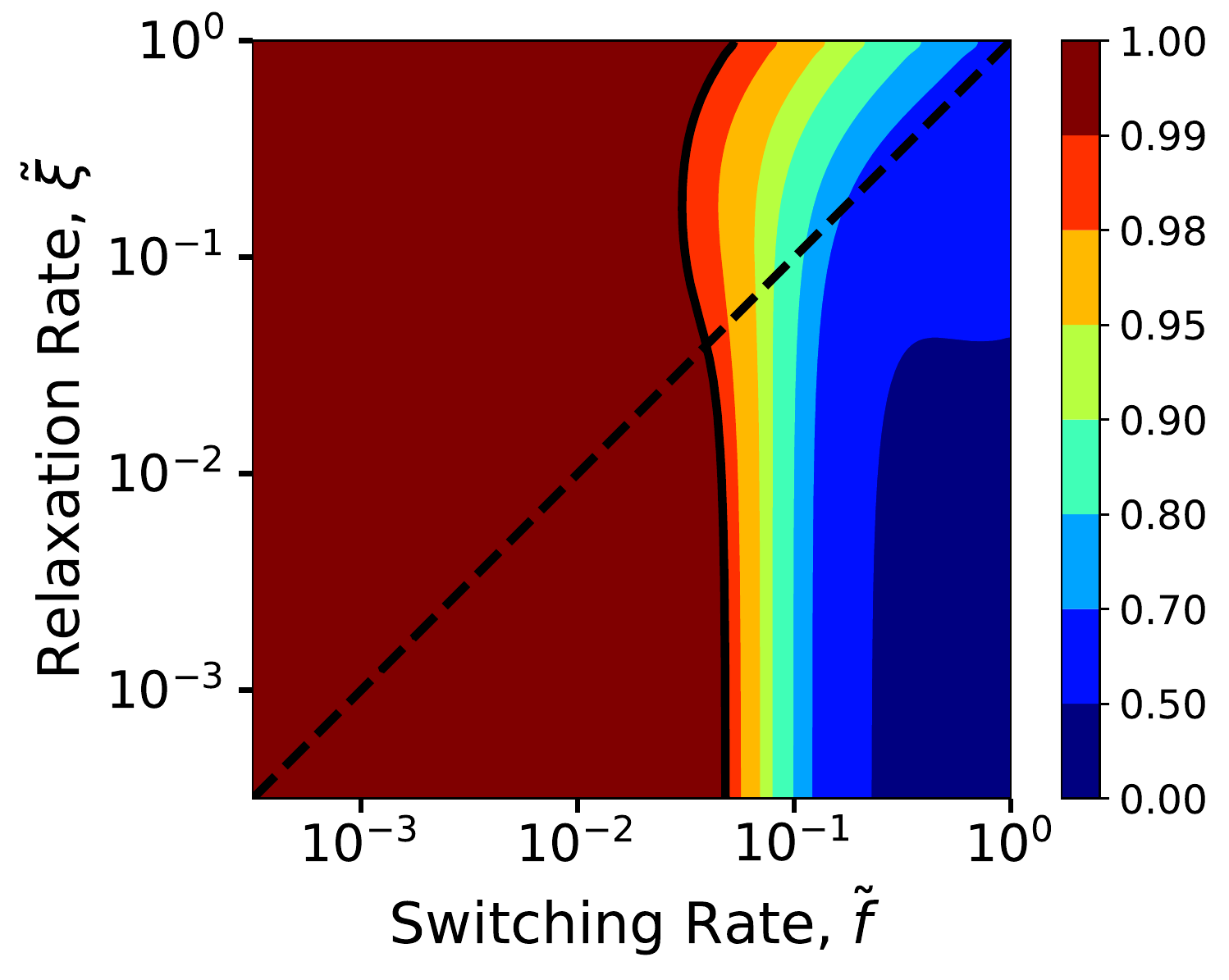}}\quad
    \subfloat[$\sscl$]{\includegraphics[width=\WW]{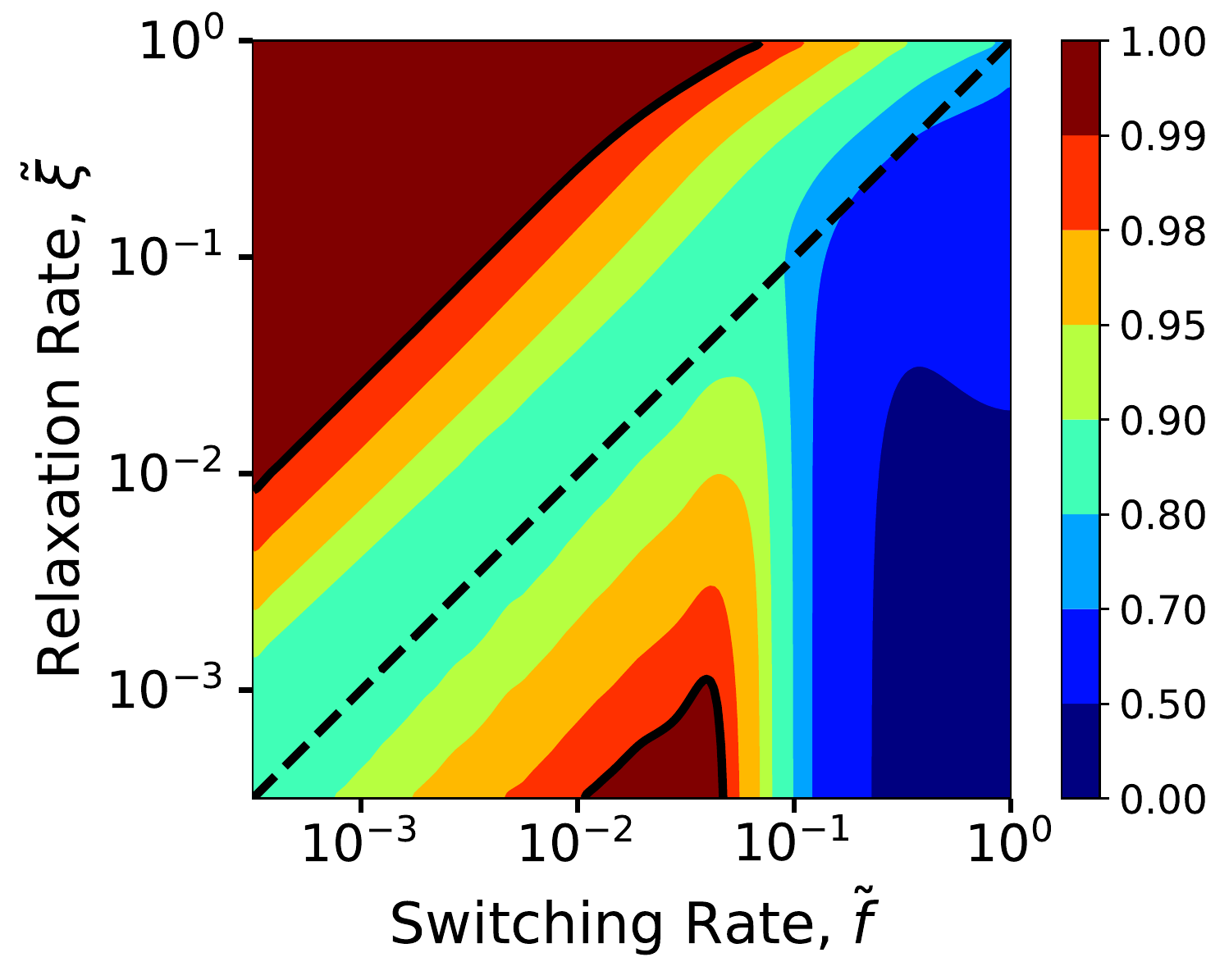}}\\
    \subfloat[$\sskt$: $\beta = 1$]{\includegraphics[width=\WW]{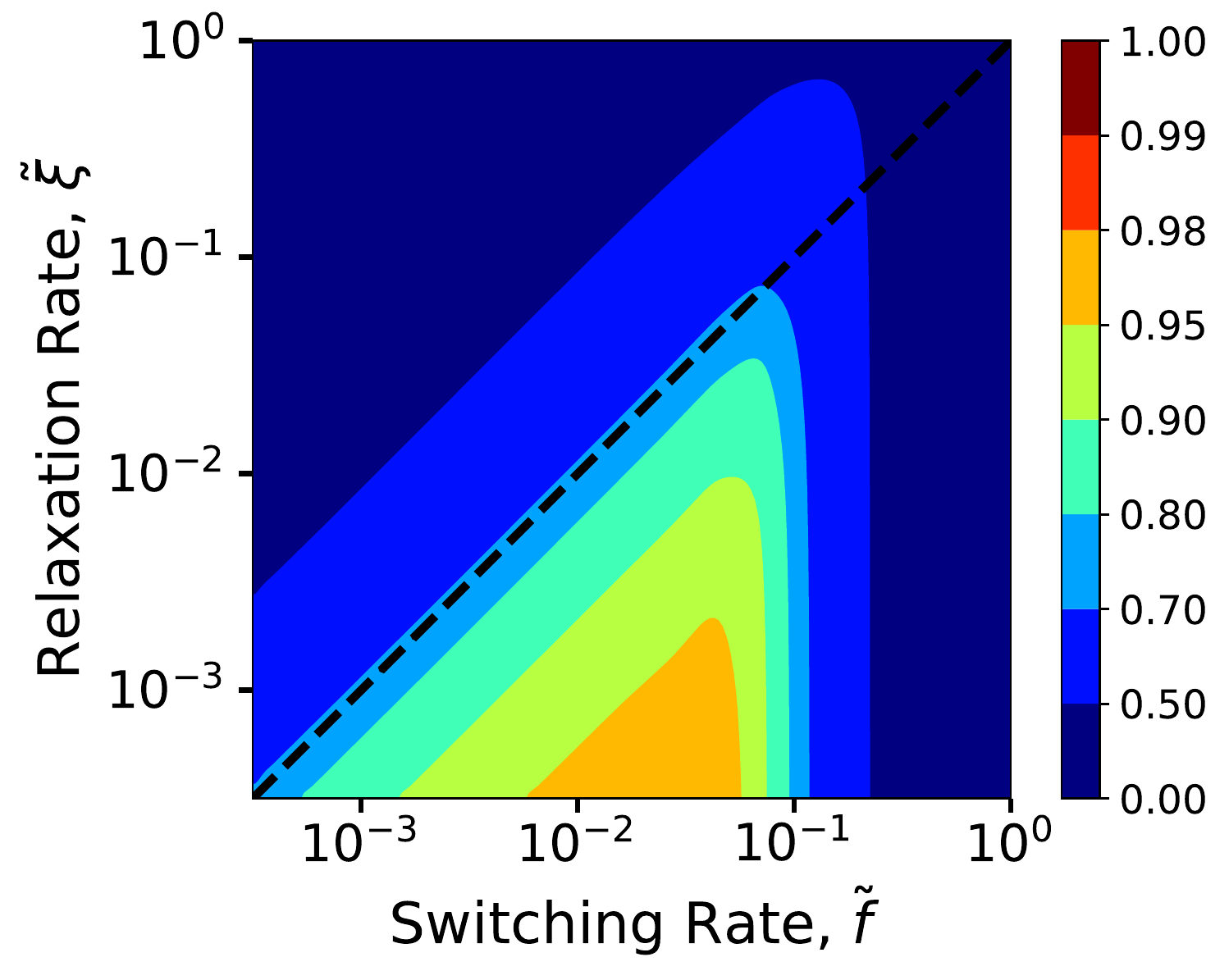}}\quad
    \subfloat[$\sskt$: $\beta = 5$]{\includegraphics[width=\WW]{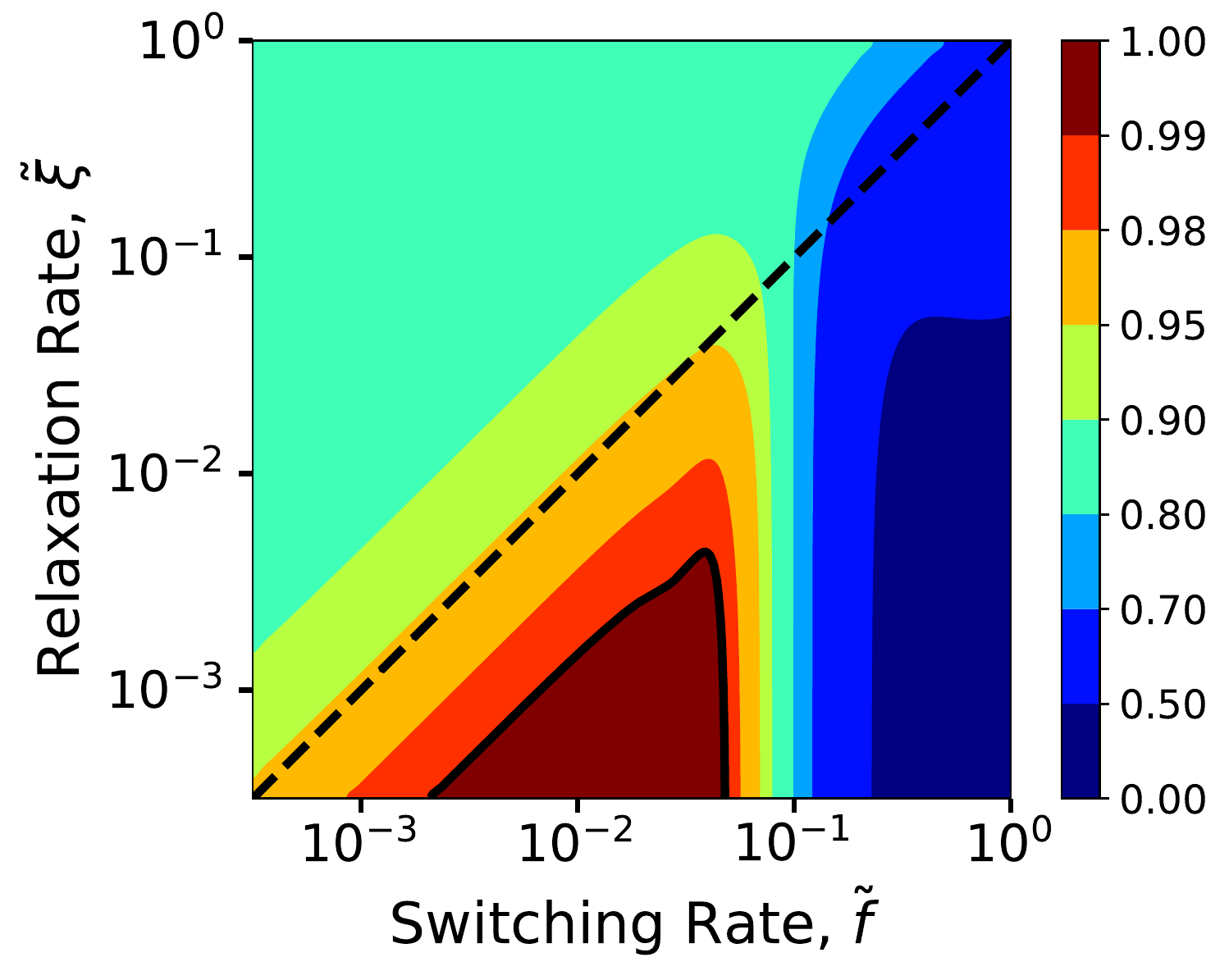}}\\
    \subfloat[$\sskt$: $\beta = 10$+]{\includegraphics[width=\WW]{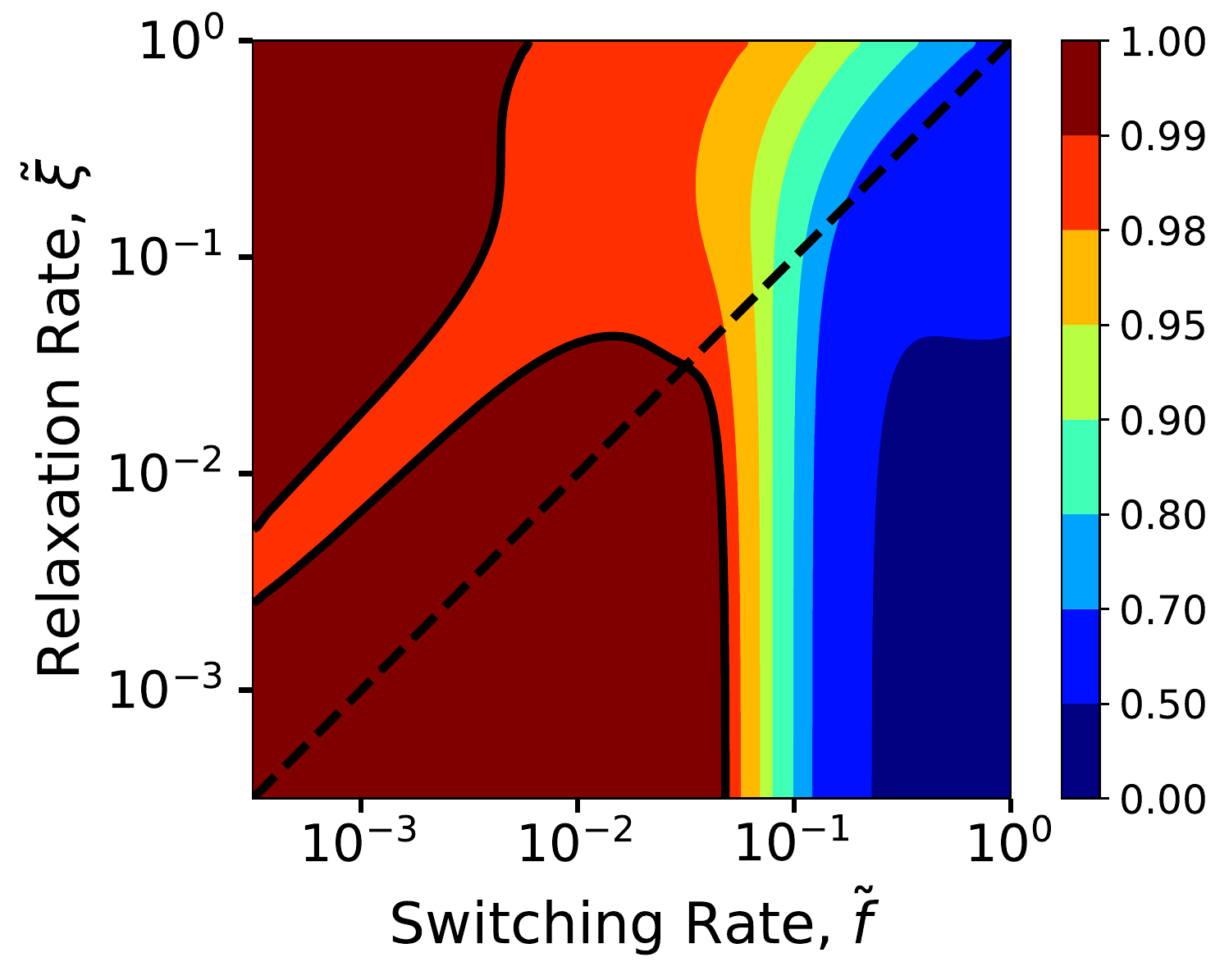}}\quad
    \subfloat[$\sskt$: $\beta = 20$]{\includegraphics[width=\WW]{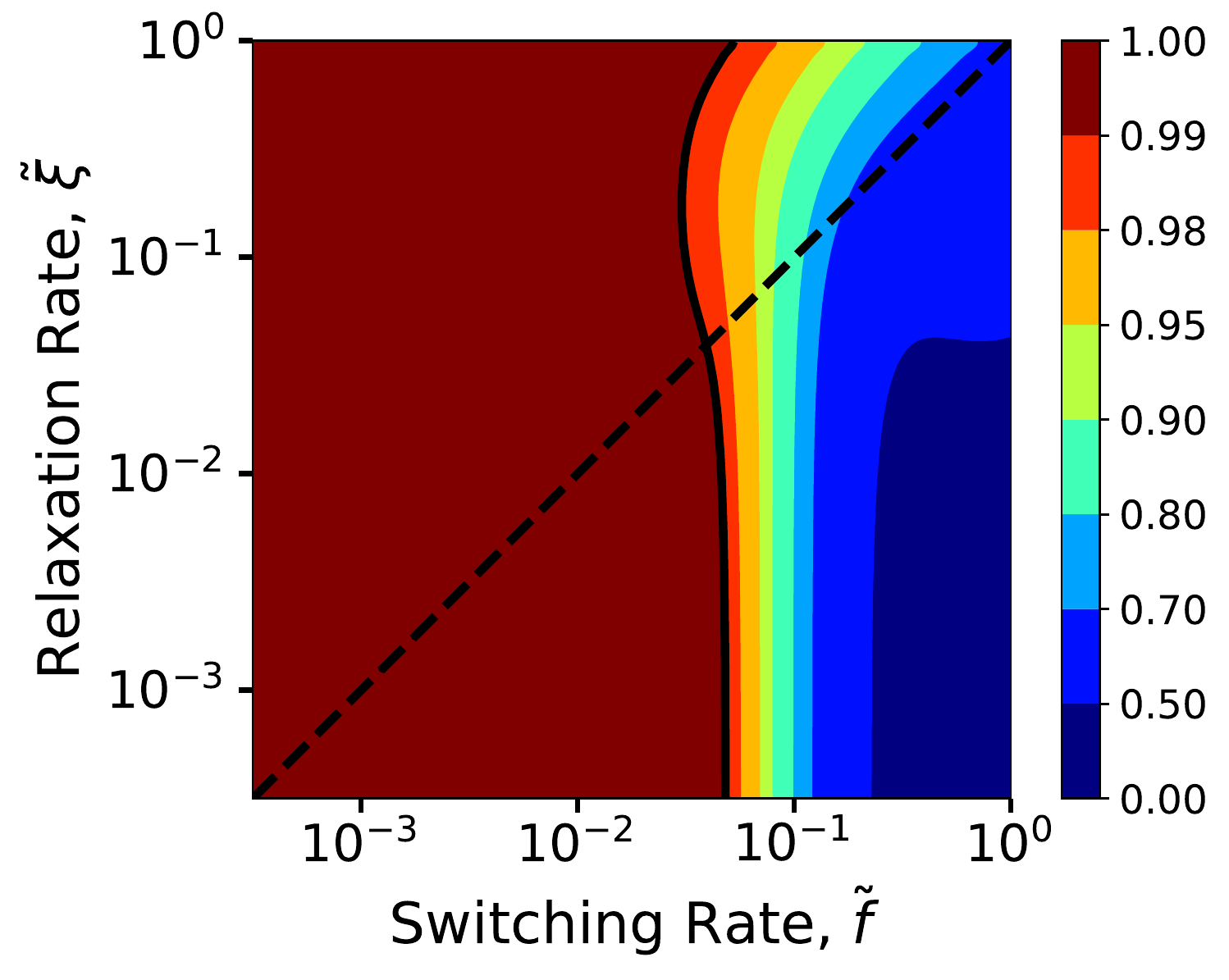}}
    \caption{Logical performance of Maj-101 for different global steady states. The dashed line represents equal switching and relaxation rates and the solid contour represents a performance of 0.99. (e) shows the behaviour immediately after the critical inverse temperature: $10+ = 10.1$}
    \label{fig: rho-ss-2d}
\end{figure}
\begin{equation}
    \etas{a}{i} = \tr \hrhos \hs{a}{i}.
\end{equation}
For the ICHA, a mean field steady state is usually employed of the form \cite{timler2002},
\begin{equation} \label{eq: mf-eta}
  \etas{a}{i} = - \tanh\pa{\tfrac{\beta}{2} |\bm{\Gamma}_i| }\frac{\Gamma_a^i}{|\bm{\Gamma}_i|},
\end{equation}
with $\bm{\Gamma}_i$ as in \cref{eqn:icha}. While it is possible to construct a steady state density operator from the local dissipation vectors by defining the higher order steady state elements \cite{mahler1998}, we have found this process to be prohibitively slow. We will present results for mean field relaxation only via the ICHA. All performance results are now in terms of the logical performance.

\subsection{Spectral Relaxation of the Density Operator}

\begin{figure}
    \includegraphics[width=.5\linewidth]{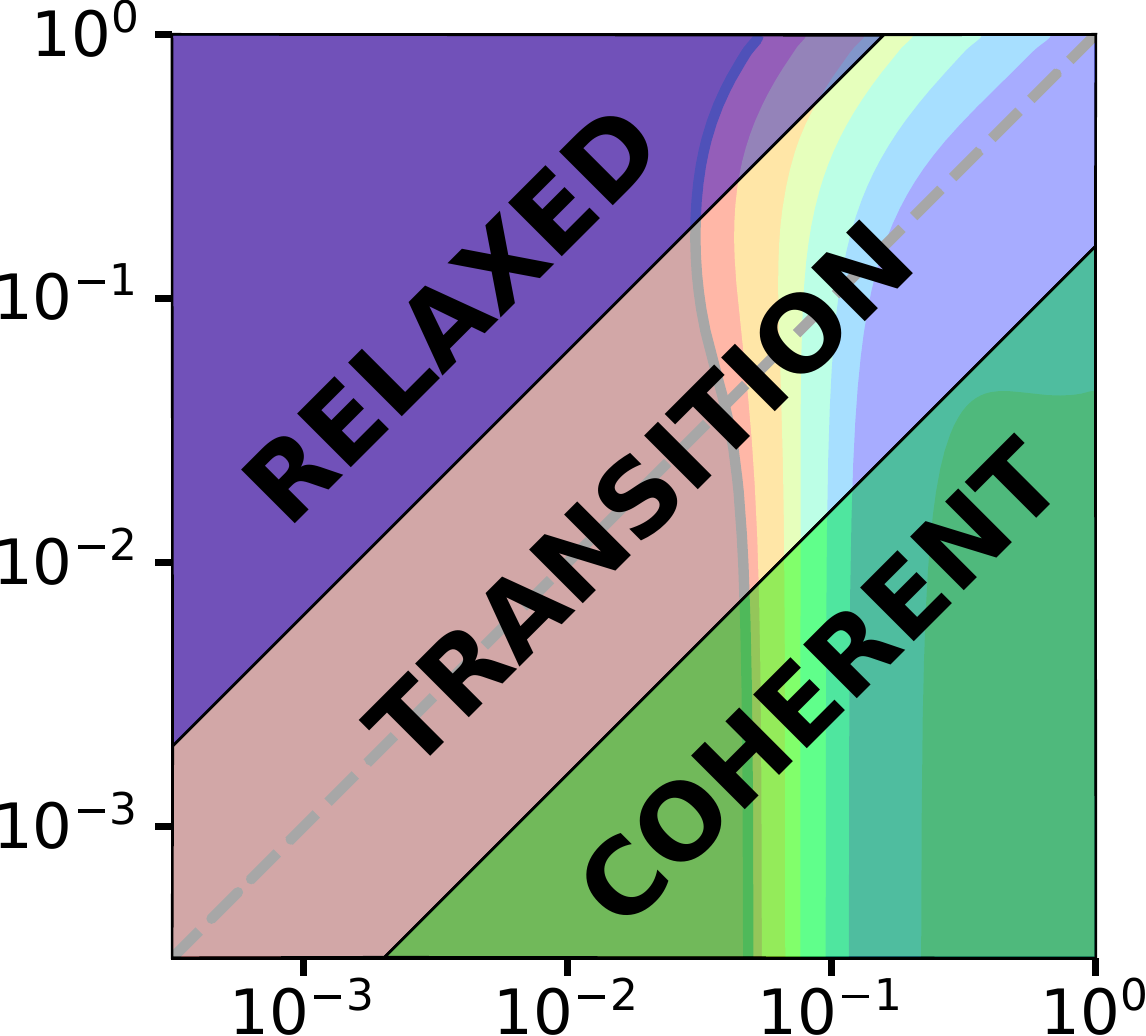}
    \caption{There are three regions of interest: the \textbf{coherent} regime, in which the system is governed by coherent dynamics; the \textbf{relaxed} regime, in which the system closely tracks the steady state; and the \textbf{transition} between these two regimes.}
    \label{fig: relax-schem}
\end{figure}

\begin{figure}
 \includegraphics[width=.65\linewidth]{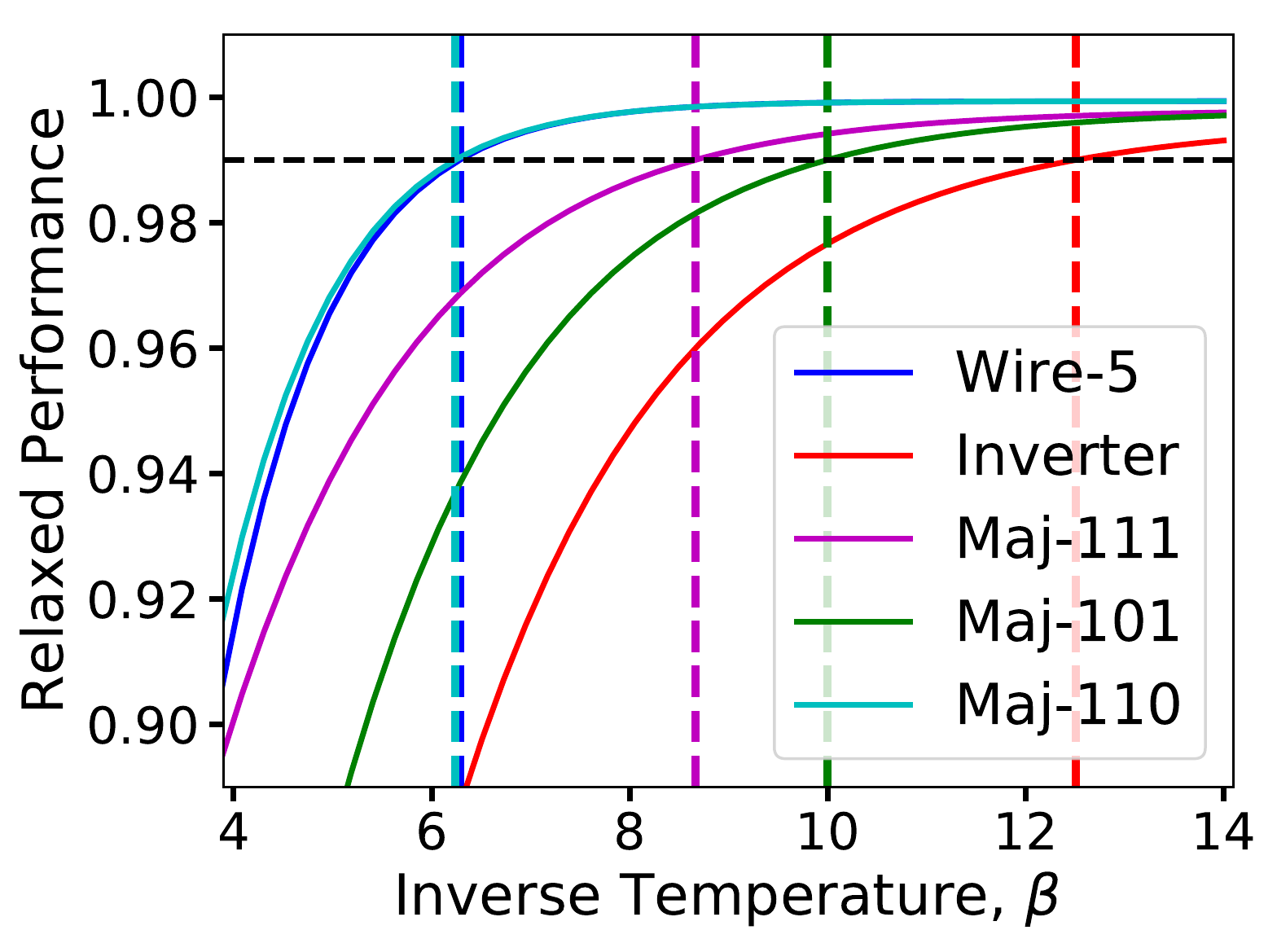}
 \caption{Logical performance in the relaxed regime for QCA components with the Boltzmann distribution steady state. $\Met_L = 0.99$ thresholds are indicated.}
 \label{fig: beta-sweep}
\end{figure}

\begin{table}
 \caption{Threshold $\beta$ value for high performance in the relaxed regime. Parameters for \cref{eq: beta*} are estimated from the spectrum of $\tfrac{1}{2}\HP$}
 \label{tab: beta-sweep}
 \begin{ruledtabular}
   \newcommand{\hsp}{\hspace{4ex}}
   \begin{tabular}{c@{\hsp}|cc@{\hsp}|cc@{\hsp}}
     Device & $d_1$ & $\Delta_1$ & \cref{eq: beta*} & \cref{fig: beta-sweep}\\
     \hline \rule{0pt}{3ex}
     Wire-5   & 5 & 1      & 6.2 & 6.3 \\
     Inverter & 1 & 0.416  & 11.1 & 12.5 \\
     Maj-111  & 1 & 0.554  & 8.3 & 8.7 \\
     Maj-101  & 2 & 0.554  & 9.6 & 10.0 \\
     Maj-110  & 5 & 1  & 6.2 & 6.2 
   \end{tabular}
 \end{ruledtabular}
\end{table}

\begin{figure}
 \newcommand{\WW}{.48\linewidth}
 \subfloat[Wire-5]{\includegraphics[width=\WW]{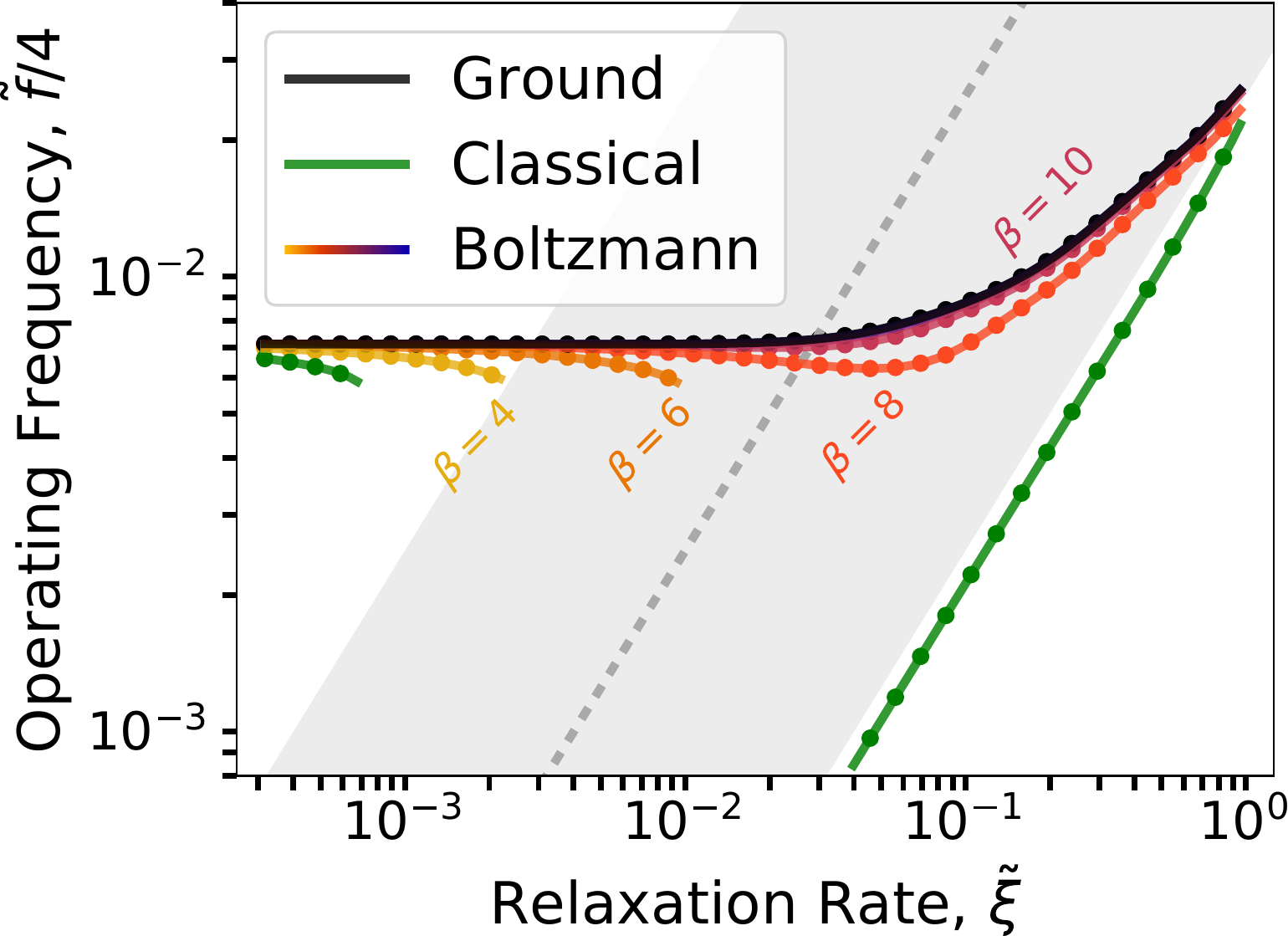}}\quad
 \subfloat[Inverter]{\includegraphics[width=\WW]{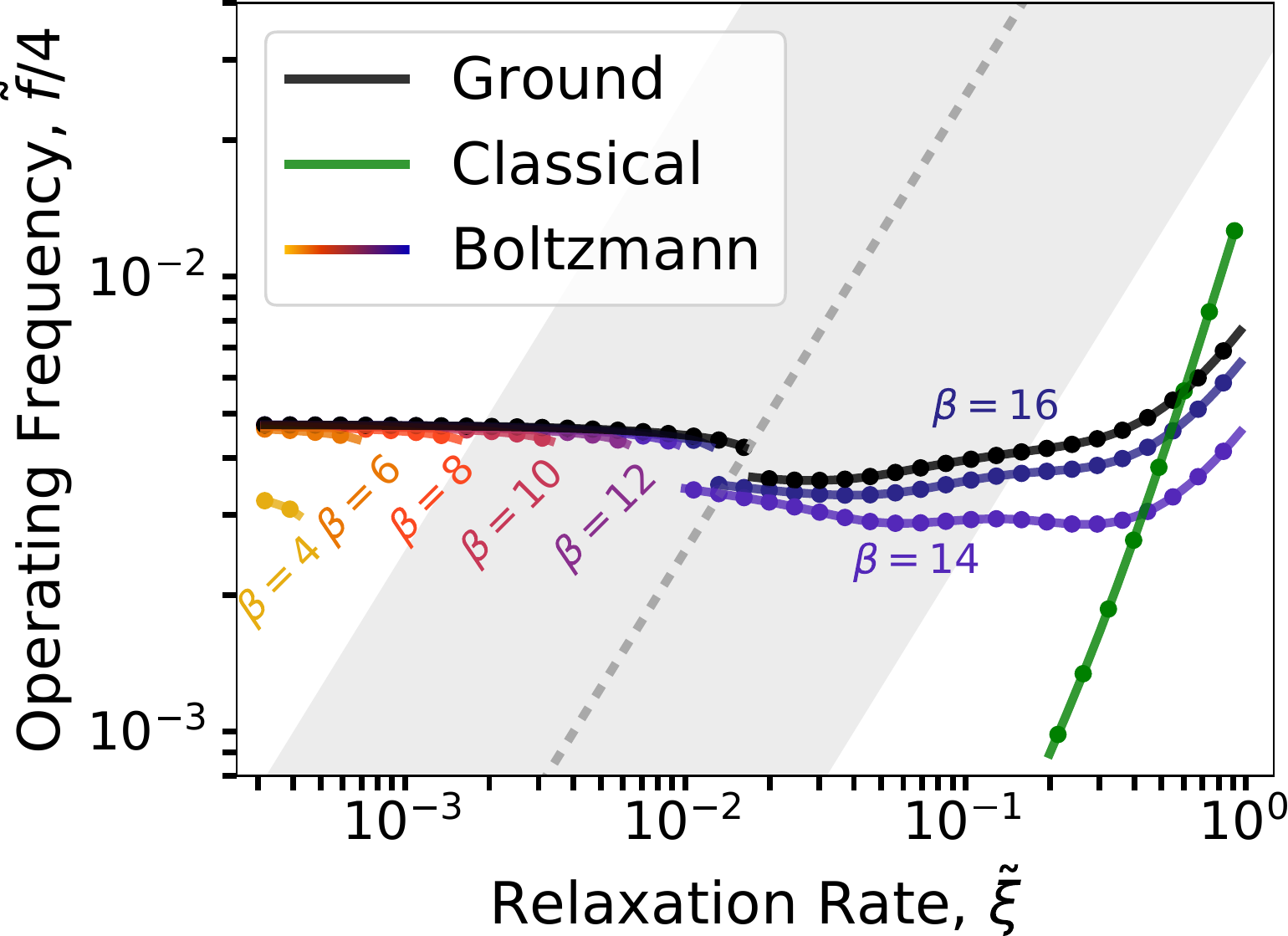}}\\
 \subfloat[Maj-111]{\includegraphics[width=\WW]{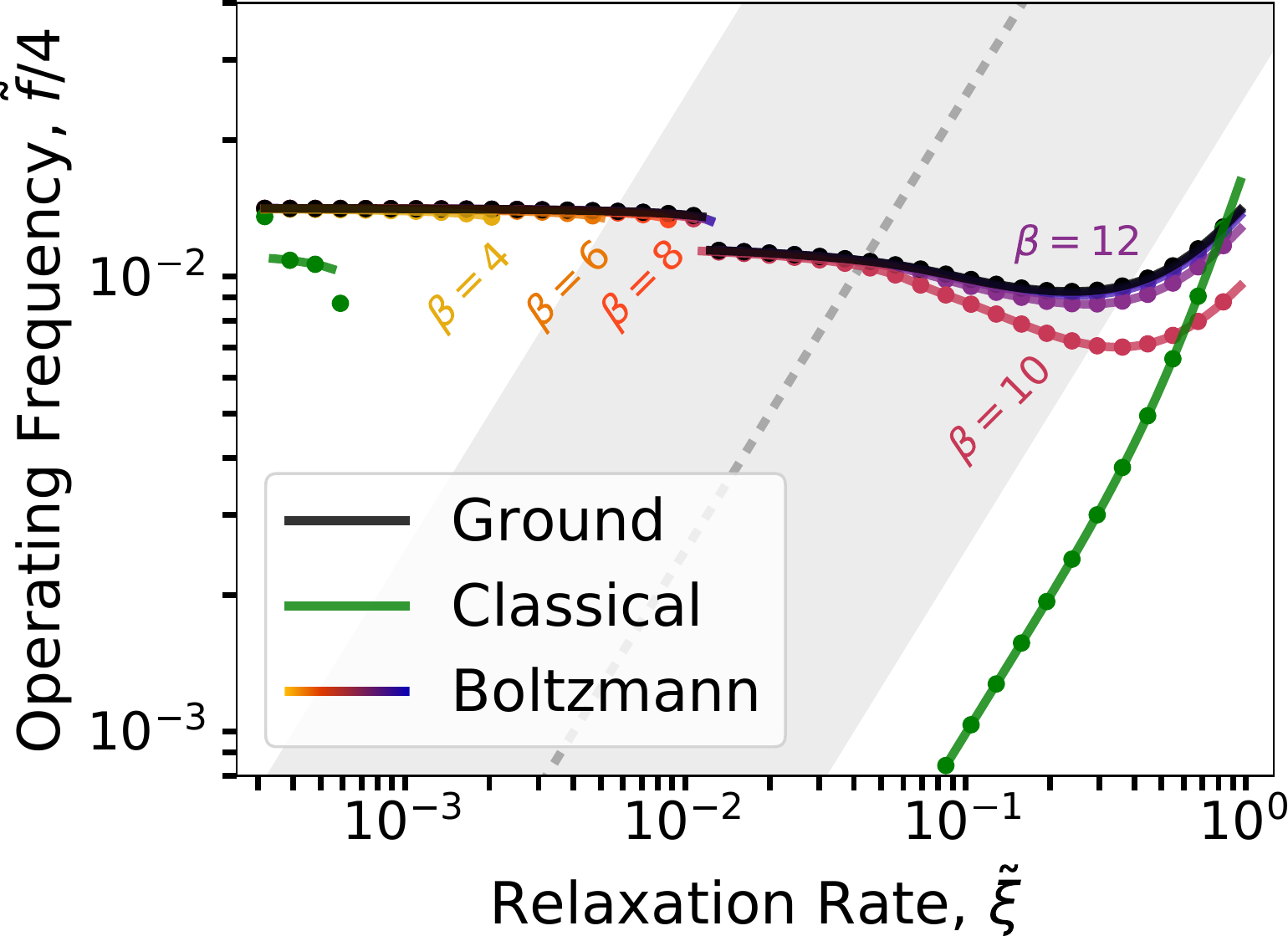}}\quad
 \subfloat[Maj-101]{\includegraphics[width=\WW]{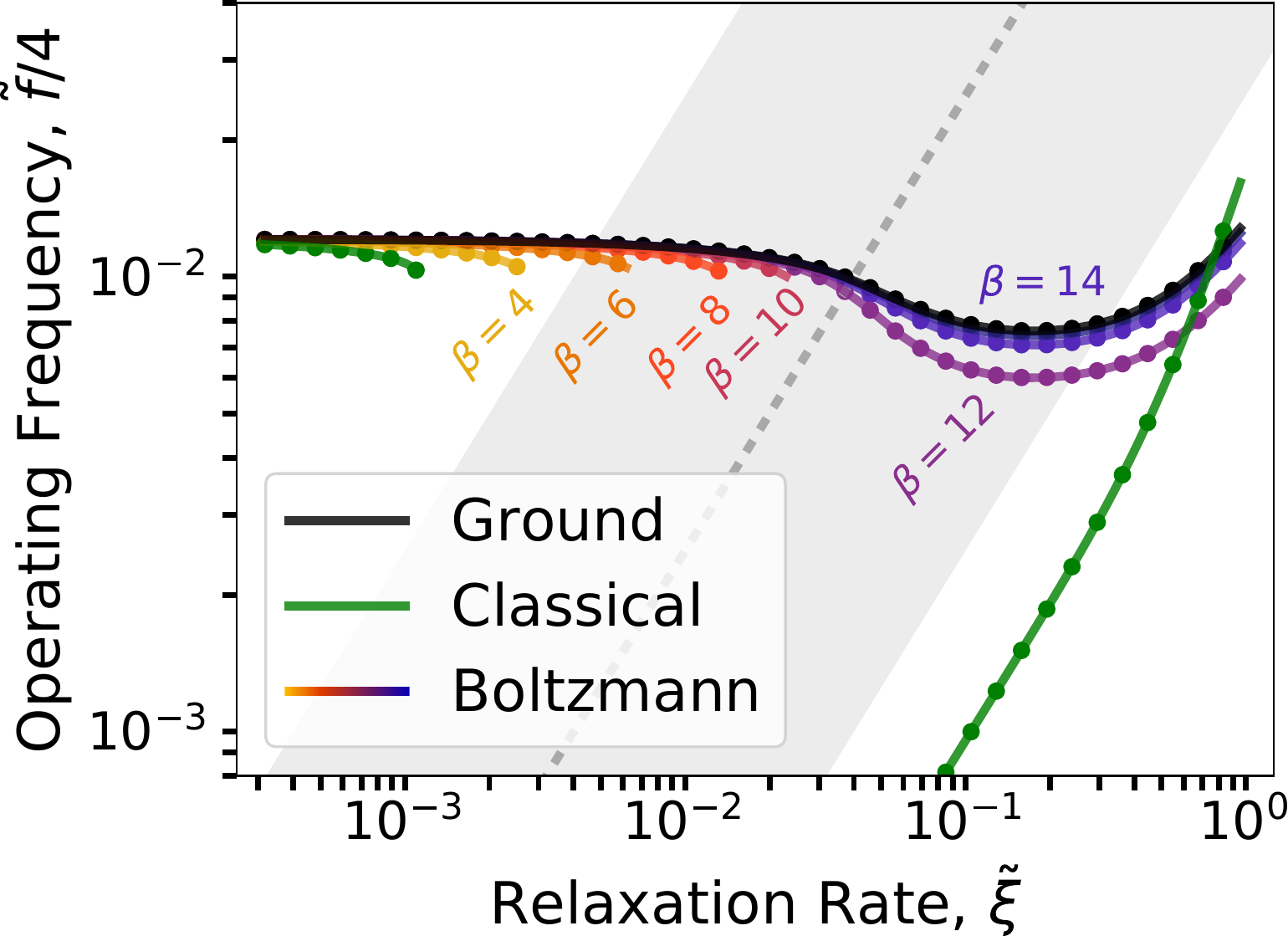}}\\
 \subfloat[Maj-110]{\includegraphics[width=\WW]{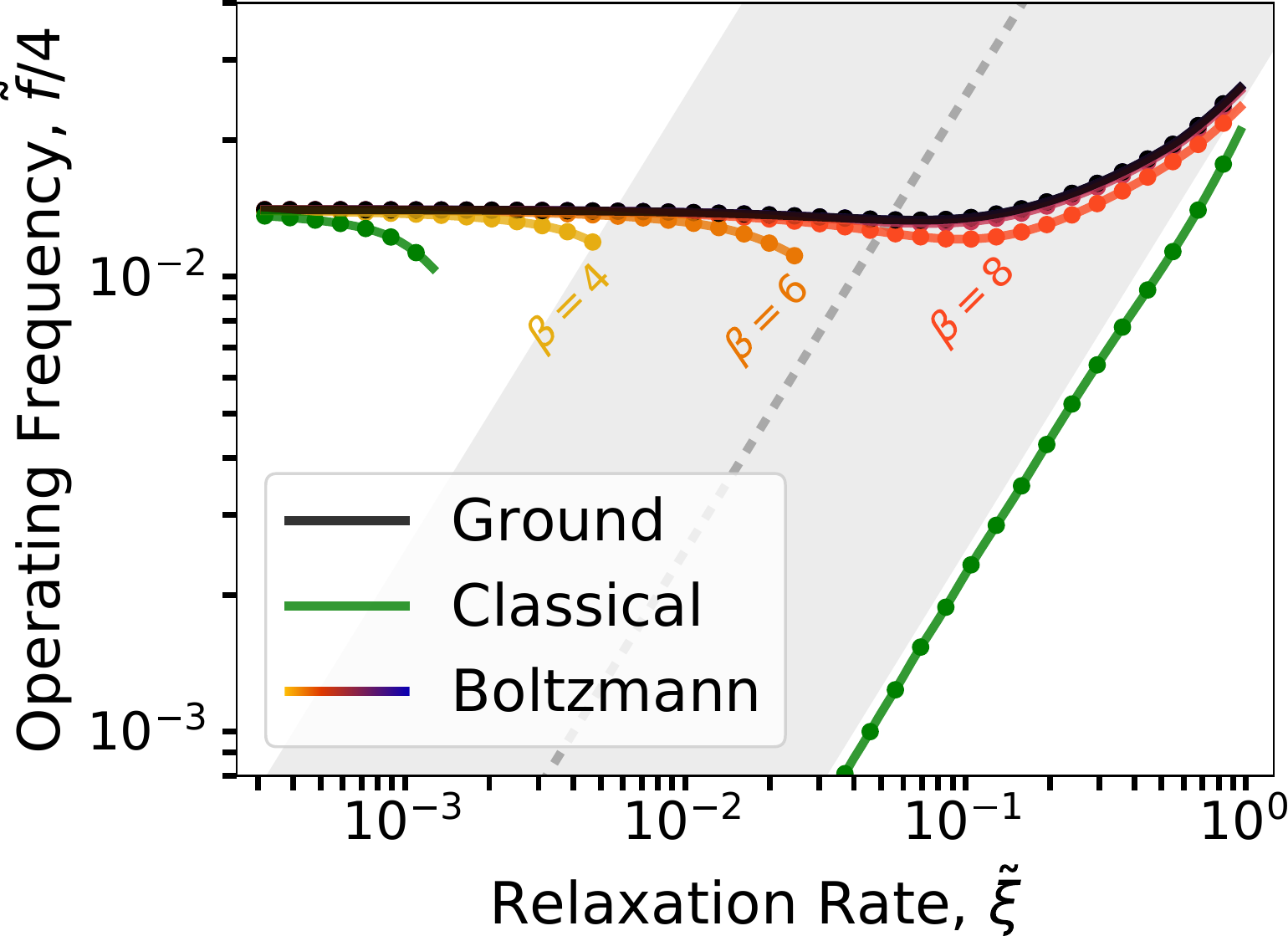}}
 \caption{Maximum switching rates for different spectral steady states. A sample set of $\beta$ values are indicated for the Boltzmann distribution. Classical steady state results are cropped for ease of view. The shaded region and dashed line indicate the approximate edges of the transition regime}
 \label{fig: spectral-ss-performance}
\end{figure}
We consider the steady states specified in \cref{eq: rho-ss} that depend only on the energy eigenspectrum. As an example, \cref{fig: rho-ss-2d} shows the logical performance for Maj-101 over a range of both the switching rate, $\runrate$, and the relaxation rate, $\dissip$. There are three regimes of interest illustrated in \cref{fig: relax-schem}. If the switching rate is large with respect to the relaxation rate, $\runrate \gtrsim 10 \dissip$, then the dynamics are approximately coherent. In this regime, the choice of steady state does not significantly affect the performance. If the relaxation rate is large with respect to the switching rate, $\dissip \gtrsim 10 \runrate$, then the system tracks the steady state and the performance is entirely governed by whether the steady state has the correct logic. For small $\atb_1$, both the ground state and classical steady states give the correct logic and thus have high performing relaxed regimes. For a Boltzmann distribution the behaviour is more complicated. The logical performance in the relaxed regime has a lower bound approximately defined by the first excited state with incorrect output logic. To first order,
\begin{equation}
   \Met_L > 1 - d_1\gibbs{\Delta_1},
\end{equation}
with $\Delta_1$ the gap between the ground state and the lowest energy incorrect state and $d_1$ its degeneracy. Here we have assumed that all energy gaps are large with respect to the thermal energy. We should expect a region of high performance in the relaxed regime for $\beta$ above approximately
\begin{equation} \label{eq: beta*}
  \beta^* \approx \brak{\log(d_1)-\log(1-0.99)} / \Delta_1.
\end{equation} 
For Maj-101 we get $\Delta_1 = 0.554$, $d_1 = 2$, and an estimated transition at $\beta^* \approx 9.6$. In \cref{fig: rho-ss-2d}(c-f) we observe a domain with $\Met_L$ above 0.99 emerge somewhere between $\beta = 5$ and $20$. The transition actually occurs just above $\beta = 10.0$ (see \cref{fig: rho-ss-2d}(e)) which is fairly close to our estimate. 
The exact value of $\Met_L$ in the limit of infinite $\dissip$ for a Boltzmann steady state is obtained as
\begin{equation} \label{eq: met-lr}
  \Met_L(\beta) = \tfrac{1}{2} (1+|\tr \sskt(1)\hs{z}{n}|)
\end{equation}
for output cell $n$. In \cref{fig: beta-sweep}, we show $\Met_L(\beta)$ for all the QCA components. \cref{tab: beta-sweep} compares the $\beta^*$ estimates using \cref{eq: beta*} with the values from \cref{fig: beta-sweep}. The most significant difference occurs for the inverter, which has a second incorrect excited state with a slightly higher energy gap of $0.554$. The contribution of this state is ignored in our estimate.

If the switching and relaxation rates are of the same order, the behaviour depends on the interplay between the coherent dynamics and the relaxation. Of the steady states considered, the most interesting behaviour occurs for $\sscl$, with a band of low performance along $\dissip \approx \runrate$. This result is easily explained by observing that the classical ground state will generally be a high energy configuration of $\HHd(0)$ and very near the ground state of $\HHd(1)$. Initially, the dissipation acts to excite the network out of the ground state; later in the clock, the dissipation acts to help drive the network back down in energy. Increasing $\dissip$ initially hurts performance until the relaxation is strong enough to overcome these initial excitations.

We summarize the performance for each QCA component by extracting the maximum operating frequency as a function of $\dissip$ for each choice of steady state. The results are shown in \cref{fig: spectral-ss-performance}. Each set of data is obtained by first finding $\runrate_{max}$ in the coherent limit and then tracking the $\Met_L = 0.99$ contour as $\dissip$ is increased. We first mention a few unsurprising results: (1) the operating frequency is approximately independent of the choice of steady state in the coherent limit; (2) the trends become parallel to $\runrate = \dissip$ in the relaxed regime, meaning performance is guaranteed as long as $\sfrac{\dissip}{\runrate}$ is sufficiently large; and (3), for Boltzmann steady states the performance is improved as the temperature is decreased. There are apparent discontinuities in some of the trends. These correspond to cases like \cref{fig: rho-ss-2d}(e), where we observe two regions of high performance. Interestingly, unless $\dissip$ is sufficiently large, we observe a decrease in maximum operating frequency for any of our spectral steady states.

\begin{figure}
 \newcommand{\WW}{.48\linewidth}
 \subfloat[Wire-5]{\includegraphics[width=\WW]{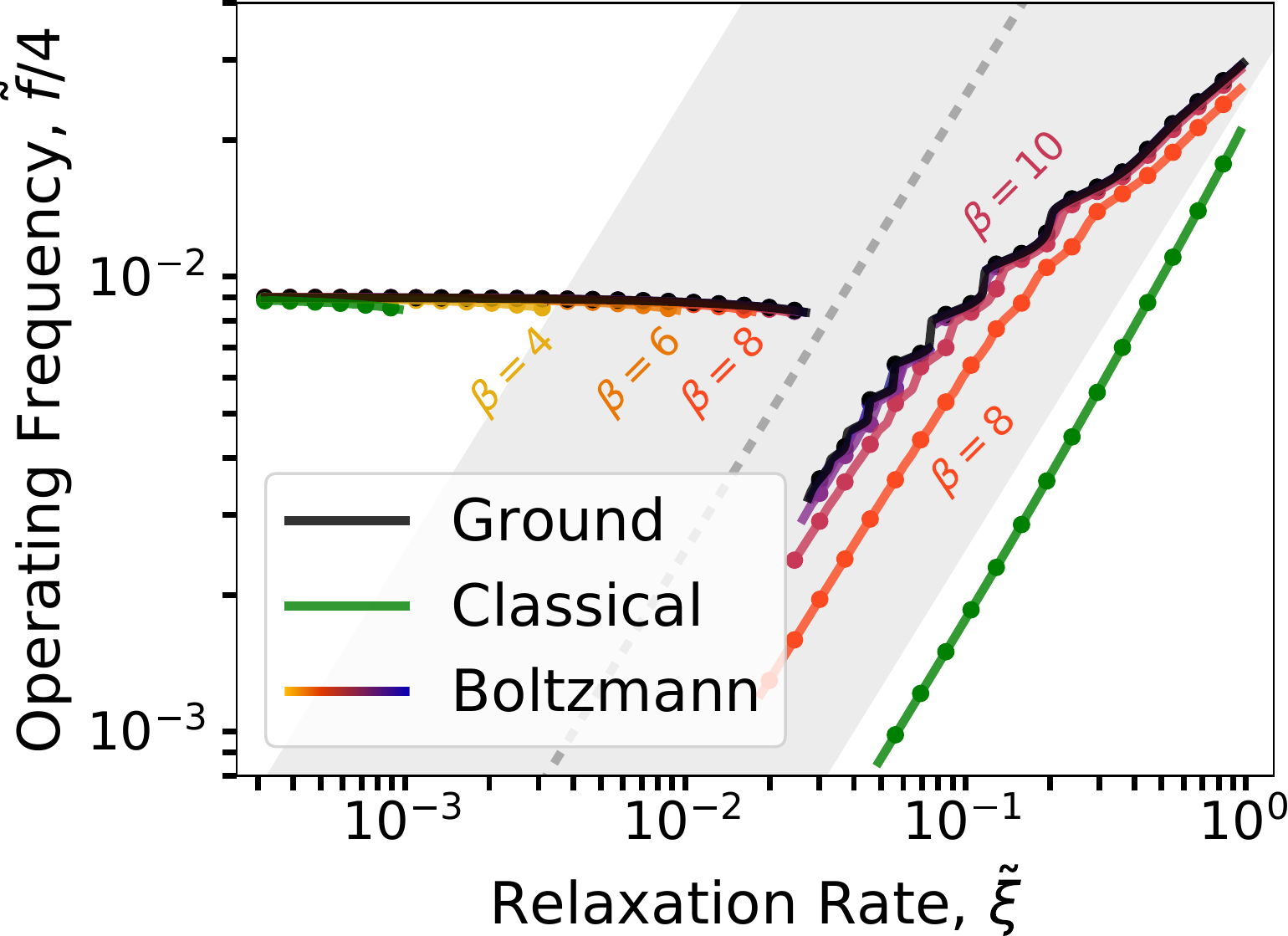}}\quad
 \subfloat[Inverter]{\includegraphics[width=\WW]{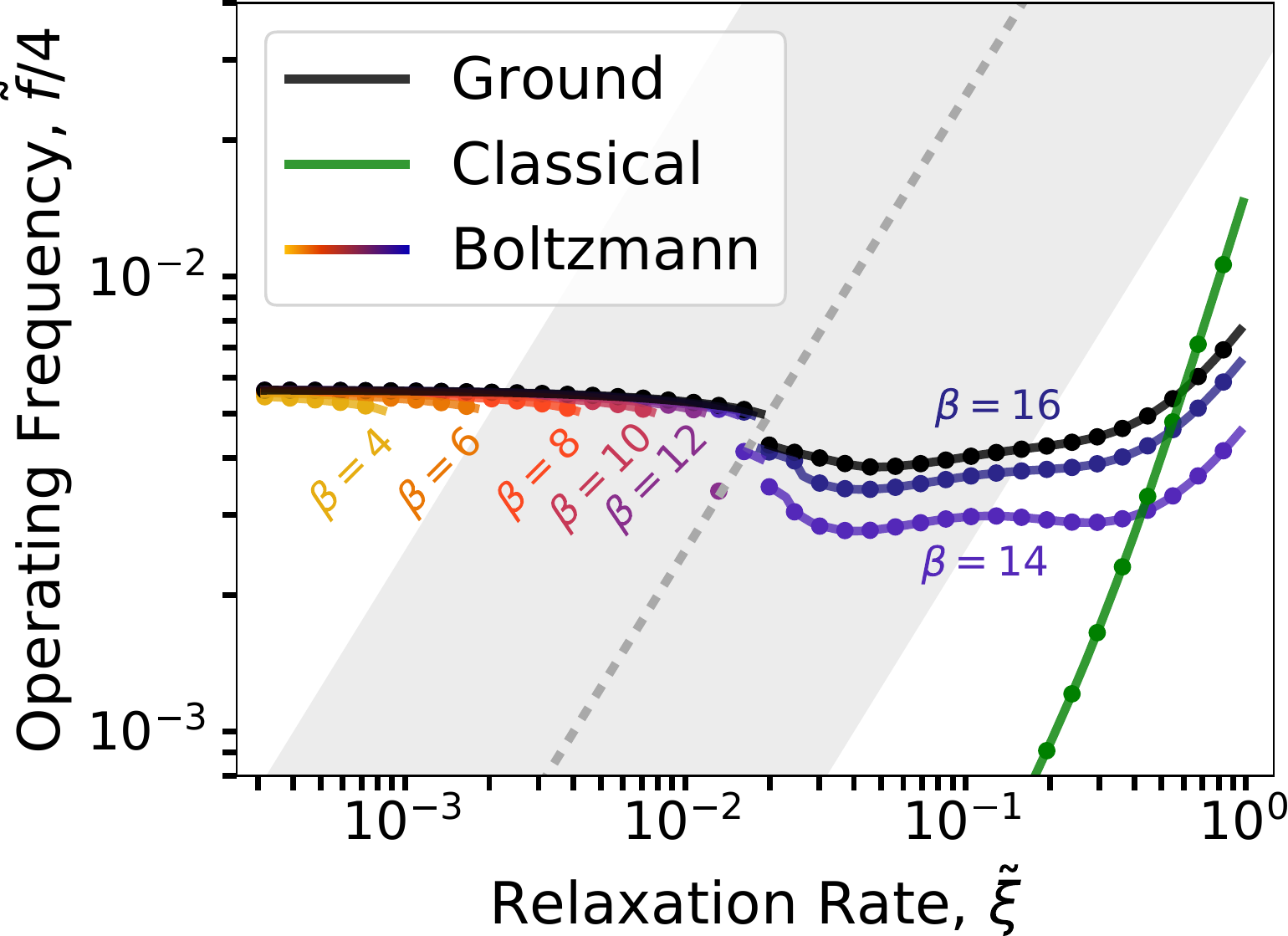}}\\
 \subfloat[Maj-111]{\includegraphics[width=\WW]{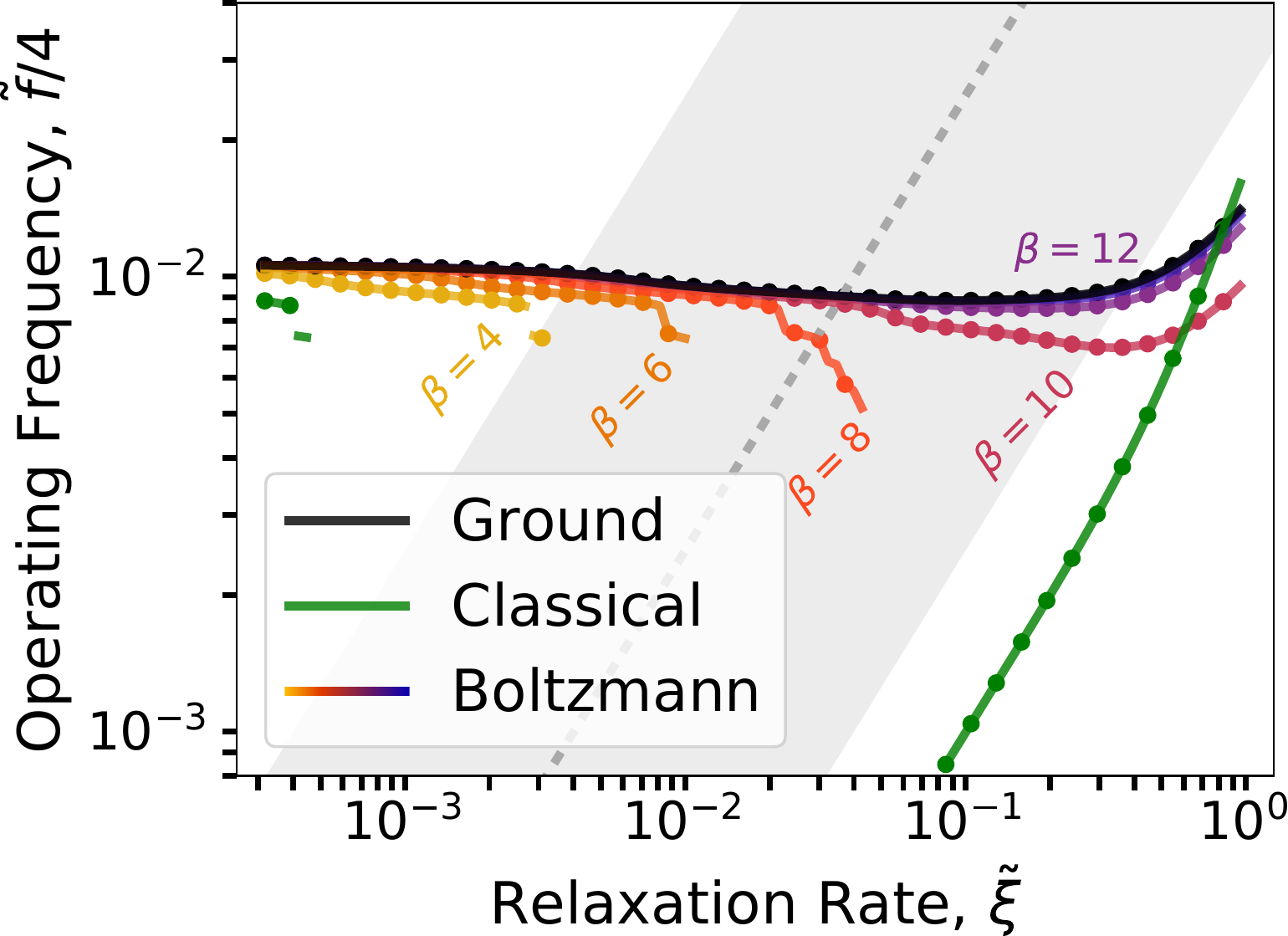}}\quad
 \subfloat[Maj-101]{\includegraphics[width=\WW]{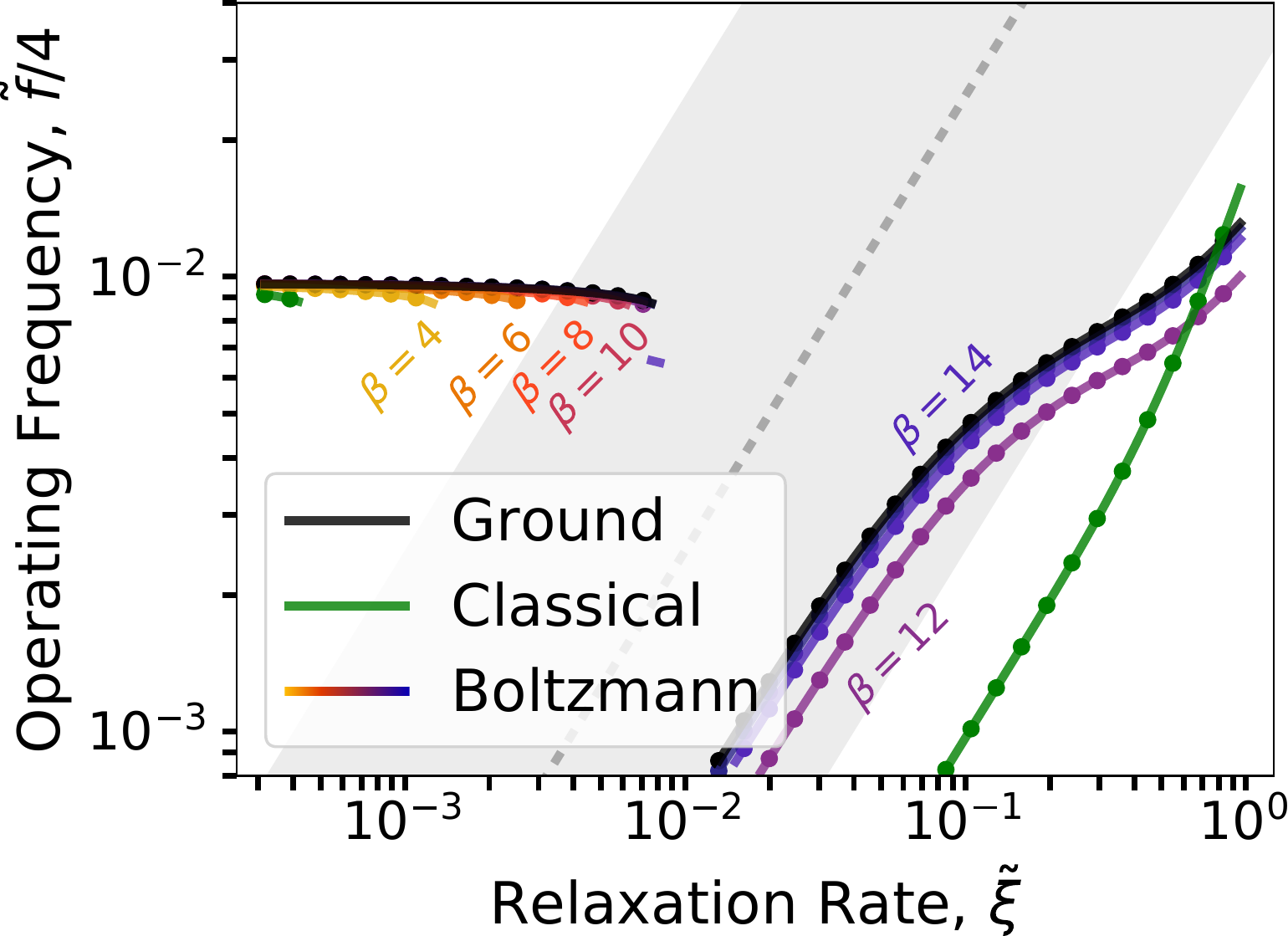}}\\
 \subfloat[Maj-110]{\includegraphics[width=\WW]{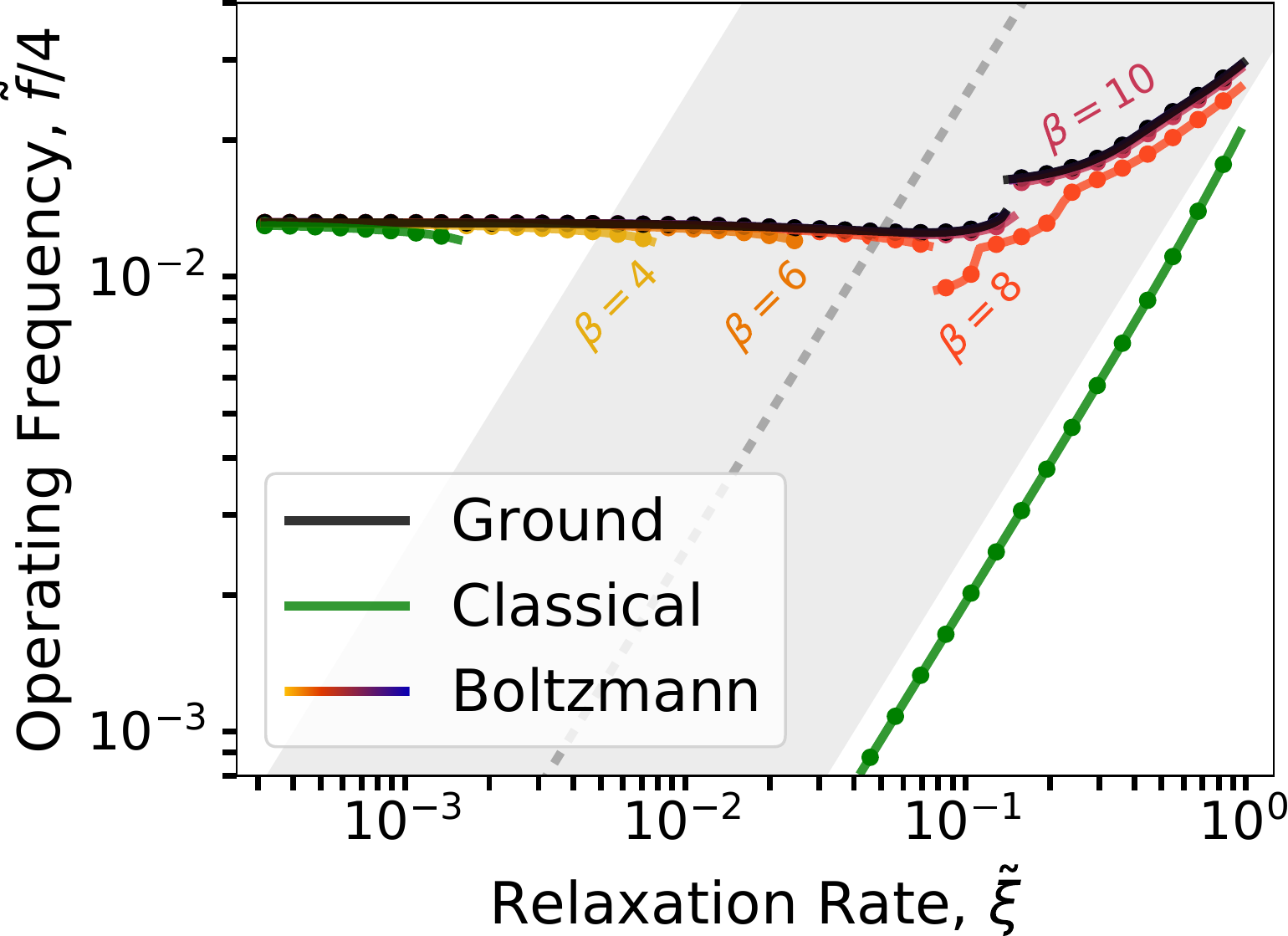}}
 \caption{ICHA: Maximum operating frequencies for the different spectral steady states. A sample set of $\beta$ values are indicated for the Boltzmann distribution.}
 \label{fig: icha-ss-performance}
\end{figure}

\subsection{Spectral Relaxation with the ICHA}

\begin{figure}
  \newcommand{\Width}{.45\linewidth}
  \subfloat[Coherence Vector: $\runrate = 2 \cdot 10^{-2}$]{\includegraphics[width=\Width]{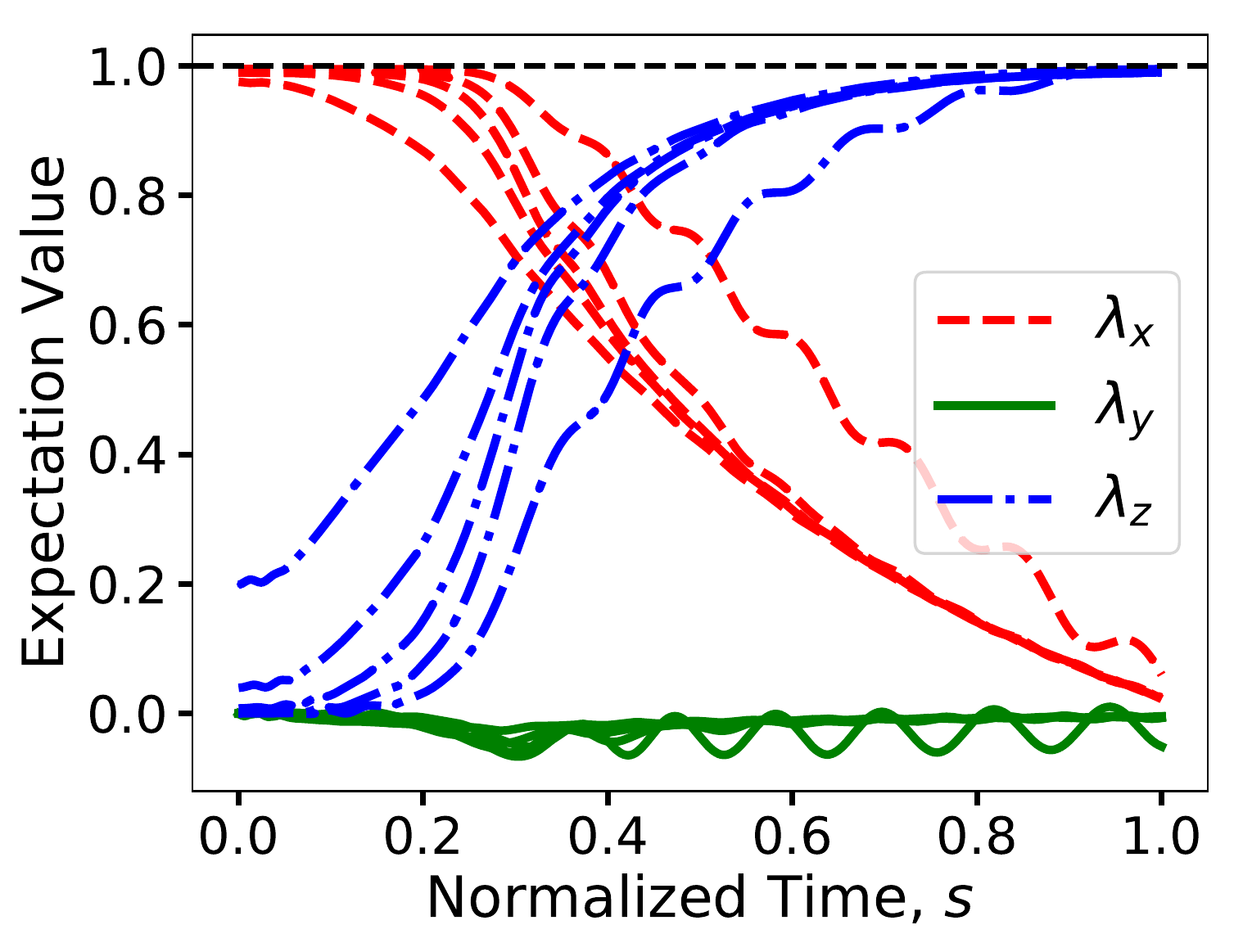}}\quad
  \subfloat[Performance Metrics]{\includegraphics[width=\Width]{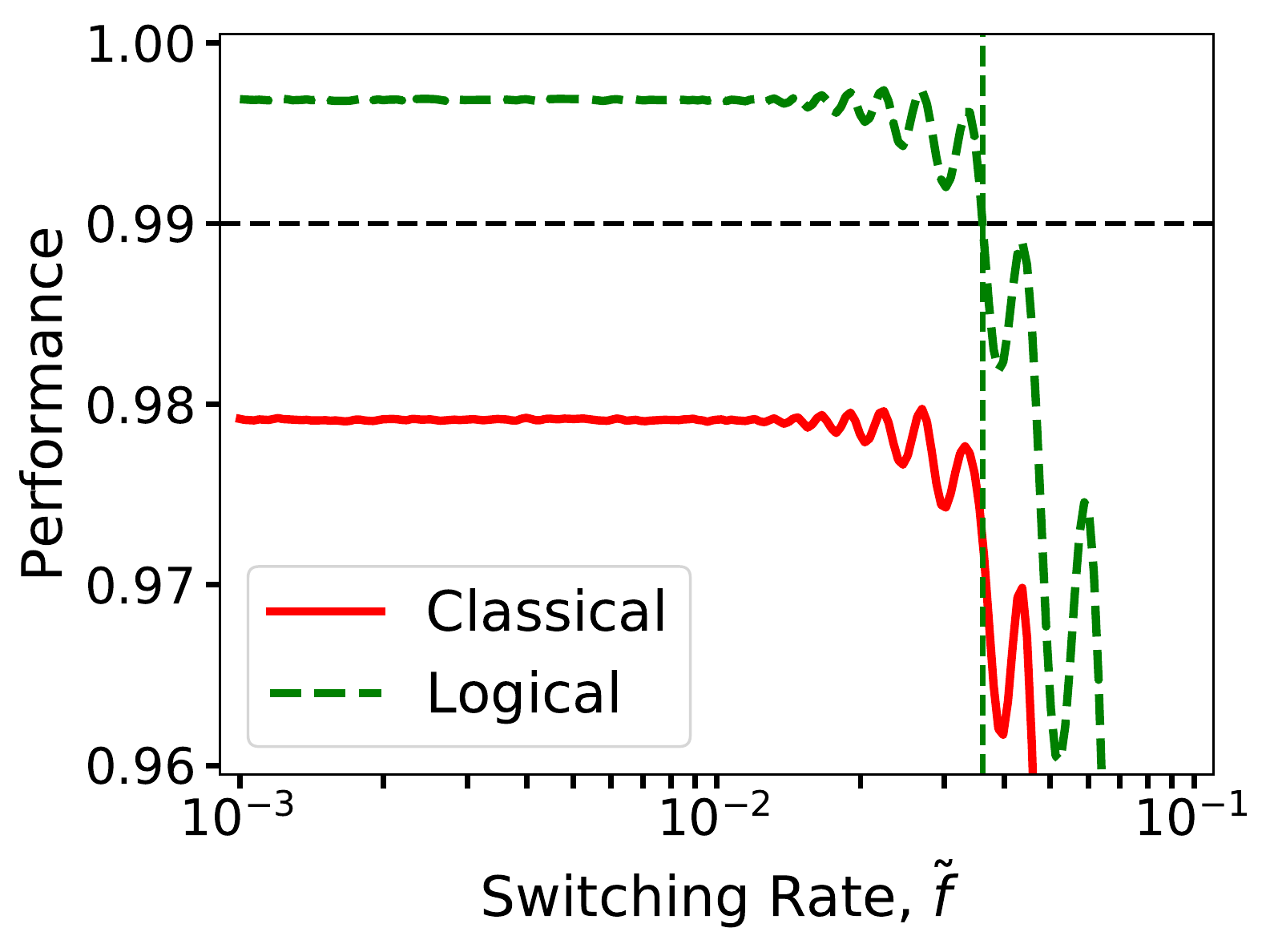}}\\
  \subfloat[$\ssgr$: $\hrho(s)$]{\includegraphics[width=\Width]{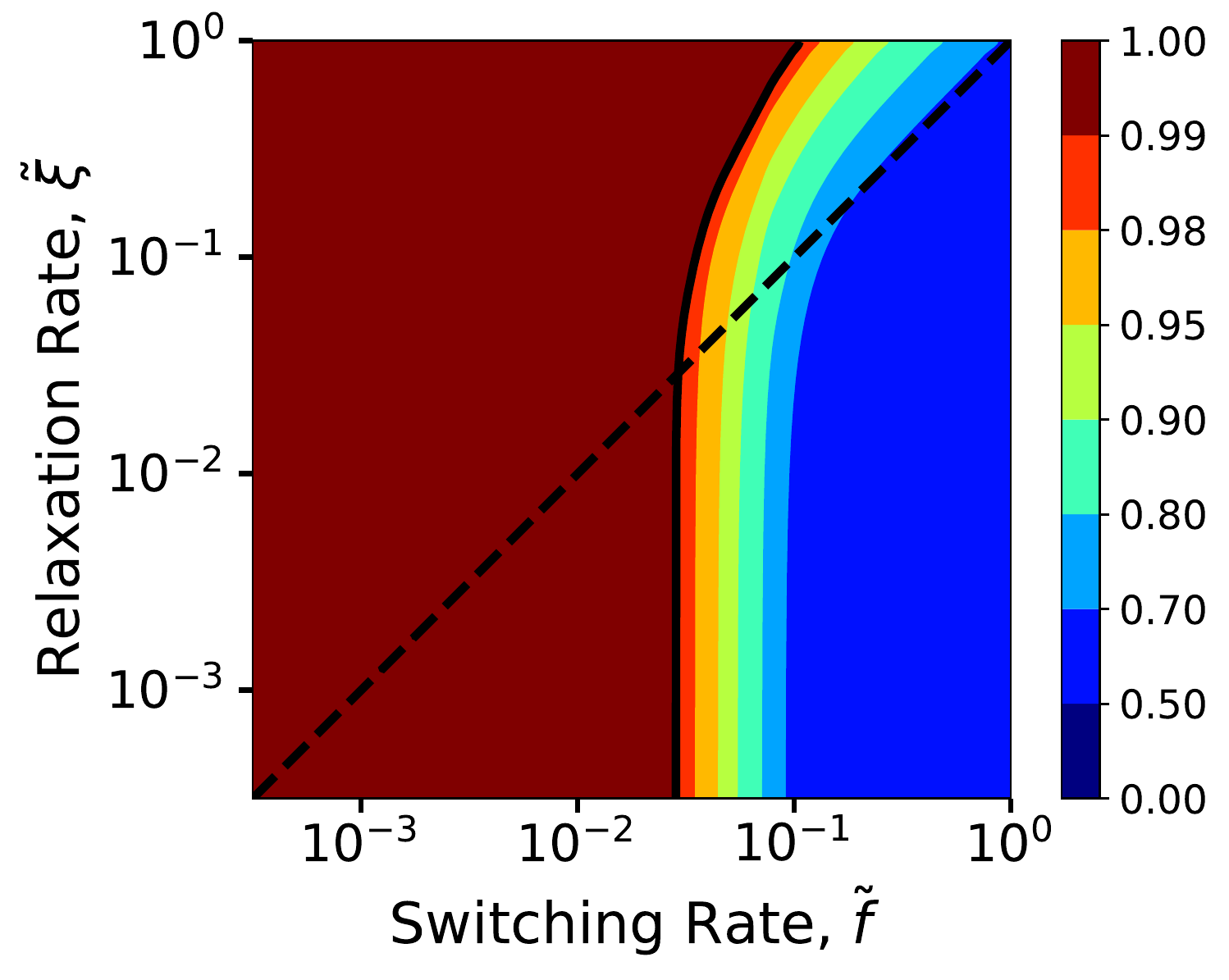}}\quad
  \subfloat[$\ssgr$: ICHA]{\includegraphics[width=\Width]{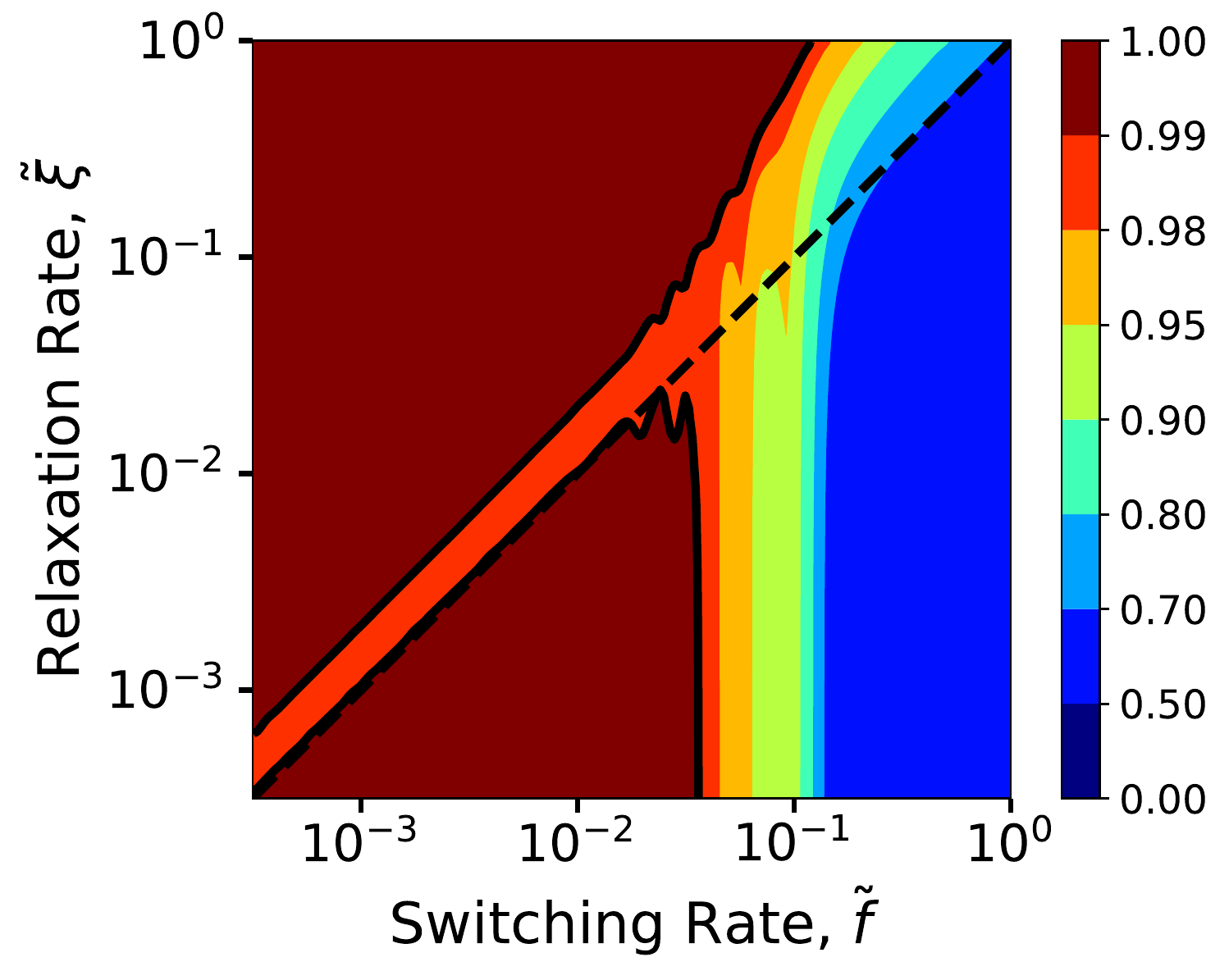}}
  \caption{ICHA simulations of Wire-5. Oscillations in the coherence vector lead to similar oscillations in the performance metrics, leading to interesting features in the maximum operating frequency contours }
  \label{fig: wire-osc}
\end{figure}

In \cref{fig: icha-ss-performance} we repeat the analysis using the ICHA. The behaviour in the relaxed regime matches what we observe in \cref{fig: spectral-ss-performance}. On inspection, the limits in the coherent regime differ slightly, with the maximum operating frequency slightly higher for wires and inverters and lower for majority gates. In the transition regime, however, we observe significantly different results, particularly for Wire-5 and Maj-101. To understand this behaviour, we look in more detail at Wire-5. We discussed the oscillations that occur with the ICHA in \cref{sec: cvf}. These oscillations are clear in \cref{fig: wire-osc}(a), which shows the coherence vector for Wire-5 for a particular choice of $\runrate$. From \cref{eqn: icha-osc}, we can predict an oscillation frequency for {Wire-5} in non-dimensional units:
\begin{equation}
  f_{osc} = \tfrac{1}{2\pi\runrate}\brak{|\lmu{z}{4}|^2 + \atb_1^2}^{\sfrac{1}{2}}.
\end{equation}
Taking $|\lmu{z}{4}| \approx 1$, we get $f_{osc} \approx 8.0$ which matches the observed oscillation. This leads to oscillations in the performance both in the coherent limit (\cref{fig: wire-osc}(b)) as well as when dissipation is added (\cref{fig: wire-osc}(d)). In \cref{fig: wire-osc}(d), we see also a band of slightly lower performance along $\dissip \approx \runrate$. That this effect appears significant is entirely a consequence of our arbitrary choice of the 0.99 threshold. This region of lower performance also exists in \cref{fig: wire-osc}(c); however, we observe slightly higher $\Met_L$ values in the low $\runrate$ regime when using the density operator approach, keeping this region above 0.99.

\subsection{Mean Field Relaxation with the ICHA}

\begin{figure}
  \includegraphics[width=.65\linewidth]{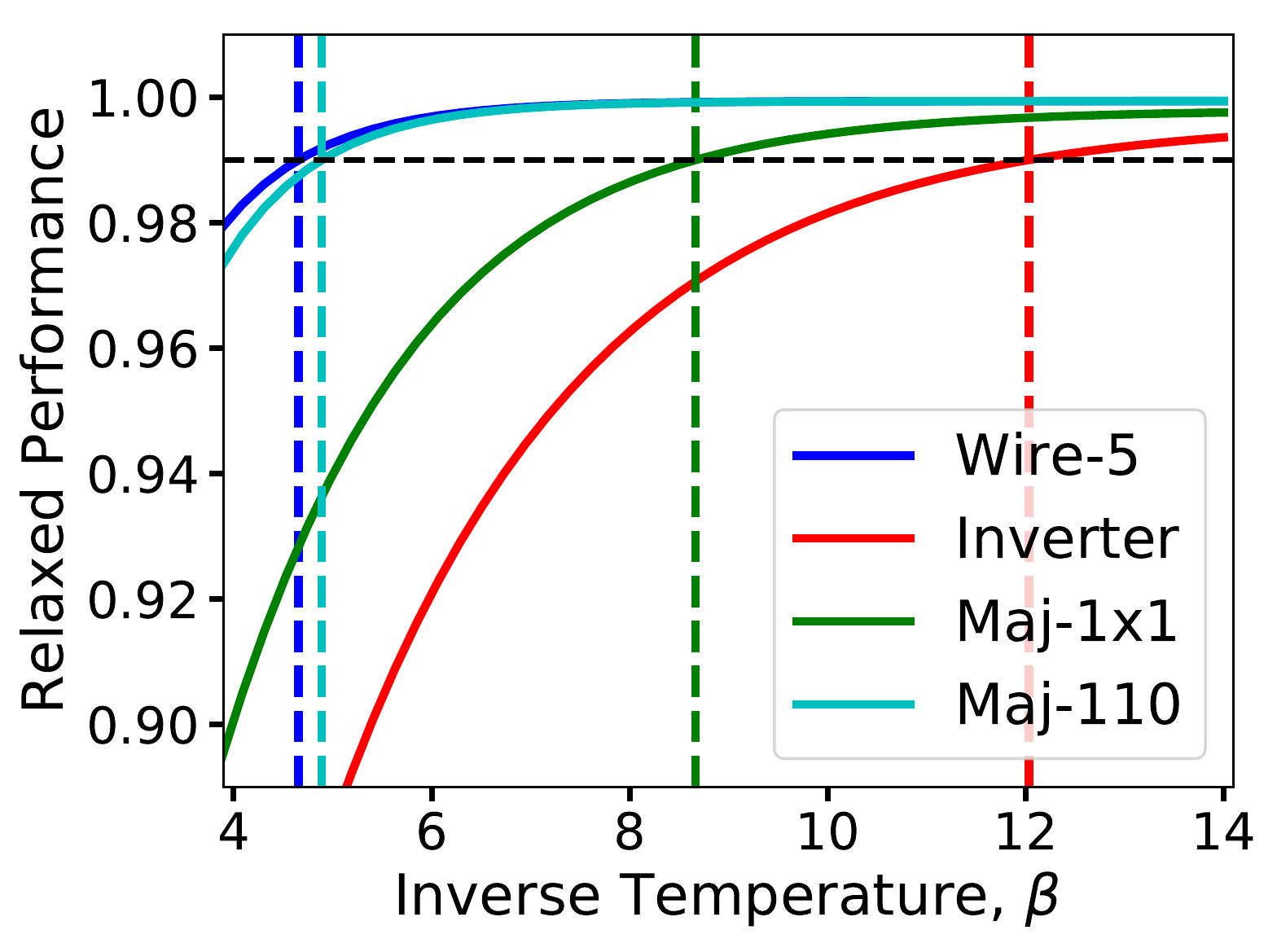}
  \caption{Logical performance in the relaxed regime for the mean field relaxation with the ICHA. Maj-101 and Maj-111 are indistinguishable for sufficiently large $\beta$.}
  \label{fig: meanlim}
\end{figure}

\begin{table}
  \caption{Comparison of $\beta^*$ values for the two temperature dependent steady states.}
  \label{table: mf-beta}
  \begin{ruledtabular}
  \begin{tabular}{c||c|c|c|c|c}
    Device & Wire-N & Inverter & Maj-111 & Maj-101 & Maj-110 \\
    \hline
    Boltzmann & 6.3 & 12.5 & 8.7 & 10.0 & 6.2\\
    Mean Field & 4.7 & 12.0 & 8.7 & 8.7 & 4.9
  \end{tabular}
  \end{ruledtabular}
\end{table}

We conclude by considering the commonly used mean field method: \cref{eq: mf-eta}. As this method is $\beta$ dependent it is natural to compare it with our analysis of the Boltzmann steady state. We can determine the $\beta^*$ sufficient to guarantee a high performing relaxed regime by making two observations: (1) in the limit of high $\dissip$ we have $\lmu{z}{n} = \eta_z^n$ and hence $\Met_L = \tfrac{1}{2} (1+ |\eta_z^n|)$ for output cell $n$; and (2), in this limit the $\bm{\Gamma_i}$ in  \cref{eq: mf-eta} becomes a function of the vector $\bm{\eta}_z$ of all the $\eta_z^i$ values. We can then obtain $\bm{\eta}_z$ as a fixed point of \cref{eq: mf-eta} by iteration starting with the ground state polarizations of $\HHd(1)$. The results of this process are shown in \cref{fig: meanlim} with a comparison of the extracted $\beta^*$ to those of the Boltzmann steady state shown in \cref{table: mf-beta}. With the exception of Maj-111, we get lower $\beta^*$ values using the mean field.

\begin{figure}
 \newcommand{\WW}{.48\linewidth}
 \subfloat[Wire-5]{\includegraphics[width=\WW]{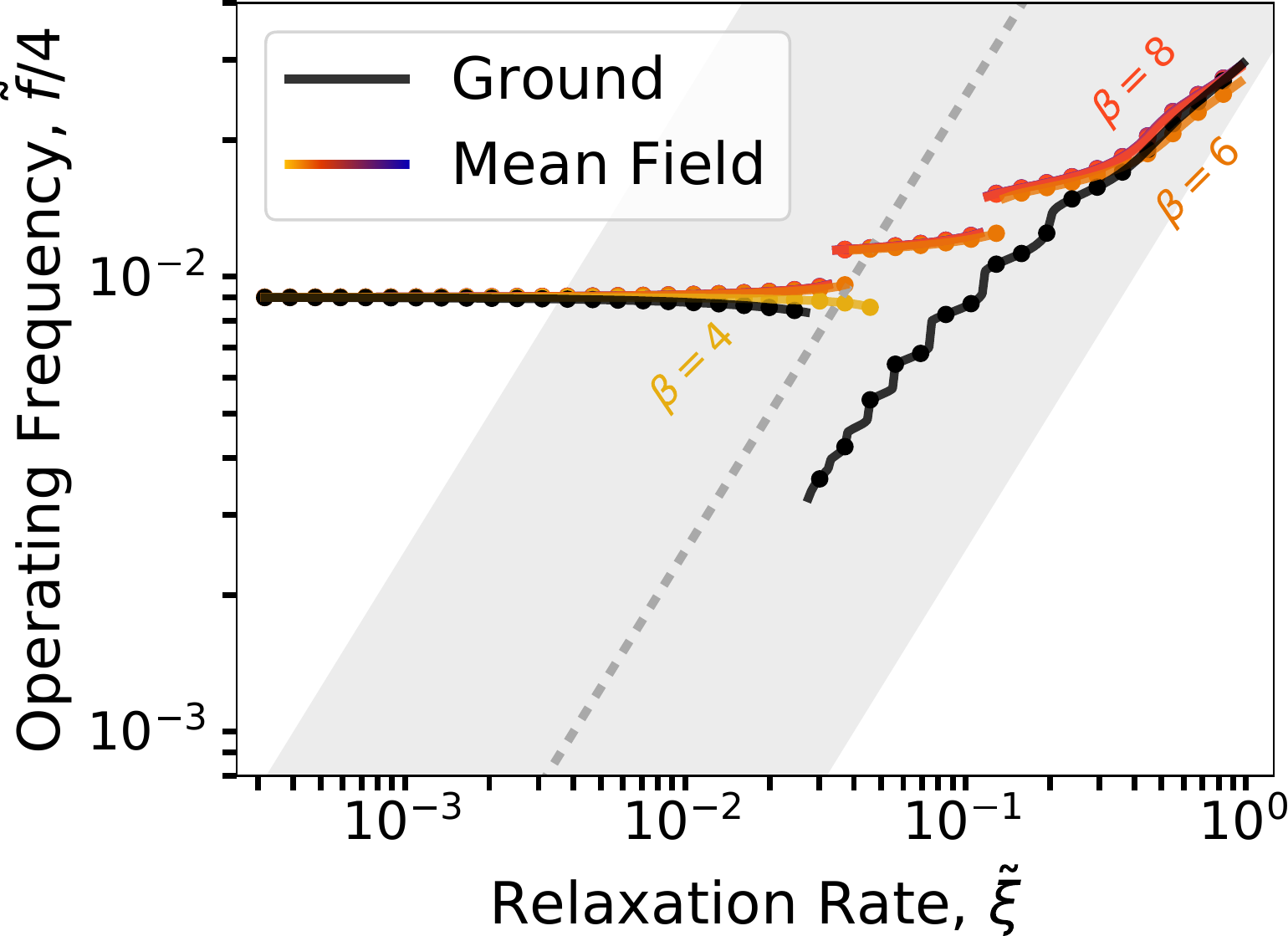}}\quad
 \subfloat[Inverter]{\includegraphics[width=\WW]{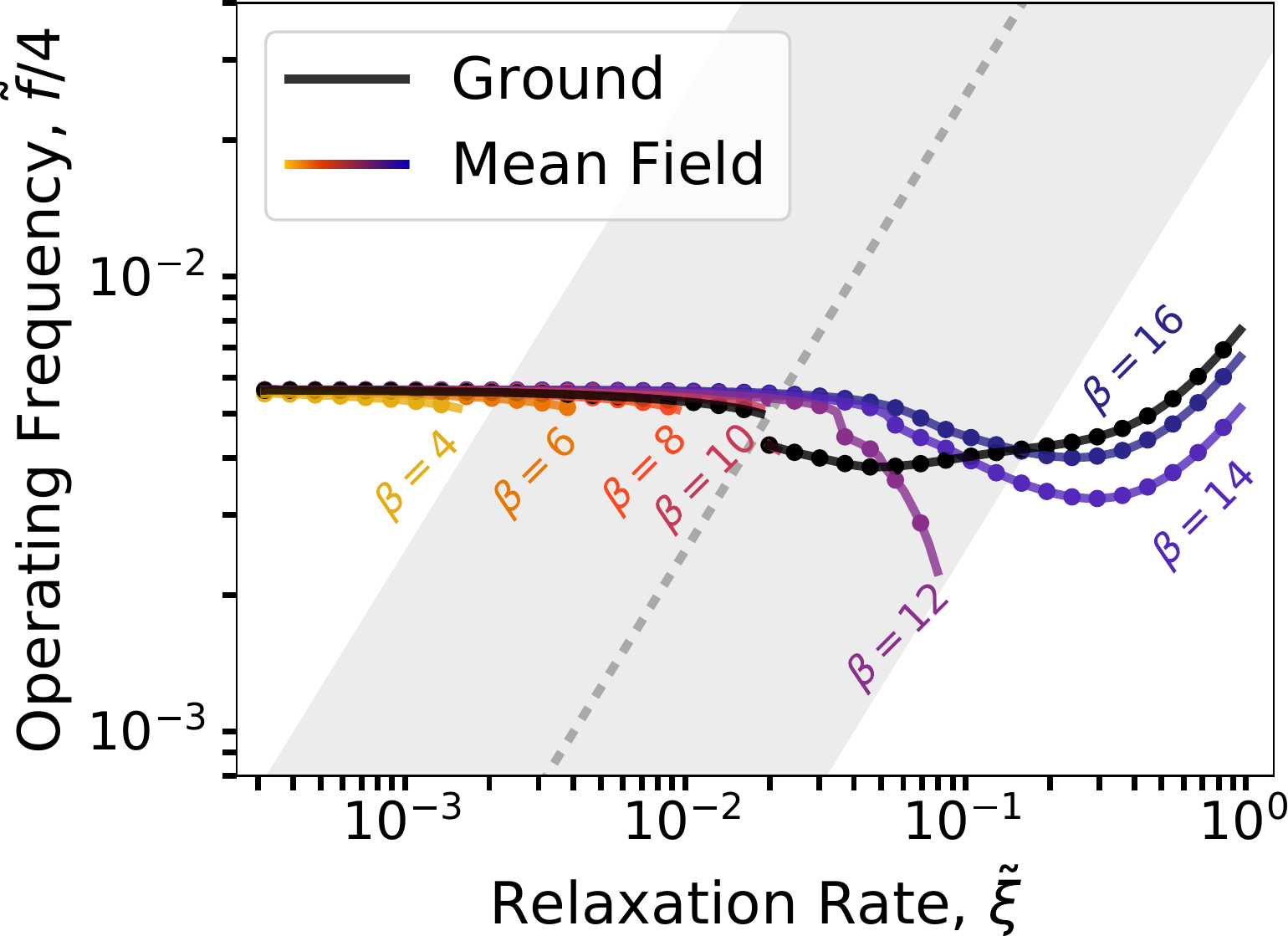}}\\
 \subfloat[Maj-111]{\includegraphics[width=\WW]{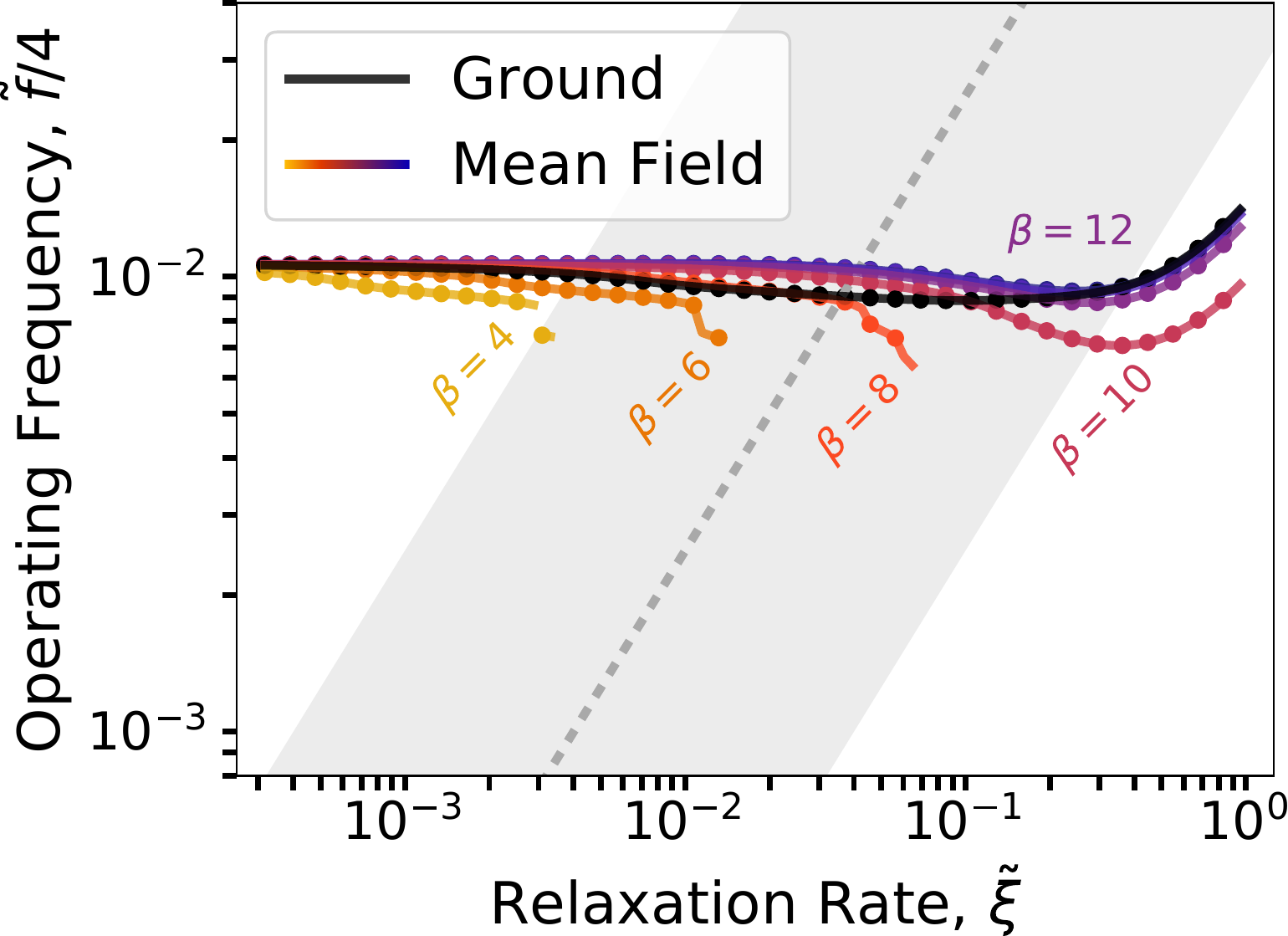}}\quad
 \subfloat[Maj-101]{\includegraphics[width=\WW]{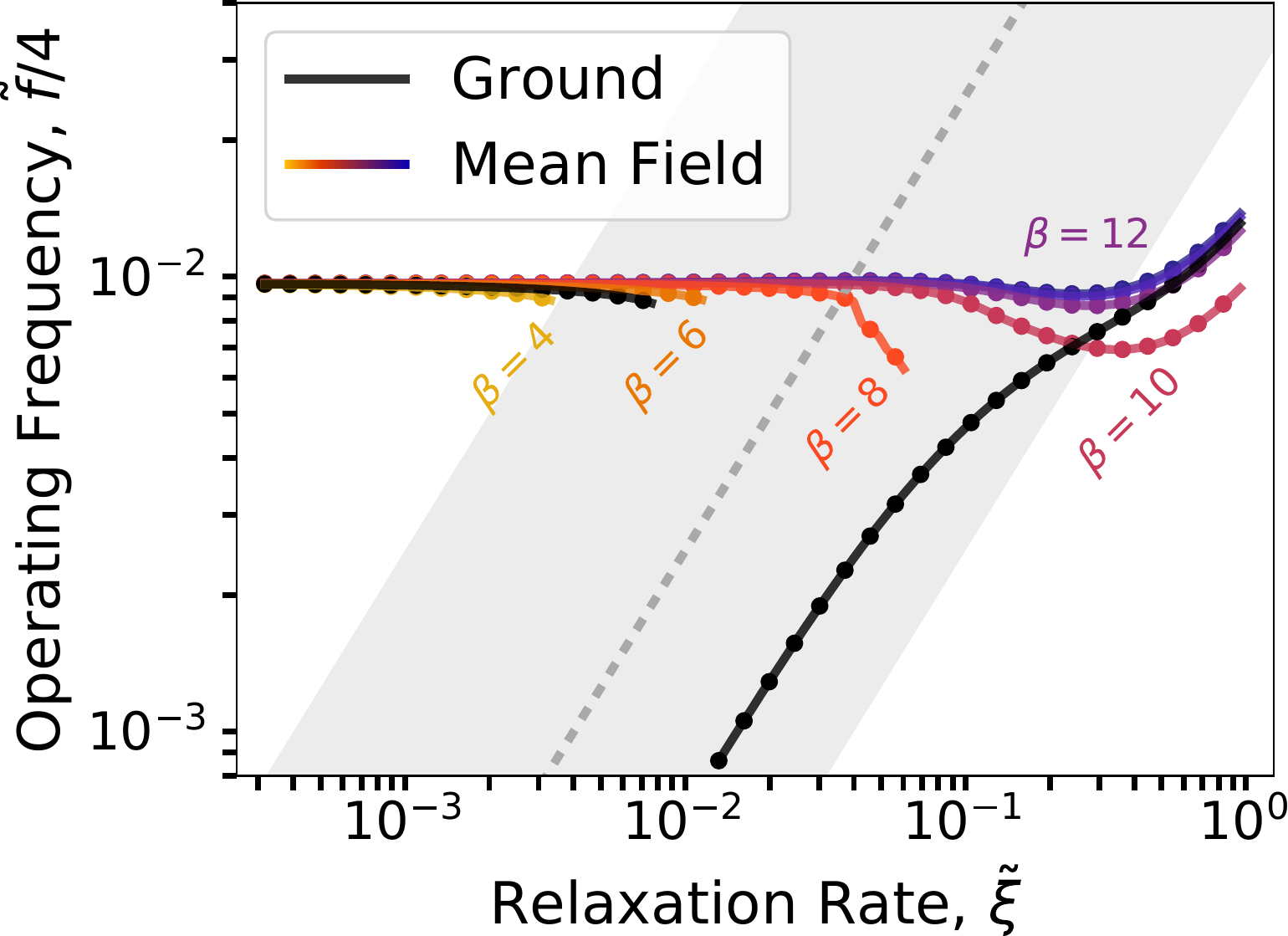}}\\
 \subfloat[Maj-110]{\includegraphics[width=\WW]{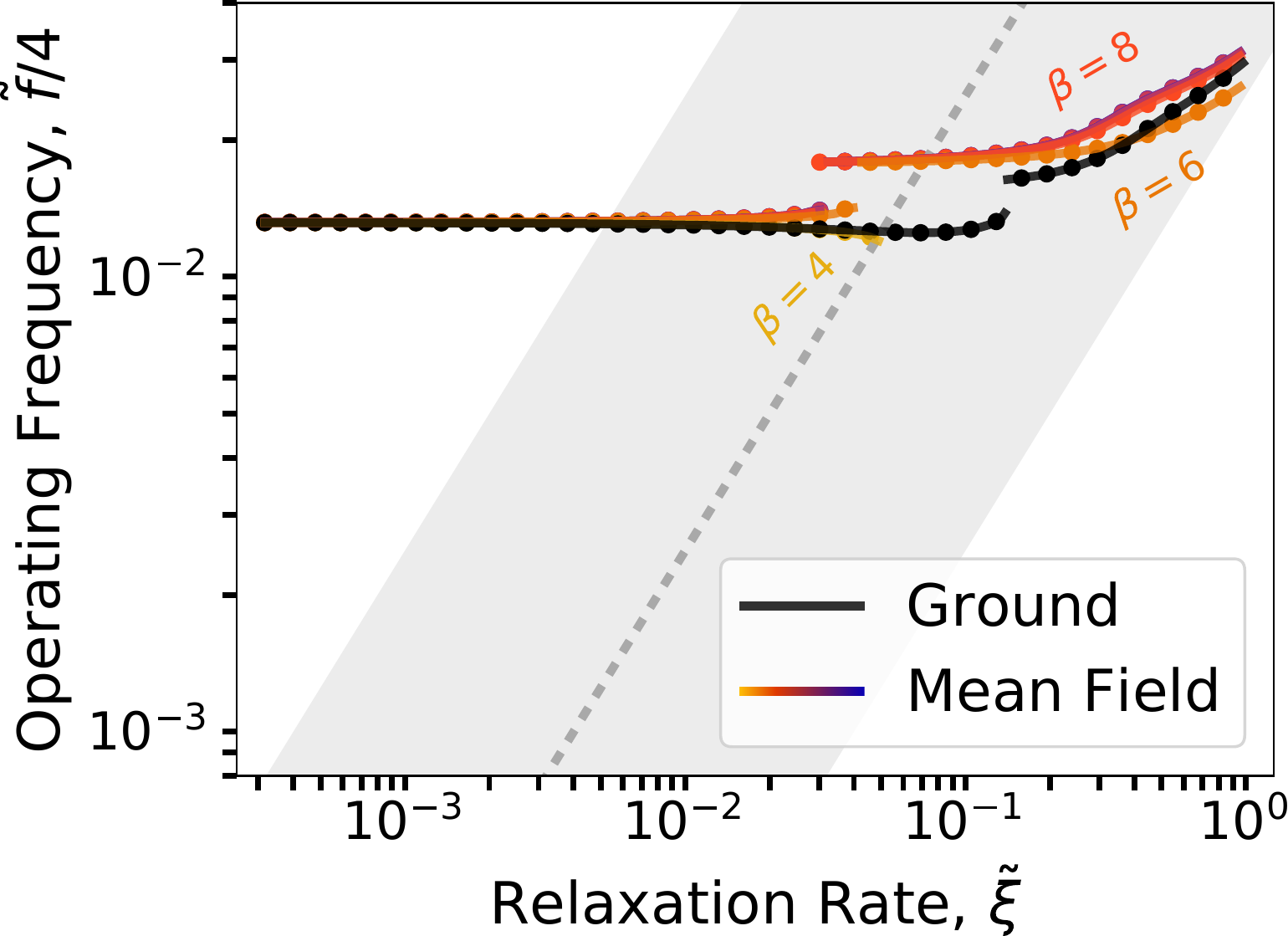}}
 \caption{Maximum operating frequencies for mean field relaxation using the ICHA. The results for the ground state are shown for reference and comparison to \cref{fig: icha-ss-performance}.}
 \label{fig: icha-mf-performance}
\end{figure}

In \cref{fig: icha-mf-performance}, we show the maximum operating frequencies for the mean field steady state. We use spectral relaxation to the ground state as reference. Qualitatively, there is much in common with the Boltzmann steady state, keeping in mind the slightly different $\beta^*$ transitions. One significant feature that was not observed in our previous results is that the mean field can give higher operating frequencies than the true ground state. This is likely a result of the tendency for the mean field to reinforce the polarizations of cells, increasing the output cell polarization and hence $\Met_L$. Though not discussed here, this effect is known to present challenges, for example, in majority gates with asymmetric inputs \cite{toth2001}.

\section{Conclusion}

Previous attempts to establish bounds on QCA clocking limit the operating frequency for two-state QCA wires to 1-3 orders of magnitude below the intrinsic frequency $f_0 = \sfrac{\Eps}{\hbar}$. We approached the issue from the perspective of achieving high device performance, defined by a 99\% likelihood of observing the correct output logic. By interpreting QCA clocking as quantum annealing, we arrive at an improved clocking schedule which allows for higher operating frequencies. In the coherent limit, we observe bounds 2-3 orders of magnitude below $f_0$ for a subset of the standard components used in QCA network design. Using an analytical solution for driven wires, we determine that wires are well described by a simple Landau-Zener model of adiabaticity, having a maximum operating frequency that falls off with the square of the wire length. This suggests an unforgiving trade-off between operating frequency and maximum clock zone size unless alternative decoherence mechanisms can be invoked\cite{blair2013}.

We investigated a simple thermal bath relaxation model for decoherence. While the choice in steady state does influence performance, the effect is relatively small unless the system is operated in a regime where the rate of relaxation dominates. Outside of this regime, we observe either of two cases. If the steady state has the correct logic in the relaxed regime, then we observe at worst an approximate factor of 2 decrease in maximum operating frequency compared to the coherent limit; otherwise, there is a maximum allowable relaxation rate beyond which high performance is not achievable. The bounds in this near-coherent regime are therefore within a factor of $~2$ of those in the coherent limit.

\begin{acknowledgments}
  We acknowledge the support of the Natural Sciences and Engineering Research Council of Canada (NSERC), [funding reference number STPGP 478838-15].
\end{acknowledgments}


\bibliography{\bibfile}

\end{document}